\documentclass[a4paper,12pt,colorlinks=true,urlcolor=blue,anchorcolor=blue,citecolor=blue,filecolor=blue,
               linkcolor=blue,menucolor=blue,linktocpage=true,pdfproducer=medialab,pagebackref=true]{article}
\pdfoutput=1
\usepackage{setspace}
\usepackage{jheppub_X}
\usepackage[T1]{fontenc}
\usepackage{graphicx}
\usepackage{amsmath,mathrsfs,booktabs}
\usepackage{amssymb}
\usepackage[dvipsnames,table]{xcolor}
\usepackage{bbm}
\usepackage{autobreak}
\usepackage{tensor}
\usepackage{bbold}
\usepackage{multicol}
\usepackage{multirow}
\usepackage{enumitem}
\usepackage{cleveref}
\crefname{table}{Table}{Tables}
\crefname{equation}{Eq.}{Eqs.}
\crefname{appendix}{App.}{Apps.}
\crefname{section}{Sec.}{Secs.}
\crefname{figure}{Fig.}{Figs.}
\usepackage[normalem]{ulem}
\usepackage{physics}
\usepackage{slashed}

\usepackage{bm}
\usepackage{setspace}
\usepackage{dsfont}
\usepackage{pifont}
\usepackage[most]{tcolorbox}
\definecolor{light-gray}{gray}{0.9}
\usepackage{orcidlink}
\usepackage[centertableaux]{ytableau}
\ytableausetup{smalltableaux}
\usepackage{longtable}
\usepackage{lieart}
\usepackage{cancel}
\usepackage{verbatim}

\newcommand{\svdots}{%
  \vbox{
    \scriptsize \baselineskip 2pt \lineskiplimit 0pt
    \hbox {.}\hbox {.}\hbox {.}\kern-0.75pt
  }%
}

\newcommand{\pd}{{\partial}}
\newcommand{\Lag}{{\mathcal{L}}}
\newcommand{\HS}{{\mathcal{H}}}
\newcommand{\I}{\text{Inv}}
\newcommand{\SPM}{R}
\newcommand{\Op}{Q}
\newcommand{\hc}{\text{h.c.}}
\newcommand{\adj}{\text{adj}}
\newcommand{\Irrep}{\text{Irrep}}
\newcommand{\tdim}{\text{dim}}
\newcommand{\highlight}[1]{{\color{brightpink} #1}}

\def\eg{\textit{e.g.}}
\def\ie{\textit{i.e.}}

\def\cf{\textit{c.f.}}

\definecolor{mygreen}{rgb}{0.0, 0.5, 0.0}

\definecolor{brightpink}{rgb}{1.0, 0.0, 0.5}

\preprint{}

\title{\Large Accidental Symmetries, Hilbert Series, and \emph{Friends}\\
}

\author[a]{Benjamín Grinstein\orcidlink{0000-0003-2447-4756},}
\author[a]{Xiaochuan Lu\orcidlink{0000-0001-8821-2574},}
\author[b]{Carlos Miró\orcidlink{0000-0003-0336-9025},}
\author[a]{and Pablo Quílez\orcidlink{0000-0002-4327-2706}}

\affiliation[a]{Department of Physics, University of California, San Diego, La Jolla, CA 92093, USA}
\affiliation[b]{Departament de Física Teòrica and Instituto de Física Corpuscular (IFIC),\\ Universitat de València -- CSIC, E-46100 Valencia, Spain}

\emailAdd{bgrinstein@ucsd.edu}
\emailAdd{xil224@ucsd.edu}
\emailAdd{carlos.miro@uv.es}
\emailAdd{pquilezlasanta@ucsd.edu}

\abstract{
Accidental symmetries in effective field theories can be established by computing and comparing Hilbert series. This invites us to study them with the tools of invariant theory.
Applying this technology, we spotlight three classes of accidental symmetries that hold to all orders for non-derivative interactions.
They are broken by derivative interactions and become ordinary finite-order accidental symmetries.
To systematically understand the origin and the patterns of accidental symmetries, we introduce a novel mathematical construct --- a (non-transitive) binary relation between subgroups that we call \textit{friendship}. Equipped with this, we derive new criteria for all-order accidental symmetries in terms of \emph{friends}, and criteria for finite-order accidental symmetries in terms of \emph{friends ma non troppo}. They allow us to verify and identify accidental symmetries more efficiently without computing the Hilbert series. We demonstrate the success of our new criteria by applying them to a variety of sample accidental symmetries, including the custodial symmetry in the Higgs sector of the Standard Model effective field theory.
}

\begin{document}
\maketitle
\flushbottom
\setcounter{page}{2}
\newpage
\begin{spacing}{1.1}
\parskip=0ex

\section{Introduction}
\label{sec:Introduction}

Symmetry principles play a crucial role in understanding the dynamics and physical predictions of Quantum Field Theories (QFTs). They have a broad application across many disciplines of theoretical and experimental physics. However, not all symmetries need to be fundamental or explicitly imposed. Some symmetries are respected \emph{accidentally} by the allowed interactions in the Lagrangian due to other constraints imposed. They are known as \emph{accidental symmetries}. From the Effective Field Theory (EFT) point of view, they typically hold to a certain order in the EFT expansion, and are broken by higher-order effective operators.

The most prominent examples of accidental symmetries in the Standard Model (SM) are perhaps the conservation of baryon number $B$ and of lepton number $L$ \cite{Weinberg:1979sa}. They are not imposed as requirements on the SM Lagrangian. However, given the SM field content and insistence on preserving the SM symmetries (the gauged $SU(3)_c\times SU(2)_L \times U(1)_Y$ together with Lorentz symmetry), no renormalizable interactions (\ie, operators up to mass dimension four) can be constructed to violate $B$ or $L$. In the language of the SM EFT (SMEFT), interactions (effective operators) that violate $L$ do not arise until mass dimension five \cite{Weinberg:1979sa}, and those violating $B$ do not arise until mass dimension six \cite{Weinberg:1979sa, Buchmuller:1985jz, Grzadkowski:2010es}. The $B$ and $L$ accidental symmetries have important phenomenological consequences in SMEFT, such as explaining the suppression of the proton decay \cite{ParticleDataGroup:2024cfk}.

Another well-known accidental symmetry in the SM and SMEFT is the custodial symmetry in the Higgs sector, where imposing the electroweak symmetry $SU(2)_L\times U(1)_Y$ leads to an enhanced symmetry $O(4) \simeq SU(2)_L \times SU(2)_R$ that is preserved by the renormalizable interactions of the Higgs fields, and broken by their derivative interactions starting at mass dimension six.\footnote{Custodial symmetry is often referred to as an \emph{approximate} accidental symmetry in the SM and SMEFT, because it is also broken by the Yukawa interactions and the gauging of the hypercharge $Y$, which are interactions at the leading order in the EFT. In this work we make a distinction between ``approximate''  and ``finite-order'' accidental symmetries, where the latter refers to symmetries broken by EFT operators \emph{beyond} the leading order.}
Custodial symmetry has been crucial in understanding some key aspects of the SM phenomenology, such as the relationship between the masses of the electroweak gauge bosons, $m_W^{}$ and $m_Z^{}$ \cite{PhysRevD.19.1277}, and the suppression of the $\rho-1$ and $T$ parameters observed in  electroweak precision tests \cite{ParticleDataGroup:2024cfk}.

Generally speaking, due to the properties of the fields and other restrictions (such as Lorentz invariance and field redefinition redundancies), sometimes imposing a smaller symmetry group $H$ can lead to a Lagrangian that respects a larger symmetry group $G$, up to an order $k$ in the EFT expansion:
\begin{equation}
H \implies G
\quad\text{for}\quad
\Lag_\text{\,EFT}
\quad\text{up to order}\; k \,.
\label{eqn:AccidentalSym}
\end{equation}
When this happens, we call $G$ an accidental symmetry obtained by imposing $H$.\footnote{
Strictly speaking, only the $G/H$ part in $G$ is ``accidental'', \ie\ preserved without explicitly imposed. However, as the cosets $G/H$ often do not form a group, for simplicity of the language, in this paper we will also use the term ``accidental symmetry'' to refer to the group $G$.}
Are there more accidental symmetries in the SM? How do we generally detect an accidental symmetry in a given EFT? How do we systematically find them all? Identifying accidental symmetries in a given theory can offer deep insights into its structure and phenomenology, so it is desired to have a systematic approach to this problem in general.

To establish an accidental symmetry in \cref{eqn:AccidentalSym}, one can enumerate the effective operators in the EFT, imposing the requirement of $G$-invariance and $H$-invariance respectively, and check if they agree up to the order $k$. A more systematic and automatic way is to compute the Hilbert series, a partition function that encodes all the effective operators in an EFT. Hilbert series is a well-established tool in invariant theory (see \eg\ \cite{Sturmfels:inv, Popov1994, bruns1998cohen}). The technology was initially introduced to particle physics to enumerate non-derivative interactions \cite{Benvenuti:2006qr, Feng:2007ur, Gray:2008yu, Jenkins:2009dy, Hanany:2010vu}, and later generalized to accommodate derivative interactions under linearly realized symmetries \cite{Lehman:2015via, Henning:2015daa, Lehman:2015coa, Henning:2015alf, Henning:2017fpj}, non-linearly realized global symmetries \cite{Henning:2017fpj, Graf:2020yxt}, and non-linearly realized gauge symmetries that are spontaneously broken \cite{Graf:2022rco}. It has also been generalized recently to handle the enumeration of group covariants \cite{Grinstein:2023njq}. With these technical advancements, Hilbert series has now been extensively applied to a variety of EFTs in particle physics \cite{Lehman:2015coa, Henning:2015alf, Liao:2016qyd, Kobach:2017xkw, Kobach:2018pie, Ruhdorfer:2019qmk, Marinissen:2020jmb, Wang:2021wdq, Yu:2021cco, Sun:2022aag, Kondo:2022wcw, Delgado:2022bho, Chang:2022crb, Bijnens:2022zqo, Delgado:2023ivp, Grojean:2023tsd, Grinstein:2023njq, Grojean:2024qdm}. In this paper, we show how to make use of Hilbert series, together with more advanced tools in invariant theory, to verify and identify accidental symmetries in a more systematic manner.

We start by highlighting a special type of accidental symmetries in \cref{sec:AllOrder} --- they hold to all orders in the EFT expansion when derivative interactions are not considered. The custodial symmetry in the Higgs sector belongs to this type: the Higgs potential $V(H, H^\dagger)$ in SMEFT preserves it to all orders. In \cref{sec:AllOrder}, we show three classes of such all-order accidental symmetries, each verified by computing and comparing the Hilbert series. Among these, one can recognize the custodial symmetry as a special case of Class II (see \cref{subsec:ClassII,subsec:Custodial}). The capability of verifying all-order accidental symmetries demonstrates the power of the Hilbert series approach --- it would be difficult to draw an all-order conclusion by a manual enumeration of effective operators.

In \cref{sec:FiniteOrder}, we move on to address ordinary accidental symmetries, which hold only up to a certain order $k$ in the EFT, as in \cref{eqn:AccidentalSym}. At higher orders, they are broken by derivative and/or non-derivative interactions. In any case, they can be established by comparing the Hilbert series. Remarkably, the all-order accidental symmetries highlighted in \cref{sec:AllOrder} may still hold up to a quite high order $k$, as even lower-order derivative interactions could preserve them.

Establishing an accidental symmetry by comparing Hilbert series is much more efficient than checking it by manually enumerating the effective operators. However, calculating a Hilbert series can be quite difficult and time consuming, still. In \cref{sec:Systematic}, we develop alternative systematic approaches to identify and verify accidental symmetries for bosonic fields, without need of calculating the Hilbert series. We introduce a new mathematical construct --- a (non-transitive) binary relation between subgroups that we call \emph{friendship}; see \cref{eqn:Friend}. Making use of this together with a theorem by Brion \cite{brion1993modules, broer1994hilbert, Grinstein:2023njq} in invariant theory, we derive new criteria for accidental symmetries. Specifically, an all-order accidental symmetry is equivalent to a friendship relation between two certain subgroups (specified in \cref{eqn:AllOrderCriterion}), while a finite-order accidental symmetry up to order $k$ is equivalent to the condition that their friendship relation only gets broken by operators of order higher than $k$ (see \cref{eqn:FiniteOrderCriterion,eqn:FiniteOrderCriterionGeneral}). In the latter case, the two subgroups are almost friends, or \emph{friends ma non troppo}.

Finally, we discuss interesting future directions for technology developments and potential phenomenological applications in \cref{sec:Outlook}. Substantial details of Hilbert series calculations are gathered in \cref{appsec:ComputingHS}.

\section{All-order accidental symmetries}
\label{sec:AllOrder}

In this section, we highlight a special type of accidental symmetries --- they hold to all orders in the EFT expansion when derivative interactions are not considered. We start by making the definition precise.

Consider a global group $G$ (which may be a product group in general) and a set of fields $\Phi(x)$ that transform as a linear representation (rep) under $G$:
\begin{equation}
\Phi \quad\longrightarrow\quad
g\, \Phi \,, \qquad
g\in G \,.
\end{equation}
We refer to the fields $\Phi(x)$ as \emph{building block fields} in this paper. Without considering derivative interactions, we are focusing on the potential part of the EFT Lagrangian
\begin{equation}
\Lag_\text{\,EFT} = -V(\Phi) + \order{\partial_\mu} \,,
\label{eqn:LagEFT}
\end{equation}
where the potential $V(\Phi)$ is an arbitrary polynomial in the fields $\Phi(x)$.\footnote{Operators built out of fermionic fields are also included in the potential, as long as they do not contain derivatives. Therefore, the polynomial algebra in $\Phi$ here is in general a Grassmann algebra; some of the $\Phi$ components might be Grassmann odd.}
Now, as opposed to imposing the full group $G$, let us impose only a subgroup of it $H\subset G$, \ie\ we only require the theory to respect $H$-invariance. Sometimes, however, due to the special properties of the groups $G$, $H$, and the building block fields $\Phi$, imposing the subgroup $H$ will automatically guarantee that the potential $V(\Phi)$ is $G$-invariant. When this happens, we say that $G$ is an  accidental symmetry of the potential $V(\Phi)$ obtained by imposing $H$:\footnote{In relativistic EFTs, Lorentz invariance is a default symmetry requirement on the EFT potential $V(\Phi)$ (as well as the full Lagrangian $\Lag_\text{\,EFT}$). For simplicity, and to keep our discussions to the point, in this section we will focus on the building block fields $\Phi(x)$ that are Lorentz scalars (even in the several ``fermionic'' cases; see explanations below). In such cases, Lorentz invariance is trivially satisfied. When it becomes non-trivial, \eg\ when derivative interactions are accommodated, it will be included as an additional symmetry factor on top of the groups $G$ and $H$; see our discussions in later sections, \eg\ \eqref{eqn:HSAccidentalLagk}.}
\begin{equation}
H \implies G
\quad\text{for}\quad
V(\Phi) \,.
\end{equation}
Since this holds for a potential that is a polynomial of arbitrary degree, we refer to this as an ``all-order'' accidental symmetry of the potential in the EFT. 

All-order accidental symmetries can be verified by computing the corresponding Hilbert series that encode all the allowed effective operators in the EFT. Specifically, we have the equivalence between the following two statements
\begin{tcolorbox}[colback=light-gray]
\begin{center}
\begin{minipage}{5.5in}\vspace{-5pt}
\begin{equation}
\hspace{-10pt}
H \implies G
\quad\text{for}\quad
V(\Phi)
\qquad\Longleftrightarrow\qquad
\HS_\I^{G,\, \Phi} (q) = \HS_\I^{H,\, \Phi} (q)  \,.
\label{eqn:HSAccidentalV}
\end{equation}
\end{minipage}
\end{center}
\end{tcolorbox}
\noindent
We refer the reader to Ref.~\cite{Grinstein:2023njq} for a review on the definition, applications, and calculation techniques of the Hilbert series. Throughout this paper, we follow the same notation system used in Ref.~\cite{Grinstein:2023njq}. In particular, $\HS_\I^{G, \Phi}(q)$ denotes the Hilbert series for the $G$-invariant $\Phi$ polynomials (with the unified grading variable $q$), while $\HS_\I^{H, \Phi}(q)$ refers to the $H$-invariant $\Phi$ polynomials. The right-hand side is the statement that all polynomials in $\Phi$ that are $H$-invariant are also $G$-invariant, so the equivalence in \cref{eqn:HSAccidentalV} is basically a tautology, but it invites us to use the language of invariant theory \cite{Sturmfels:inv, Popov1994, bruns1998cohen} in addressing the question of accidental symmetries.

For the rest of this section, we show three classes of examples of all-order accidental symmetries for the potential term in the EFT Lagrangian. We verify each of them using \cref{eqn:HSAccidentalV} and computing the Hilbert series explicitly. We emphasize that this is by no means an exhaustive classification. On the contrary, these are just a few examples that we have been able to identify, and we suspect there are many more.

\subsection{Class I: $SU(N) \implies SU(N)\times U(1)$}
\label{subsec:ClassI}

We begin with a simple class of all-order accidental symmetries, which are summarized in \cref{tab:ClassI}. In this class, we have $G=SU(N)\times U(1)$ and the subgroup $H\subset G$ is the $SU(N)$ factor in $G$. The building block fields $\Phi(x)$ are bosonic fields. They consist of two irreducible representations (irreps) under the group $G$: $\Phi=(\phi, \phi^*)$, where $\phi$ transforms as fundamental under the $SU(N)$ factor with charge $+1$ under the $U(1)$ factor, while $\phi^*$ is the complex conjugate of $\phi$. The $H$ irrep components of $\Phi(x)$ follow straightforwardly: we have $\varphi=\phi$ transforming as a fundamental irrep under $H$ and $\varphi^*=\phi^*$ transforming as an anti-fundamental irrep. Note that we have introduced the new names $(\varphi, \varphi^*)$ to emphasize that they are viewed as $H$ irreps.

\begin{table}[t]
\renewcommand{\arraystretch}{1.3}
\setlength{\arrayrulewidth}{.2mm}
\setlength{\tabcolsep}{1.25em}
\centering
\ytableausetup{boxsize=0.5em}
\begin{tabular}{ccc}
\toprule
Class I & $G=SU(N)\times U(1)$ & $H=SU(N)$ \\
\midrule
\multirow{2}{*}{$\Phi(x)$}
& $\phi\; \;\sim\; (\ydiagram{1} \;,\; +1)$ & $\varphi\; \;\sim\; \ydiagram{1}$ \\
& $\phi^* \;\sim\; (\overline{\ydiagram{1}} \;,\; -1)$ & $\varphi^* \;\sim\; \overline{\ydiagram{1}}$ \\
\midrule
Primary Invariants & $\phi^\dagger \phi$ & $\varphi^\dagger \varphi$ \\
\bottomrule
\end{tabular}
\caption{All-order accidental symmetry Class I. $G=SU(N)\times U(1)$ and the subgroup $H\subset G$ is the $SU(N)$ factor in $G$. The building block fields $\Phi(x)$ are bosonic fields. Under the group $G$, they consist of two irreps: $\Phi=(\phi, \phi^*)$, where $\phi$ transforms as a fundamental rep with charge $+1$, and $\phi^*$ is its complex conjugate; there is a single primary invariant $\phi^\dagger \phi$. Under the subgroup $H$, $\Phi=(\varphi, \varphi^*)$, a fundamental irrep and its complex conjugate; there is a single primary invariant $\varphi^\dagger \varphi$.}
\label{tab:ClassI}
\end{table}

From the set up in \cref{tab:ClassI}, we see that all $H$-invariant polynomials are generated by the combination $\varphi^\dagger \varphi$; this is called a primary invariant in the context of invariant theory \cite{Grinstein:2023njq}. On the other hand, note that we have
\begin{equation}
\varphi^\dagger \varphi = \phi^\dagger \phi \,,
\label{eqn:PrimaryI}
\end{equation}
which is also a $G$-invariant combination. Therefore, all $H$-invariant polynomials in $\Phi$ are guaranteed to be $G$-invariant: the extra $U(1)$ factor in $G$ is an accidental symmetry that holds to all orders in the EFT potential.

To verify this class of all-order accidental symmetries, we can compute the Hilbert series for the $G$-invariant and $H$-invariant polynomials in $\Phi$ respectively, and compare them. Assigning each irrep a distinct grading variable (denoted by the same symbols as the fields, in slight abuse of notation), we obtain the multi-graded Hilbert series as
\begin{subequations}\label{eqn:HSIDistinct}
\begin{align}
\HS_\I^{SU(N)\times U(1),\, (\phi,\, \phi^*)} (\phi,\, \phi^*) &= \frac{1}{1 - \phi \phi^*} \,, \label{eqn:HSIG} \\[5pt]
\HS_\I^{SU(N),\, (\varphi,\, \varphi^*)} (\varphi,\, \varphi^*) &= \frac{1}{1 - \varphi \varphi^*} \,. \label{eqn:HSIH}
\end{align}
\end{subequations}
Details of computing these Hilbert series are presented in \cref{appsubsubsec:ClassISingleBoson}. Recall that $\phi=\varphi$ and $\phi^*=\varphi^*$ are for the same fields named differently, so the above two Hilbert series agree. We can present this agreement in the form of \cref{eqn:HSAccidentalV} with a unified grading scheme $\phi = \phi^* = \varphi = \varphi^* = q$:
\begin{equation}
\HS_\I^{SU(N)\times U(1),\, (\phi,\, \phi^*)} (q)
= \HS_\I^{SU(N),\, (\varphi,\, \varphi^*)} (q)
= \frac{1}{1-q^2} \,.
\label{eqn:HSIAgree}
\end{equation}
We emphasize that while one can usually obtain an intuitive understanding of an all-order accidental symmetry by enumerating and comparing the primary invariants, as we did above in \cref{eqn:PrimaryI}, a more reliable way is to verify it by computing and comparing the Hilbert series.
After all, relying on intuitive understanding can run afoul of, for example,  misidentification of primary or secondary invariants.

\subsubsection*{Fermionic case}

For this class of examples, it is important that the building block fields $\Phi(x)$ are bosonic. Consider the case of fermionic building blocks\footnote{Note that, in this context, by fermionic field we just mean Grassmann odd, \ie\, anti-commuting; we are not enlarging them into spinors under the Lorentz group.}
and, to clearly distinguish them, use the following notation:
\begin{equation}
\Phi(x) \;\longrightarrow\; \Psi(x) \,,\qquad
(\phi,\, \phi^*) \;\longrightarrow\; (\psi,\, \psi^*) \,,\qquad
(\varphi,\, \varphi^*) \;\longrightarrow\; (\chi,\, \chi^*) \,.
\end{equation}
The two Hilbert series will become finite polynomials, which we expect to be
\begin{subequations}\label{eqn:HSIFermionDisagree}
\begin{align}
\HS_\I^{SU(N)\times U(1),\, (\psi,\, \psi^*)}(q) &= 1 + q^2 + q^4 + \cdots + q^{2N} \,, \label{eqn:HSIFermionG} \\[5pt]
\HS_\I^{SU(N),\, (\chi,\, \chi^*)} (q) &= 1 + q^2 + q^4 + \cdots + q^{2N} + 2q^N \,. \label{eqn:HSIFermionH}
\end{align}
\end{subequations}
They are no longer equal. The extra terms $2q^N$ in the $H=SU(N)$ case correspond to the baryon-like $SU(N)$ invariants:
\begin{subequations}\label{eqn:PrimaryIBaryon}
\begin{align}
\chi_{[1}^{} \chi_2^{} \cdots \chi_{N]}^{}
&= \frac{1}{N!}\, \epsilon^{i_1 i_2 \cdots i_N}\, \chi_{i_1}^{} \chi_{i_2}^{} \cdots \chi_{i_N}^{} \,, \\[5pt]
\chi_{[1}^* \chi_2^* \cdots \chi_{N]}^*
&= \frac{1}{N!}\, \epsilon^{i_1 i_2 \cdots i_N}\, \chi_{i_1}^* \chi_{i_2}^* \cdots \chi_{i_N}^* \,,
\end{align}
\end{subequations}
which are non-vanishing when $(\chi,\,\chi^*)$ are fermionic fields,\footnote{These baryon-like combinations are Grassmann odd for odd $N$; they can be forbidden by imposing an additional $\mathbb{Z}_2$ symmetry. In the cases of actual EFTs, where the fermionic building block fields are Lorentz spinors, these combinations are not Lorentz scalars and cannot be in the potential interactions.}
 and they would vanish for bosonic fields.
We also note that apart from the extra $2q^N$ terms, \cref{eqn:HSIFermionDisagree} is a truncation of \cref{eqn:HSIAgree} due to the fermionic nature of the building blocks. We will come back to discuss this in \cref{sec:Systematic}.

\subsubsection*{Multiple flavors}

Our Class I all-order accidental symmetries can be generalized to the case of multiple flavors, that is, the case that the building block includes several fields, all transforming according to the same  representations that appear in the single flavor case above, as summarized in \cref{tab:ClassIk1k2}. Under the group $G = SU(N)\times U(1)$, the building block fields $\Phi(x)$ consist of $k_1<N$ flavors of fundamental irreps with charge $+1$ ($\phi_i$ with $i=1,2,\cdots, k_1$), and $k_2<N$ flavors of anti-fundamental irreps with charge $-1$ ($\phi_j^*$ with $j=1,2,\cdots, k_2$). The corresponding $H$ irreps are $\varphi_i = \phi_i$ (fundamental irreps) and $\varphi_j^* = \phi_j^*$ (anti-fundamental irreps). All $H$-invariant polynomials in $\Phi$ are generated by the primary invariants
\begin{equation}
\varphi_j^\dagger \varphi_i = \phi_j^\dagger \phi_i \,,
\label{eqn:PrimaryIk1k2}
\end{equation}
which are also $G$-invariant combinations. Again, we get the extra $U(1)$ factor in $G$ as an all-order accidental symmetry without imposing it.

\begin{table}[t]
\renewcommand{\arraystretch}{1.3}
\setlength{\arrayrulewidth}{.2mm}
\setlength{\tabcolsep}{1.25em}
\centering
\ytableausetup{boxsize=0.5em}
\begin{tabular}{ccc}
\toprule
Class I, Multiple Flavors $(k_1,k_2)$ & $G=SU(N)\times U(1)$ & $H=SU(N)$ \\
\midrule
\multirow{2}{*}{$\Phi(x)$}
& $\phi_i\; \;\sim\; (\ydiagram{1} \;,\; +1)$ & $\varphi_i\; \;\sim\; \ydiagram{1}$ \\
& $\phi_j^* \;\sim\; (\overline{\ydiagram{1}} \;,\; -1)$ & $\varphi_j^* \;\sim\; \overline{\ydiagram{1}}$ \\
\midrule
Primary Invariants & $\phi_j^\dagger \phi_i$ & $\varphi_j^\dagger \varphi_i$ \\
\bottomrule
\end{tabular}
\caption{All-order accidental symmetry Class I with multiple flavors. $G=SU(N)\times U(1)$ and the subgroup $H\subset G$ is the $SU(N)$ factor in $G$. The building block fields $\Phi(x)$ are bosonic fields. Under the group $G$, they consist of $k_1<N$ flavors of $\phi_i$ (the fundamental irrep with charge $+1$) and $k_2<N$ flavors of the complex conjugate irrep $\phi_j^*$. The primary invariants are $\phi_j^\dagger \phi_i$ with $i=1,2,\cdots,k_1$ and $j=1,2,\cdots,k_2$. Under the subgroup $H$, $\Phi$ consist of $k_1$ flavors of the fundamental irrep $\varphi_i$ and $k_2$ flavors of the anti-fundamental irrep $\varphi_j^*$. The primary invariants are $\varphi_j^\dagger \varphi_i$.}
\label{tab:ClassIk1k2}
\end{table}

One can verify the multiple flavor case of Class I all-order accidental symmetries by computing the corresponding Hilbert series. Assigning each irrep a distinct grading variable, we obtain the multi-graded Hilbert series as\footnote{The computation of \cref{eqn:HSIk1k2G} is shown in \cref{appsubsubsec:ClassIMultiple}. The result in  \cref{eqn:HSIk1k2H} is expected, which we will prove with an indirect approach in \cref{subsubsec:FriendsClassI}.}
\begin{subequations}\label{eqn:HSIk1k2Distinct}
\begin{align}
\HS_\I^{SU(N)\times U(1),\, (k_1\times \phi,\, k_2\times \phi^*)} (\phi_i, \phi_j^*) &= \prod_{i=1}^{k_1} \prod_{j=1}^{k_2} \frac{1}{ 1 - \phi_i \phi_j^*}
\qquad\text{for}\qquad
k_1 \,,\, k_2 < N \,, \label{eqn:HSIk1k2G} \\[5pt]
\HS_\I^{SU(N),\, (k_1\times \varphi,\, k_2\times \varphi^*)} (\varphi_i, \varphi_j^*) &= \prod_{i=1}^{k_1} \prod_{j=1}^{k_2} \frac{1}{ 1 - \varphi_i \varphi_j^*}
\qquad\text{for}\qquad
k_1 \,,\, k_2 < N \,. \label{eqn:HSIk1k2H}
\end{align}
\end{subequations}
Now unifying the grading variables $\phi_i = \phi_j^* = \varphi_i = \varphi_j^* = q$, \cref{eqn:HSIk1k2Distinct} is put in the form of \cref{eqn:HSAccidentalV}:
\begin{equation}
\HS_\I^{SU(N)\times U(1),\, (k_1\times \phi,\, k_2\times \phi^*)} (q)
= \HS_\I^{SU(N),\, (k_1\times \varphi,\, k_2\times \varphi^*)} (q)
= \frac{1}{(1-q^2)^{k_1k_2}} \,.
\label{eqn:HSIkik2Agree}
\end{equation}
This is the multiple flavor generalization of the result in \cref{eqn:HSIAgree}.

Our Class I all-order accidental symmetries stop working if $k_1\ge N$ and/or $k_2\ge N$, due to the appearance of the baryon-like $H=SU(N)$ invariants, \ie\, fully antisymmetric contractions of $N$ fundamentals (or anti-fundamentals) with an $\epsilon$ tensor. These are the same combinations given in \cref{eqn:PrimaryIBaryon}, but using multiple flavors of bosonic fields instead of a single flavor of fermionic field; they arise when we have $k_1\ge N$ or $k_2\ge N$. This is also reflected by the Hilbert series. For example, when $k_1=N$ and $k_2<N$, we expect\footnote{\cref{eqn:HSINk2G,eqn:HSINNG} are derived in \cref{appsubsubsec:ClassIMultiple}. For \cref{eqn:HSINk2H,eqn:HSINNH}, we have verified them up to $N\le 6$.}
\begin{subequations}\label{eqn:HSINk2Distinct}
\begin{align}
\HS_\I^{SU(N)\times U(1),\, (N\times \phi,\, k_2\times \phi^*)} (\phi_i, \phi_j^*) &= \left( \prod_{i=1}^{N} \prod_{j=1}^{k_2} \frac{1}{ 1 - \phi_i \phi_j^*} \right) \,, \label{eqn:HSINk2G} \\[5pt]
\HS_\I^{SU(N),\, (N\times \varphi,\, k_2\times \varphi^*)} (\varphi_i, \varphi_j^*) &= \left( \prod_{i=1}^{N} \prod_{j=1}^{k_2} \frac{1}{ 1 - \varphi_i \varphi_j^*} \right) \frac{1}{1 - \varphi_1 \cdots  \varphi_N} \,. \label{eqn:HSINk2H}
\end{align}
\end{subequations}
While for $k_1=k_2=N$, we expect
\begin{subequations}\label{eqn:HSINNDistinct}
\begin{align}
\hspace{-10pt}
\HS_\I^{SU(N)\times U(1),\, N\times (\phi,\, \phi^*)} (\phi_i, \phi_j^*) &= \prod_{i,j=1}^N \frac{1}{ 1 - \phi_i \phi_j^*} \,, \label{eqn:HSINNG} \\[5pt]
\HS_\I^{SU(N),\, N\times (\varphi,\, \varphi^*)} (\varphi_i, \varphi_j^*) &= \frac{1-\varphi_1\varphi_1^*\cdots \varphi_N \varphi_N^*}{(1- \varphi_1\cdots \varphi_N)(1-\varphi_1^*\cdots \varphi_N^*)} \prod_{i,j=1}^N \frac{1}{ 1 - \varphi_i \varphi_j^*} \,. \label{eqn:HSINNH}
\end{align}
\end{subequations}
The non-trivial numerator in \cref{eqn:HSINNH} reflects the fact that the product of a baryon and anti-baryon invariants can be expressed as polynomials of the bilinear primary invariants; it appears with a minus sign to avoid double counting once the expression is expanded as a series. We refer the reader to Ref.~\cite{Grinstein:2023njq} for the interpretation of the Hilbert series numerator and the distinction between primary and secondary invariants. We see that the equation on the right side of \eqref{eqn:HSAccidentalV} is no longer satisfied in these cases.

\subsection{Class II: $SU(N)\times U(1) \implies SO(2N)$}
\label{subsec:ClassII}

In the previous subsection, we saw that the potential in a theory with a pair of fields, one transforming as the fundamental and the other as the anti-fundamental irreps under $SU(N)$, $(\varphi \sim \ydiagram{1} \,,\, \varphi^* \sim \overline{\ydiagram{1}})$, actually presents a larger all-order symmetry that arises accidentally: $SU(N)\times U(1)$. Does the theory have an even larger symmetry?

\begin{table}[t]
\renewcommand{\arraystretch}{1.3}
\setlength{\arrayrulewidth}{.2mm}
\setlength{\tabcolsep}{1.25em}
\centering
\ytableausetup{boxsize=0.5em}
\begin{tabular}{ccc}
\toprule
Class II & $G=SO(2N)$ & $H=SU(N)\times U(1)$ \\
\midrule
\multirow{2}{*}{$\Phi(x)$}
& \multirow{2}{*}{$\phi\; \;\sim\; \ydiagram{1}$} & $\varphi\; \;\sim\; (\ydiagram{1} \;,\; +1)$ \\
& & $\varphi^* \;\sim\; (\overline{\ydiagram{1}} \;,\; -1)$ \\
\midrule
Primary Invariants & $\phi^T \phi$ & $\varphi^\dagger \varphi$ \\
\bottomrule
\end{tabular}
\caption{All-order accidental symmetry Class II. $G=SO(2N)$ and the subgroup $H=SU(N)\times U(1)$; see text for details on how $H$ is embedded in $G$. The building block fields $\Phi(x)$ are bosonic fields. Under the group $G$, it is a vector irrep $\phi$; there is a single primary invariant $\phi^T \phi$. Under the subgroup $H$, $\Phi$ decompose into two irreps: $\Phi=(\varphi, \varphi^*)$, where $\varphi$ is a fundamental rep with charge $+1$, and $\varphi^*$ is its complex conjugate; there is a single primary invariant $\varphi^\dagger \varphi$. See text for the relation between $\phi$ and $(\varphi, \varphi^*)$.}
\label{tab:ClassII}
\end{table}

As argued before, to preserve the $SU(N)$ invariance, $\varphi$ and $\varphi^*$ need to come together and all $SU(N)$ invariants are generated by the combination $\varphi^\dagger \varphi$. Now, writing
\begin{subequations}
\begin{align}
\varphi   &= \Re\varphi + i\Im\varphi \,, \\[5pt]
\varphi^* &= \Re\varphi - i\Im\varphi \,,
\end{align}
\end{subequations}
we see that the above combination is equal to
\begin{equation}
\varphi^\dagger \varphi = (\Re\varphi)^T (\Re\varphi) + (\Im\varphi)^T (\Im\varphi) \,,
\end{equation}
which is invariant under an $SO(2N)$ rotation\footnote{It actually preserves an $O(2N)$ symmetry, but we will focus on the $SO(2N)$ branch of it in this class of examples.}
of the $2N$-dimensional real vector:
\begin{equation}
\phi = \mqty(\Re\varphi \\ \Im\varphi) \,.
\label{eqn:phivarphi}
\end{equation}
This leads to our Class II all-order accidental symmetries
\begin{equation}
SU(N)\times U(1) \implies SO(2N) \quad\text{for}\quad
V(\Phi)\,,
\end{equation}
which is summarized in \cref{tab:ClassII}. The special case of this class at $N=2$ is the famous custodial symmetry for the Higgs sector of the Standard Model, which we will discuss in more detail in \cref{subsec:Custodial}.

In this class, we have $G=SO(2N)$ and the building block fields $\Phi(x)$ are bosonic fields that transform as a single vector irrep $\phi$. The subgroup $H\subset G$ is $SU(N)\times U(1) \simeq U(N)$, under which $\Phi(x)$ decomposes into two irreps: $\Phi = (\varphi, \varphi^*)$, where $\varphi$ is a fundamental rep with charge $+1$, and $\varphi^*$ is its complex conjugate. The embedding of $H$ in $G$ is given as follows. The group $H$ has $(N^2-1)$ generators for the $SU(N)$ factor, and one generator for the $U(1)$ factor. These generators are $N\times N$ hermitian matrices:
\begin{equation}
\left( t_H^a \right)^\dagger = t_H^a = \Re(t_H^a) + i\Im(t_H^a) \,,
\end{equation}
where $\Re(t_H^a)$ is real symmetric and $\Im(t_H^a)$ is real antisymmetric. A group element in $H$ is given by $h(\alpha) = \exp(i\alpha^a t_H^a)$; its action on $\varphi$ is generated by repeatedly acting the exponent:
\begin{equation}
i\alpha^a t_H^a\, \varphi = i\alpha^a \big[ \Re(t_H^a) + i\Im(t_H^a) \big] (\Re\varphi + i\Im\varphi) \,.
\end{equation}
This can be translated into an action on the $2N$-dimensional real vector $\phi$ in \cref{eqn:phivarphi}:
\begin{equation}
i\alpha^a t_{G(H)}^a\, \phi
= \mqty(\Re(i\alpha^a t_H^a\, \varphi) \\ \Im(i\alpha^a t_H^a\, \varphi) )
= i\alpha^a i\mqty( \Im(t_H^a) & \Re(t_H^a) \\ -\Re(t_H^a) & \Im(t_H^a) ) \phi \,.
\end{equation}
Here ``$t_{G(H)}^a$'' denotes the generators of group $G$ when restricted to the subgroup $H$. They can be read off from the above, which specifies the embedding of $H$ in $G$:
\begin{equation}
t_H^a \quad\longrightarrow\quad
t_{G(H)}^a = i \mqty(\Im(t_H^a) & \Re(t_H^a) \\ -\Re(t_H^a) & \Im(t_H^a) ) \,.
\label{eqn:EmbeddingII}
\end{equation}
Note that $t_{G(H)}^a$ given above are antisymmetric, purely imaginary, and traceless:
\begin{equation}
\tr \big[ t_{G(H)}^a \big] = 2i \tr \big[ \Im(t_H^a) \big] = 0 \,.
\end{equation}
These imply that they are indeed a subset of the $SO(2N)$ generators.

From the set up in \cref{tab:ClassII} and \cref{eqn:phivarphi} we see that the primary invariants are
\begin{equation}
\varphi^\dagger \varphi = \phi^T \phi \,.
\label{eqn:PrimaryII}
\end{equation}
To verify this class, we again compute the Hilbert series. Assigning each irrep a distinct grading variable, we obtain the multi-graded Hilbert series as (see \cref{appsubsubsec:ClassIISingle} for details)
\begin{subequations}\label{eqn:HSIIDistinct}
\begin{align}
\HS_\I^{SO(2N),\, \phi} (\phi) &= \frac{1}{1 - \phi^2} \,, \label{eqn:HSIIG}\\[5pt]
\HS_\I^{SU(N)\times U(1),\, (\varphi,\, \varphi^*)} (\varphi, \varphi^*) &= \frac{1}{1 - \varphi \varphi^*} \,. \label{eqn:HSIIH}
\end{align}
\end{subequations}
They are the same Hilbert series written in different grading variables (\cf\ \cref{eqn:PrimaryII}). We can put them in the form of \cref{eqn:HSAccidentalV} by unifying the grading variables $\phi = \varphi = \varphi^* = q$:
\begin{equation}
\HS_\I^{SO(2N),\, \phi} (q) = \HS_\I^{SU(N)\times U(1),\, (\varphi,\, \varphi^*)} (q) = \frac{1}{1-q^2} \,.
\label{eqn:HSIIAgree}
\end{equation}

Much as in the case of Class I, our Class II all-order accidental symmetries fail when $\Phi(x)$ are fermionic building blocks. In addition, it also fails when there are multiple flavors (see \cref{appsubsubsec:ClassIIMultiple} for details):
\begin{subequations}\label{eqn:HSIIkDistinct}
\begin{alignat}{2}
\HS_\I^{SO(2N),\, k\times \phi} (\phi_i) &= \prod_{1\le i\le j \le k} \frac{1}{1 - \phi_i \phi_j}
&&\qquad\text{for}\qquad
k < N \,, \label{eqn:HSIIkG} \\[5pt]
\HS_\I^{SU(N)\times U(1),\, k\times (\varphi,\, \varphi^*)} (\varphi_i, \varphi_j^*) &= \prod_{i,j=1}^k \frac{1}{ 1 - \varphi_i \varphi_j^*}
&&\qquad\text{for}\qquad
k < N \,. \label{eqn:HSIIkH}
\end{alignat}
\end{subequations}
Clearly, when $k>1$, the two Hilbert series no longer agree:
\begin{subequations}\label{eqn:HSIIkDisagree}
\begin{alignat}{2}
\HS_\I^{SO(2N),\, k\times \phi} (q) &= \frac{1}{ (1 - q^2)^{k(k+1)/2}}
&&\qquad\text{for}\qquad
k < N \,, \\[5pt]
\HS_\I^{SU(N)\times U(1),\, k\times (\varphi,\, \varphi^*)} (q) &= \frac{1}{ (1 - q^2)^{k^2}}
&&\qquad\text{for}\qquad
k < N \,.
\end{alignat}
\end{subequations}
The reason for the mismatch at $k>1$ is that among the $k^2$ primary invariants $\varphi_i^\dagger \varphi_j$ for the group $H=SU(N)\times U(1)$, only the $k(k+1)/2$ symmetric ones respect the larger symmetry $G=SO(2N)$:
\begin{equation}
\varphi_{(i}^\dagger \varphi_{j)}^{}
= \frac12 \left( \varphi_i^\dagger \varphi_j + \hc \right)
= \phi_i^T \phi_j = \phi_j^T \phi_i \,,
\qquad\text{with}\qquad
\phi_i = \mqty( \Re\varphi_i \\ \Im\varphi_i ) \,,
\label{eqn:PrimarySymmetric}
\end{equation}
while the antisymmetric combinations do not:
\begin{equation}
\varphi_{[i}^\dagger \varphi_{j]}^{} = \frac12 \left( \varphi_i^\dagger \varphi_j - \hc \right) \,,
\label{eqn:PrimaryAsymmetric}
\end{equation}
as reflected by the Hilbert series in \cref{eqn:HSIIkDistinct,eqn:HSIIkDisagree}.

\subsection{Class III: $SU(2N-1)\times U(1) \implies SU(2N)$}
\label{subsec:ClassIII}

Our third class of all-order accidental symmetries are summarized in \cref{tab:ClassIII}. In this class, we have $G=SU(2N)$ and the building block fields $\Phi(x)$ are bosonic fields that transform as an antisymmetric two-index irrep $\phi\sim \ydiagram{1,1}$ .\footnote{In this class of examples, our building block fields $\Phi(x)$ are complex without its conjugate. In terms of practical applications, this class could be accidental symmetries of a superpotential in supersymmetric EFTs.}
The subgroup $H\subset G$ is $SU(2N-1)\times U(1)$, where the $SU(2N-1)$ factor is a regular embedding in the first $(2N-1)\times(2N-1)$ block of $SU(2N)$, while the $U(1)$ factor is generated by the following traceless generator in $G=SU(2N)$:
\begin{equation}
Q = \text{diag} \left( +1 \;,\; \cdots \;,\; +1 \;,\; -2N+1 \right) \,.
\end{equation}
Under the subgroup $H$, $\Phi=\phi$ decompose into two irreps: $\varphi_1$ that transforms as the irrep $\ydiagram{1,1}$ of $SU(2N-1)$ with charge $+2$ under the $U(1)$ factor, and $\varphi_2$ that transforms as a fundamental irrep of $SU(2N-1)$ with charge $-2N+2$ under the $U(1)$. Specifically, we have the relation
\begin{subequations}
\begin{align}
\varphi_1^{ij} &= \phi^{ij} \,, \\[5pt]
\varphi_2^i &= \phi^{i (2N)} = - \phi^{(2N)i} \,,
\end{align}
\end{subequations}
with $i, j = 1, \cdots, 2N-1$. As stated in \cref{tab:ClassIII}, all the $H$-invariant polynomials in $\Phi$ are generated by the single primary invariant
\begin{align}
\varphi_1^{[12} \cdots \varphi_1^{(2N-3)(2N-2)} \varphi_2^{2N-1]}
&= \frac{1}{(2N-1)!}\, \epsilon_{i_1 i_2 \cdots i_{2N-3} i_{2N-2} i_{2N-1}}\,
\varphi_1^{i_1 i_2} \cdots \varphi_1^{i_{2N-3} i_{2N-2}} \varphi_2^{i_{2N-1}} \notag\\[5pt]
&= \frac{1}{(2N)!}\, \epsilon_{i_1 i_2 \cdots i_{2N-1} i_{2N}}\, \phi^{i_1 i_2} \cdots \phi^{i_{2N-1} i_{2N}} \notag\\[5pt]
&= \phi^{[12} \cdots \phi^{(2N-1)(2N)]} \,.
\label{eqn:PrimaryIII}
\end{align}
The right-hand side is proportional to the Pfaffian of the even-dimensional antisymmetric matrix $\phi$:
\begin{equation}
\phi^{[12} \cdots \phi^{(2N-1)(2N)]} \;\propto\; \text{pf}\, (\phi) \propto \sqrt{\det\phi} \,,
\end{equation}
which is also $G$-invariant. In fact, the $U(1)$ factor in the subgroup $H$ is optional in this class, because the $SU(2N-1)$ invariance in $H$ already forces everything to be generated by the above primary invariant.

\begin{table}[t]
\renewcommand{\arraystretch}{1.3}
\setlength{\arrayrulewidth}{.2mm}
\setlength{\tabcolsep}{1.25em}
\centering
\ytableausetup{boxsize=0.5em}
\begin{tabular}{ccc}
\toprule
Class III & $G=SU(2N)$ & $H=SU(2N-1)\times U(1)$ \\
\midrule
\multirow{2}{*}{$\Phi(x)$}
& \multirow{2}{*}{$\phi\; \;\sim\; \ydiagram{1,1}$} & $\varphi_1\; \;\sim\; \big(\, \ydiagram{1,1} \;,\; +2 \,\big)$ \\
& & \quad $\varphi_2 \;\sim\;\; (\ydiagram{1} \;,\; -2N+2)$ \\
\midrule
Primary Invariants & $\phi^{[12} \cdots \phi^{(2N-1)(2N)]}$ & 
$\varphi_1^{[12} \cdots \varphi_1^{(2N-3)(2N-2)} \varphi_2^{2N-1]}$ \\
\bottomrule
\end{tabular}
\caption{All-order accidental symmetry Class III. $G=SU(2N)$ and the subgroup $H=SU(2N-1)\times U(1)$; see text for how $H$ is embedded in $G$. The building block fields $\Phi(x)$ are bosonic fields. They form an antisymmetric two-index irrep $\phi\sim\ydiagram{1,1}$ under the group $G$. Under the subgroup $H$, $\Phi$ decomposes into two irreps: $\varphi_1 \sim \ydiagram{1,1}$ with charge $+2$, and a fundamental irrep $\varphi_2$ with charge $-2N+2$. Specifically, we have $\varphi_1^{ij} = \phi^{ij}$ and $\varphi_2^i = \phi^{i(2N)}=-\phi^{(2N)i}$ with $i,j=1,\cdots,2N-1$. There is a single primary invariant $\phi^{[12} \cdots \phi^{(2N-1)(2N)]} = \varphi_1^{[12} \cdots \varphi_1^{(2N-3)(2N-2)} \varphi_2^{2N-1]}$. The $U(1)$ factor in $H$ is optional in this class, as we  explain in the text.}
\label{tab:ClassIII}
\end{table}

To verify this class of all-order accidental symmetries, we again compute the Hilbert series. Assigning each irrep a distinct grading variable, we obtain the multi-graded Hilbert series as
\begin{subequations}\label{eqn:HSIIIDistinct}
\begin{align}
\HS_\I^{SU(2N),\, \phi} (\phi) &= \frac{1}{1 - \phi^N} \,, \label{eqn:HSIIIG} \\[5pt]
\HS_\I^{SU(2N-1)\times U(1),\, (\varphi_1,\, \varphi_2)} (\varphi_1, \varphi_2) &= \frac{1}{1 - \varphi_1^{N-1} \varphi_2} \,, \label{eqn:HSIIIH1} \\[5pt]
\HS_\I^{SU(2N-1),\, (\varphi_1,\, \varphi_2)} (\varphi_1, \varphi_2) &= \frac{1}{1 - \varphi_1^{N-1} \varphi_2} \,. \label{eqn:HSIIIH2}
\end{align}
\end{subequations}
Detailed steps of computing these Hilbert series are presented in \cref{appsubsubsec:ClassIIIG,appsubsubsec:ClassIIIH1,appsubsubsec:ClassIIIH2}. We see that they all agree. We put them in the form of the equation on the right side of \eqref{eqn:HSAccidentalV} by unifying the grading variables $\phi = \varphi_1 = \varphi_2 = q$:
\begin{equation}
\HS_\I^{SU(2N),\, \phi} (q) = \HS_\I^{SU(2N-1)\times U(1),\, (\varphi_1,\, \varphi_2)} (q)
= \HS_\I^{SU(2N-1),\, (\varphi_1,\, \varphi_2)} (q) 
= \frac{1}{1-q^N} \,.
\label{eqn:HSIIIAgree}
\end{equation}
As the $U(1)$ factor in $H$ is optional, our Class III all-order accidental symmetries can also be summarized as
\begin{equation}
SU(2N-1) \implies
SU(2N-1)\times U(1) \implies
SU(2N) \quad\text{for}\quad
V(\Phi)\,.
\end{equation}

\subsubsection*{Fermionic case}

There is also a fermionic version of Class III all-order accidental symmetries, as summarized in \cref{tab:ClassIIIFermion}. The groups $G=SU(2N)$ and $H=SU(2N-1)\times U(1)$ are the same as in the bosonic case. The fermionic building block fields $\Psi(x)$ transform as a fundamental irrep $\psi$ under $G$. They decompose into two irreps under $H$: $\chi_1$ transforming as a fundamental irrep with charge $+1$, and $\chi_2$ transforming as a singlet irrep with charge $-2N+1$:
\begin{subequations}
\begin{align}
\chi_1^i &= \psi^i
\qquad\text{for}\quad
1 \le i \le 2N-1 \,, \\[5pt]
\chi_2 &= \psi^{2N} \,.
\end{align}
\end{subequations}
We see that all $H$-invariant polynomials in $\Psi$ are generated by the primary invariant
\begin{equation}
\chi_1^1 \cdots \chi_1^{2N-1} \chi_2 = \psi^1 \cdots \psi^{2N-1} \psi^{2N} \,,
\label{eqn:PrimaryIIIFermion}
\end{equation}
which is also $G$-invariant. Note that in this fermionic case, the two combinations, $\chi_1^1 \cdots \chi_1^{2N-1}$ and $\chi_2$, are separately invariant under the $SU(2N-1)$ factor in $H$. Therefore, the $U(1)$ factor in $H$ is necessary to ensure $G$-invariance.

\begin{table}[t]
\renewcommand{\arraystretch}{1.3}
\setlength{\arrayrulewidth}{.2mm}
\setlength{\tabcolsep}{1.25em}
\centering
\ytableausetup{boxsize=0.5em}
\begin{tabular}{ccc}
\toprule
Class III, Fermionic Case & $G=SU(2N)$ & $H=SU(2N-1)\times U(1)$ \\
\midrule
\multirow{2}{*}{$\Psi(x)$}
& \multirow{2}{*}{$\psi\; \;\sim\; \ydiagram{1}$} & $\chi_1\; \;\sim\; \big(\, \ydiagram{1} \;,\; +1 \,\big)$ \\
& & \quad $\chi_2 \;\sim\;\; (\mathbf{1} \;,\; -2N+1)$ \\
\midrule
Primary Invariants & $\psi^1 \cdots \psi^{2N-1} \psi^{2N}$ & 
$\chi_1^1 \cdots \chi_1^{2N-1} \chi_2$ \\
\bottomrule
\end{tabular}
\caption{All-order accidental symmetry Class III, fermionic case. $G=SU(2N)$ and the subgroup $H=SU(2N-1)\times U(1)$ is embedded in $G$ the same way as in the bosonic case in \cref{tab:ClassIII}. The building block fields $\Psi(x)$ are fermionic fields. Under the group $G$, they form a fundamental irrep $\psi$. Under the subgroup $H$, $\Psi$ decompose into two irreps: a fundamental irrep $\chi_1$ with charge $+1$, and a singlet irrep $\chi_2$ with charge $-2N+1$. Specifically, $\chi_1^i = \psi^i$ with $1\le i \le 2N-1$ and $\chi_2 = \psi^{2N}$. There is a single primary invariant $\psi^1 \cdots \psi^{2N-1} \psi^{2N} = \chi_1^1 \cdots \chi_1^{2N-1} \chi_2$. In this fermionic case, the $U(1)$ factor in $H$ is not optional.}
\label{tab:ClassIIIFermion}
\end{table}

To verify this fermionic class of all-order accidental symmetries, one again computes the Hilbert series. Assigning each irrep a distinct grading variable, we obtain (see \cref{appsubsubsec:ClassIIIFermion} for details)
\begin{subequations}\label{eqn:HSIIIFermionDistinct}
\begin{align}
\HS_\I^{SU(2N),\, \psi} (\psi) &= 1 + \psi^{2N} \,, \label{eqn:HSIIIFermionG} \\[5pt]
\HS_\I^{SU(2N-1)\times U(1),\, (\chi_1,\, \chi_2)} (\chi_1, \chi_2) &= 1 + \chi_1^{2N-1} \chi_2  \,. \label{eqn:HSIIIFermionH}
\end{align}
\end{subequations}
We see that they agree. Unifying the grading variables $\psi = \chi_1 = \chi_2 = q$, we put them in the form of \cref{eqn:HSAccidentalV}:
\begin{equation}
\HS_\I^{SU(2N),\, \psi} (q) = \HS_\I^{SU(2N-1)\times U(1),\, (\chi_1,\, \chi_2)} (q)
= 1 + q^{2N} \,.
\label{eqn:HSIIIFermionAgree}
\end{equation}
Here, we note again that the Hilbert series is truncated due to the fermionic nature of the building block fields. We will discuss this more in \cref{sec:Systematic}.

To some extent, the bosonic irreps of Class III accidental symmetries summarized in \cref{tab:ClassIII} can be viewed as composite ``bosonizations'' of the fermionic irreps in \cref{tab:ClassIIIFermion}. Specifically, we can make the identification
\begin{equation}
\phi^{ij} \sim \psi^{[i} \psi^{j]} \,,\qquad
\varphi_1^{ij} \sim \chi_1^{[i} \chi_1^{j]} \,,\qquad
\varphi_2^i \sim \chi_1^i \chi_2 \,.
\end{equation}
One can check the agreement on the charges of these irreps. However, we emphasize that this composite picture is not precise, because the allowed effective operators in the fermionic case will still truncate due to the fermionic nature of the fields. This difference is reflected by the truncated Hilbert series in \cref{eqn:HSIIIFermionDistinct}, as compared to those in \cref{eqn:HSIIIDistinct}.

\section{Finite-order accidental symmetries}
\label{sec:FiniteOrder}

In \cref{sec:AllOrder}, we identified three classes of accidental symmetries that hold to all orders in the EFT expansion when derivative interactions are not considered, \ie, when we focus on the potential term $V(\Phi)$ in the EFT. We verified them by computing the Hilbert series and checking against the criterion  \eqref{eqn:HSAccidentalV}.

One may wonder how these analyses based only on the potential term (\ie, polynomials in the fields $\Phi(x)$) can be useful for actual EFTs, where one cannot simply ignore the derivative interactions. As we will explain in this section, all-order accidental symmetries respected by the potential term $V(\Phi)$ will be broken by derivative interactions. However, this breaking typically arises beyond the leading order in the EFT, and as a result they become finite-order accidental symmetries. In such cases, the Hilbert series is still a reliable tool for verifying and analyzing them. To demonstrate this, we first discuss some examples of finite-order accidental symmetries within the potential interactions $V(\Phi)$ in \cref{subsec:Nonderivative}. We then explain how to understand the impacts of derivative interactions in \cref{subsec:Custodial}, taking the custodial violation in SMEFT as an illustrative example. Finally in \cref{subsec:General}, we provide a general criterion of using Hilbert series to verify finite-order accidental symmetries of the full Lagrangian $\Lag_\text{\,EFT}$, irrespective of whether the symmetry-breaking term originates from the potential or the derivative interactions.

\subsection{Accidental symmetry breaking by non-derivative interactions}
\label{subsec:Nonderivative}

Non-derivative interactions can break an accidental symmetry, making it only hold up to a certain order in the EFT potential $V(\Phi)$. To verify this kind of finite-order accidental symmetries, one can readily generalize the criterion in \cref{eqn:HSAccidentalV}, by comparing the two Hilbert series up to a given order:
\begin{tcolorbox}[colback=light-gray]
\begin{center}
\begin{minipage}{5.5in}\vspace{-20pt}
\begin{multline}
H \implies G
\quad\text{for}\quad
V(\Phi)
\quad\text{up to order}\; k
\\[5pt]
\Longleftrightarrow\qquad
\HS_\I^{H,\, \Phi} (q) = \HS_\I^{G,\, \Phi} (q) + \order{q^{k+1}}  \,.
\label{eqn:HSAccidentalVk}
\end{multline}
\end{minipage}\vspace{3pt}
\end{center}
\end{tcolorbox}
\noindent
Note that the $\order{q^{k+1}}$ terms are always positive, as all $G$-invariants are $H$-invariants.

\subsubsection*{Class I with multiple flavors}
One example was already mentioned in \cref{subsec:ClassI} --- the Class I all-order accidental symmetries will break when there are too many flavors in the building block fields, due to the appearance of the baryon-like $H=SU(N)$ invariants. However, these baryon combinations have at least $N$ powers of the building block fields $\Phi$, so they only start to appear at order $q^N$. Therefore, we expect the accidental symmetry to hold up to order $q^{N-1}$. To verify this finite-order accidental symmetry, we can check against the criterion above in \cref{eqn:HSAccidentalVk}: taking the Hilbert series presented in \cref{eqn:HSINk2Distinct,eqn:HSINNDistinct}, we see that disagreements between $\HS_\I^{G,\, \Phi}(q)$ and $\HS_\I^{H,\, \Phi}(q)$ indeed only start from order $q^N$.

\subsubsection*{A product subgroup}

As another example, let us look at the finite-order accidental symmetry summarized in \cref{tab:Rectangular}. In this case, we consider a direct product group $H=SU(3)\times SU(2)$, and a pair of bosonic building block fields $\Phi=(\varphi, \varphi^*)$, where $\varphi$ is a $3\times 2$ complex matrix, transforming as a bifundamental irrep $\varphi\sim (\mathbf{3} \,,\, \mathbf{2})$ under $H$:
\begin{equation}
\varphi \quad\longrightarrow\quad
g_3\, \varphi\, g_2^\dagger \,,
\qquad\text{with}\qquad
g_3 \in SU(3) \,,\quad
g_2 \in SU(2) \,,
\label{eqn:BilinearTrans}
\end{equation}
and $\varphi^*$ is the complex conjugate of this matrix.

\begin{table}[t]
\renewcommand{\arraystretch}{1.3}
\setlength{\arrayrulewidth}{.2mm}
\setlength{\tabcolsep}{1.25em}
\centering
\ytableausetup{boxsize=0.5em}
\begin{tabular}{ccc}
\toprule
Finite-order Example & $G=SU(6)$ & $H=SU(3)\times SU(2)$ \\
\midrule
\multirow{2}{*}{$\Phi(x)$}
& $\phi\; \;\sim\; \ydiagram{1}$ & $\varphi\; \;\sim\; (\mathbf{3} \,,\, \mathbf{2})$ \\
& $\phi^* \;\sim\; \overline{\ydiagram{1}}$ & $\varphi^* \;\sim\; (\mathbf{\bar 3} \,,\, \mathbf{2})$ \\
\midrule
Primary Invariants & $\phi^\dagger \phi$ & $\tr(\varphi^\dagger \varphi)$ \;,\; $\tr(\varphi^\dagger \varphi \varphi^\dagger \varphi)$ \\
\bottomrule
\end{tabular}
\caption{Finite-order accidental symmetry with $G=SU(6)$ and the product subgroup $H=SU(3)\times SU(2)\subset G$. See text for the details of the embedding. The building block fields $\Phi(x)$ are bosonic fields. Under the group $G$, they consist of two irreps: $\Phi=(\phi, \phi^*)$, with $\phi$ a fundamental irrep and $\phi^*$ the complex conjugate; there is a single primary invariant $\phi^\dagger \phi$. Under the subgroup $H$, $\Phi=(\varphi, \varphi^*)$, with $\varphi$ a $3\times 2$ complex matrix transforming as a bifundamental $\varphi \sim (\mathbf{3} \,,\, \mathbf{2})$, and $\varphi^*$ the complex conjugate matrix; there are two primary invariants: $\tr(\varphi^\dagger \varphi)$ and $\tr(\varphi^\dagger \varphi \varphi^\dagger \varphi)$.}
\label{tab:Rectangular}
\end{table}

One can alternatively list the six components of the matrix $\varphi$ into a complex vector $\phi$, such as $\phi = (\varphi_{11}, \varphi_{12}, \varphi_{21}, \varphi_{22}, \varphi_{31}, \varphi_{32})^T$. In this six-dimensional vector space, the transformations in \cref{eqn:BilinearTrans} correspond to the tensor product matrix $g_6$:
\begin{equation}
\phi \quad\longrightarrow\quad
g_6\, \phi \,,
\qquad\text{with}\qquad
g_6 = g_3 \otimes g_2^* \,,
\end{equation}
where ``$\otimes$'' denotes the Kronecker product of matrices, and the resulting $g_6$ is a $6\times 6$ special unitary matrix
\begin{subequations}
\begin{align}
g_6^\dagger\, g_6 &= (g_3^\dagger\, g_3) \otimes (g_2^T g_2^*) = 1 \,, \\[5pt]
\det (g_6) &= [\det (g_3)]^2 [\det(g_2^*)]^3 = 1 \,.
\end{align}
\end{subequations}
Therefore, the group $H$ is a subgroup of the $G=SU(6)$ that acts linearly on $\phi$.

All the $G$-invariants are generated by the norm of $\phi$:
\begin{equation}
\phi^\dagger \phi = \tr(\varphi^\dagger \varphi) \,.
\label{eqn:PrimaryRectangularG}
\end{equation}
For $H$-invariants, however, there is one additional primary invariant:\footnote{Trace of more powers of $\varphi^\dagger\varphi$ are redundant by Cayley-Hamilton theorem.}
\begin{equation}
\tr(\varphi^\dagger \varphi \varphi^\dagger \varphi) \,,
\label{eqn:PrimaryRectangularH}
\end{equation}
and hence we do not expect the two Hilbert series to be equal. However, as this additional primary invariant contains four powers of $\Phi$, $H\implies G$ is a finite-order accidental symmetry for $V(\Phi)$ that holds up to order $q^3$.

One can verify this finite-order accidental symmetry by computing the Hilbert series. We obtain the multi-graded results as\footnote{More generally, for a rectangular setup $m>n$, we have
\begin{equation}
\HS_\I^{SU(m)\times SU(n),\, (\varphi,\, \varphi^*)} (\varphi, \varphi^*) = \frac{1}{(1 - \varphi \varphi^*)(1-(\varphi\varphi^*)^2) \cdots (1-(\varphi\varphi^*)^n)} \,.
\end{equation}
}
\begin{subequations}\label{eqn:HSRectangular}
\begin{align}
\HS_\I^{SU(6),\, (\phi,\, \phi^*)} (\phi, \phi^*) &= \frac{1}{1 - \phi \phi^*} \,, \\[5pt]
\HS_\I^{SU(3)\times SU(2),\, (\varphi,\, \varphi^*)} (\varphi, \varphi^*) &= \frac{1}{(1 - \varphi \varphi^*)(1- (\varphi\varphi^*)^2)} \,,
\end{align}
\end{subequations}
and indeed they differ due to the additional primary invariant in \cref{eqn:PrimaryRectangularH}. Unifying the grading variables $\phi = \phi^* = \varphi = \varphi^* = q$, we get
\begin{subequations}\label{eqn:HSRectangularq}
\begin{align}
\HS_\I^{SU(6),\, (\phi,\, \phi^*)} (q) &= \frac{1}{1 - q^2} \,, \\[5pt]
\HS_\I^{SU(3)\times SU(2),\, (\varphi,\, \varphi^*)} (q) &= \frac{1}{(1 - q^2)(1-q^4)} \,,
\end{align}
\end{subequations}
and clearly
\begin{equation}
\HS_\I^{SU(3)\times SU(2)} (q) = \HS_\I^{SU(6),\, (\phi,\, \phi^*)} (q) + \order{q^4} \,.
\label{eqn:HSRectangularAgree}
\end{equation}
This demonstrates the criterion in \eqref{eqn:HSAccidentalVk}.

\subsection{Custodial violation from derivative interactions}
\label{subsec:Custodial}

In this subsection, we move on to study the effects of derivative interactions on accidental symmetries. Let us see an explicit example for illustration. We take the $N=2$ case of the Class II all-order accidental symmetries discussed in \cref{subsec:ClassII}:
\begin{equation}
SU(2) \times U(1) \implies SO(4)
\quad\text{for}\quad
V(\Phi) \,.
\label{eqn:Custodial}
\end{equation}
This can be applied to the Higgs sector of the SM or SMEFT. Under the electroweak symmetry group $SU(2)_L \times U(1)_Y$, the Higgs field $H(x)$ transforms as $(\mathbf{2}, +\frac12)$. For the purpose of determining the allowed effective operators, one can rescale the hypercharge $+\frac12 \to +1$. With this, the building block fields $H(x)$ and $H^*(x)$ transform precisely as the $(\varphi,\, \varphi^*)$ pair in \cref{tab:ClassII}, and the general discussions in \cref{subsec:ClassII} will apply. In particular, polynomials in $H(x)$ and $H^*(x)$ that are $SU(2)_L\times U(1)_Y$ invariant are generated by the combination
\begin{equation}
|H|^2 = H^\dagger H = \frac12 \left( \phi_1^2 + \phi_2^2 + \phi_3^2 + \phi_4^2 \right) \,,
\qquad\text{with}\qquad
H = \frac{1}{\sqrt{2}} \mqty( \phi_1 + i\phi_2 \\ \phi_3 + i\phi_4 ) \,,
\label{eqn:phiH}
\end{equation}
which enjoys a larger $SO(4)$ symmetry acting on the four-dimensional real vector $\phi \equiv (\phi_1, \phi_2, \phi_3, \phi_4)^T$ --- this is the famous custodial symmetry.\footnote{Upon electroweak symmetry breaking, this enhanced $SO(4)\simeq SU(2)_L\times SU(2)_R$ symmetry is broken to the diagonal $SU(2)_V$, which is more often referred to as the custodial symmetry in the literature. In this paper, we use the term to refer to the $SO(4)\simeq SU(2)_L\times SU(2)_R$ group in the unbroken phase.}
Therefore, imposing $SU(2)_L\times U(1)_Y$ invariance forces the Higgs potential $V(H, H^\dagger)$ in SMEFT to respect the custodial $SO(4)$ symmetry to all orders.

However, the custodial symmetry is broken by derivative interactions in the Higgs sector of SMEFT.\footnote{The Yukawa and gauge interactions in SMEFT also break the custodial symmetry. Here we focus on the effects of $\pd_\mu$ only, to keep our discussions to the point.}
For example, at mass dimension six, there are two independent effective operators of type $H^4\pd^2$ (\ie, containing two powers of the derivatives and four powers of the Higgs fields $H, H^*$)  that are electroweak invariant \cite{Grzadkowski:2010es}:
\begin{equation}
\Op_{H\Box} = - (\pd_\mu |H|^2 ) (\pd^\mu |H|^2) \,,\qquad
\Op_{HD} = (H^\dagger \pd_\mu H)^* (H^\dagger \pd^\mu H) \,.
\label{eqn:QHboxQHD}
\end{equation}
The first operator $\Op_{H\Box}$ respects the custodial symmetry, while the second operator $\Op_{HD}$ does not (and is hence constrained by the $T$ parameter in  electroweak precision tests \cite{Grinstein:1991cd}). One can understand the custodial violation of $\Op_{HD}$ from the rewriting
\begin{equation}
\Op_{HD} = \frac14 \left( H^\dagger \pd_\mu H + \hc \right)^2 - \frac14 \left( H^\dagger \pd_\mu H - \hc \right)^2 \,.
\label{eqn:OpHDRewriting}
\end{equation}
The first term $H^\dagger \pd_\mu H + \hc = \pd_\mu (H^\dagger H)$ is custodial symmetric but the second term is not. The situation here is precisely the same as in our discussions around \cref{eqn:PrimarySymmetric,eqn:PrimaryAsymmetric}. When derivatives are considered, one can view $(\pd_\mu H \,,\, \pd_\mu H^*)$ as a second flavor of the $(\varphi,\, \varphi^*)$ pair in \cref{tab:ClassII}. Therefore, we are dealing with (at least) two flavors:
\begin{equation}
\left( \varphi_1 \,,\, \varphi_1^* \right) = \left( H \,,\, H^* \right)
\,,\qquad
\left( \varphi_2 \,,\, \varphi_2^* \right) = \left( \pd_\mu H \,,\, \pd_\mu H^* \right) \,.
\label{eqn:TwoFlavors}
\end{equation}
From the discussions around \cref{eqn:PrimarySymmetric,eqn:PrimaryAsymmetric}, we learned that as soon as we have two flavors, the Class II all-order accidental symmetries will no longer hold. With the identification in \cref{eqn:TwoFlavors}, we recognize the custodial violating term in \cref{eqn:OpHDRewriting} precisely as the antisymmetric combination in \cref{eqn:PrimaryAsymmetric}:
\begin{equation}
\frac12 \left( H^\dagger \pd_\mu H - \hc \right)
= \frac12 \left( \varphi_1^\dagger \varphi_2 - \hc \right)
= \varphi_{[1}^\dagger \varphi_{2]}^{} \,,
\end{equation}
while the custodial preserving term in \cref{eqn:OpHDRewriting} is the symmetric combination in \cref{eqn:PrimarySymmetric}.

As seen from the discussion above, the origin of the custodial violation by derivative interactions can be understood as having multiple flavors. In fact, \cref{eqn:TwoFlavors} is not quite accurate --- $(\pd_\mu H,\, \pd_\mu H^*)$ technically consist of four new flavors of the $(\varphi,\, \varphi^*)$ pair in \cref{tab:ClassII}, because $\pd_\mu$ has four components. Together they form a vector irrep of the Lorentz group, and imposing Lorentz invariance also played a role in determining the allowed effective operators listed in \cref{eqn:QHboxQHD}.

More generally, at higher orders in SMEFT, derivative interactions will get more flavors of the $(\varphi,\, \varphi^*)$ pair in \cref{tab:ClassII} involved, such as $(\pd_\mu \pd_\nu H,\, \pd_\mu \pd_\nu H^*)$. Eventually, one needs to include the full tower of the ``Single Particle Modules'' (SPMs) \cite{Henning:2017fpj} built out of the Higgs fields $H$, $H^*$, and the derivatives:
\begin{equation}
\SPM_H = \mqty( H \\ \pd_\mu H \\ \pd_{(\mu} \pd_{\nu)} H \\ \vdots) \,,\qquad
\SPM_{H^*} = \mqty( H^* \\ \pd_\mu H^* \\ \pd_{(\mu} \pd_{\nu)} H^* \\ \vdots) \,.
\label{eqn:SPMH}
\end{equation}
Here the parentheses around the indices imply that we are only keeping the traceless symmetric components. Antisymmetric components vanish because the partial derivatives commute. The trace components are eliminated in the SPM to take care of the so-called equations of motion (EOM) redundancies.\footnote{The origin of the EOM redundancy is the field redefinition equivalence in the Lagrangian formulation of EFTs, that allows one to eliminate operators that contain EOM factors, order by order in the EFT expansion, starting from the next-to-leading order. It should be added, however, that  there is no EOM redundancy for operators at the leading order of the EFT, and one needs to restore the trace components in the SPM when enumerating them.}
To determine the allowed effective operators that are independent, one considers all the polynomials in these SPM components, and then impose the $SU(2)_L\times U(1)_Y$ invariance, as well as the Lorentz invariance. On top of this, one also needs to address the integration by parts (IBP) redundancies, \ie, to remove operators that are total derivatives.

Although derivative interactions in the Higgs sector will break the custodial symmetry, we note that the breaking only starts at mass dimension six. At the renormalizable level (\ie\ up to mass dimension four), the allowed Lagrangian terms are
\begin{equation}
H^\dagger H \,,\qquad
(H^\dagger H)^2 \,,\qquad
(\pd_\mu H^\dagger) (\pd^\mu H) \,,
\end{equation}
which are all invariant under the custodial symmetry, even with derivative interactions included. This can be generally understood from the fact that derivatives have positive mass dimensions. Because of this (and the requirement of Lorentz invariance), accidental symmetry breaking from derivative interactions typically starts only from a certain non-trivial EFT order, below which the accidental symmetry still holds.

\subsection{General criterion for finite-order accidental symmetries}
\label{subsec:General}

Let us summarize a few general lessons that we have learned so far about accidental symmetries:
\begin{itemize}
\item When we focus on non-derivative interactions only, \ie, the potential interactions $V(\Phi)$ in the EFT, there may exist accidental symmetries that hold to all orders. Several classes of examples were provided in \cref{sec:AllOrder}. All-order accidental symmetries can be verified with the criterion in \eqref{eqn:HSAccidentalV}.
\item Some accidental symmetries will hold only up to a certain order in the EFT potential $V(\Phi)$. This was discussed in \cref{subsec:Nonderivative}. One class of them arises when we have too many flavors of irreps in the building block fields, which turns an all-order accidental symmetry into a finite-order one. Finite-order accidental symmetries in $V(\Phi)$ can be verified with the criterion in \eqref{eqn:HSAccidentalVk}.
\item Derivative interactions will also break all-order accidental symmetries. Their effects can be understood as getting infinitely many flavors of irreps involved; these are the tower of irreps in the SPMs of the fields (see \eg\ \cref{eqn:SPMH} for the Higgs fields). This was explained in \cref{subsec:Custodial}.
\item When an all-order accidental symmetry is broken by derivative interactions, the breaking typically starts from a non-trivial order in the EFT expansion, because derivatives come with positive orders in the EFT power counting. Therefore, an all-order accidental symmetry respected by the potential term $V(\Phi)$ becomes a finite-order accidental symmetry when the full Lagrangian $\Lag_\text{\,EFT}$ is considered.
\end{itemize}

In general, to verify a finite-order accidental symmetry in the full Lagrangian $\Lag_\text{\,EFT}$, we can use a criterion similar to that in \eqref{eqn:HSAccidentalVk}:
\begin{tcolorbox}[colback=light-gray]
\begin{center}
\begin{minipage}{5.5in}\vspace{-20pt}
\begin{multline}
H \implies G
\quad\text{for}\quad
\Lag_\text{\,EFT}
\quad\text{up to order}\; k
\\[8pt]
\Longleftrightarrow\quad
\HS_{\I,\text{ IBP}}^{\text{Lorentz}\times H,\, \SPM_\Phi} (q) = \HS_{\I,\text{ IBP}}^{\text{Lorentz}\times G,\, \SPM_\Phi} (q) + \order{q^{k+1}}  \,.
\label{eqn:HSAccidentalLagk}
\end{multline}
\end{minipage}\vspace{3pt}
\end{center}
\end{tcolorbox}
\noindent
Compared to criterion \eqref{eqn:HSAccidentalVk}, the building block fields $\Phi$ are now replaced by their SPMs $\SPM_\Phi$, a tower made of the fields and their derivatives, much like the $\SPM_H$ and $\SPM_{H^*}$ in \cref{eqn:SPMH}.
Again, not all the derivative components are included in $\SPM_\Phi$ --- we need to eliminate the components that correspond to the EOM factors. They depend on the type of the fields. The SPM in \cref{eqn:SPMH} was specifically for scalar fields. We refer the reader to Ref.~\cite{Henning:2017fpj} for rigorous definitions of the SPMs for different type of fields, including scalars, Weyl fermions, field strengths, as well as Goldstone bosons.

It is also understood that when computing the Hilbert series in \eqref{eqn:HSAccidentalLagk}, one needs to impose the Lorentz invariance and address the IBP redundancies. We refer the reader to Ref.~\cite{Henning:2017fpj} for an elaboration on how to do these carefully, which is crucial for correctly computing the Hilbert series for actual EFTs.

We emphasize that the criterion in \eqref{eqn:HSAccidentalLagk} works for verifying general finite-order accidental symmetries, irrespective of whether the symmetry breaking originates from the potential or the derivative interactions. Similar with the criterion in \eqref{eqn:HSAccidentalV}, the equivalence in \eqref{eqn:HSAccidentalVk} and \eqref{eqn:HSAccidentalLagk} are essentially tautologies --- restating the meaning of the Hilbert series. But again they allow us to study accidental symmetries with the powerful tools of invariant theory, which opens up the possibility for more systematic approaches, as we will discuss in \cref{sec:Systematic}. For the rest of this subsection, we give a few  examples to demonstrate criterion  \eqref{eqn:HSAccidentalLagk}, focusing on the scenario that symmetry breaking comes from derivative interactions.

\subsubsection*{Example: custodial symmetry}

To demonstrate the criterion in \eqref{eqn:HSAccidentalLagk}, let us apply it to checking the custodial symmetry in the Higgs sector of SMEFT. We use the SPMs $\SPM_H$, $\SPM_{H^*}$ for the Higgs fields in \cref{eqn:SPMH}, and the SPM $\SPM_\phi$ for the custodial $SO(4)$ vector field $\phi$ introduced around \cref{eqn:phiH}. We further impose the Lorentz invariance, and properly take care of the EOM and IBP redundancies. In the end, we obtain the following Hilbert series up to mass dimension nine:
\begin{subequations}\label{eqn:HSCustodial}
\begin{align}
\HS_{\I,\text{ IBP}}^{\text{Lorentz}\times SO(4),\, \SPM_\Phi} (\phi, \pd)
&= 1 + \phi^2 + \Big( \phi^4 + \phi^2 \pd^2 \Big)\notag\\[5pt]
&\quad + \Big( \phi^6 + \phi^4\pd^2 \Big)
\notag\\[5pt]
&\quad
+ \Big( \phi^8 + \phi^6 \pd^2 + 2\, \phi^4\pd^4 \Big) 
\notag\\[5pt]
&\quad
+ \order{\text{dim-10}}
\,, \\[10pt]
\HS_{\I,\text{ IBP}}^{\text{Lorentz}\times SU(2)_L\times U(1)_Y,\, \SPM_\Phi} (H, H^*, \pd)
&= 1 + HH^* + \Big[ (HH^*)^2 + HH^*\pd^2 \Big]
\notag\\[5pt]
&\quad
+ \Big[ (HH^*)^3 + 2\, (HH^*)^2 \pd^2 \Big]
\notag\\[5pt]
&\quad
+ \Big[ (HH^*)^4 + 2\, (HH^*)^3 \pd^2 + 3\, (HH^*)^2 \pd^4 \Big]
\notag\\[5pt]
&\quad
+ \order{\text{dim-10}} \,.
\end{align}
\end{subequations}
Here, the EFT power counting is governed by the canonical mass dimensions, set by $[\phi]=[H]=[H^*]=[\pd_\mu]=1$, and we are using ``$\pd\,$'' as a grading variable to track the power of derivatives $\pd_\mu$. Unifying grading variables $\phi=H=H^*=\pd=q$, we get
\begin{subequations}\label{eqn:HSCustodialq}
\begin{align}
\HS_{\I,\text{ IBP}}^{\text{Lorentz}\times SO(4),\, \SPM_\Phi} (q)
&= 1 + q^2 + 2\, q^4 + 2\, q^6 + 4\, q^8 + \order{q^{10}}
\,, \\[10pt]
\HS_{\I,\text{ IBP}}^{\text{Lorentz}\times SU(2)_L\times U(1)_Y,\, \SPM_\Phi} (q)
&= 1 + q^2 + 2\, q^4 + 3\, q^6 + 6\, q^8 + \order{q^{10}} \,.
\end{align}
\end{subequations}
We see that these two Hilbert series do satisfy the criterion in \eqref{eqn:HSAccidentalLagk}; their difference starts at $\order{q^6}$. From the multi-graded Hilbert series in \cref{eqn:HSCustodial}, we also see that their disagreements are due to derivative interactions.

\subsubsection*{Example: Class I single flavor with derivative interactions}

As another example of symmetry breaking by derivative interactions, we consider the single flavor case of the Class I all-order accidental symmetries summarized in \cref{tab:ClassI}. From our general discussions above, we know that including derivative interactions will break this all-order accidental symmetry, turning it into a finite-order one. For the discussion  below, let us specify to the cases of $N=3$ and $N=4$, and check what happens explicitly.

As before, we compute the Hilbert series by using the SPMs $\SPM_\Phi = (\SPM_\phi, \SPM_{\phi^*}) = (\SPM_\varphi, \SPM_{\varphi^*})$, imposing Lorentz invariance, and properly taking care of the EOM and IBP redundancies. In the end, for $N=3$, we obtain the following results up to mass dimension ten:
\begin{subequations}\label{eqn:HSISPMN3}
\begin{align}
\HS_{\I,\text{ IBP}}^{\text{Lorentz}\times SU(3)\times U(1),\, \SPM_\Phi} &= 1 + \phi \phi^* 
+ \Big[ (\phi \phi^*)^2 + \phi \phi^* \pd^2 \Big]
+ \Big[ (\phi \phi^*)^3 + 2 (\phi \phi^*)^2 \pd^2 \Big]
\notag\\[5pt]
&\quad
+ \Big[ (\phi \phi^*)^4 + 2 (\phi \phi^*)^3 \pd^2 + 3 (\phi \phi^*)^2 \pd^4 \Big]
\notag\\[5pt]
&\quad
+ \Big[ (\phi \phi^*)^5 + 2 (\phi \phi^*)^4 \pd^2 + 11 (\phi \phi^*)^3 \pd^4 + 4 (\phi \phi^*)^2 \pd^6 \Big]
\notag\\[5pt]
&\quad
+ \order{\text{dim-11}} 
\notag\\[5pt]
&\hspace{-40pt}
= 1 + q^2 + 2\, q^4 + 3\, q^6 + 6\, q^8 + 18\, q^{10} + \order{q^{11}} \,, \\[12pt]
\HS_{\I,\text{ IBP}}^{\text{Lorentz}\times SU(3),\, \SPM_\Phi} &= 1 + \varphi \varphi^* 
+ \Big[ (\varphi \varphi^*)^2 + \varphi \varphi^* \pd^2 \Big]
+ \Big[ (\varphi \varphi^*)^3 + 2 (\varphi \varphi^*)^2 \pd^2 \Big]
\notag\\[5pt]
&\quad
+ \Big[ (\varphi \varphi^*)^4 + 2 (\varphi \varphi^*)^3 \pd^2 + 3 (\varphi \varphi^*)^2 \pd^4 \Big]
\highlight{+ \Big[ \left( \varphi^3 + \varphi^{*3} \right) (\varphi \varphi^*) \pd^4 \Big]}
\notag\\[5pt]
&\quad
+ \Big[ (\varphi \varphi^*)^5 + 2 (\varphi \varphi^*)^4 \pd^2 + 11 (\varphi \varphi^*)^3 \pd^4 + 4 (\varphi \varphi^*)^2 \pd^6 \highlight{ + (\varphi^6 + \varphi^{*6}) \pd^4} \Big]
\notag\\[5pt]
&\quad
+ \order{\text{dim-11}}
\notag\\[5pt]
&\hspace{-40pt}
= 1 + q^2 + 2\, q^4 + 3\, q^6 + 6\, q^8 + \highlight{2\, q^9} + (18 \highlight{+2})\, q^{10} + \order{q^{11}} \,.
\end{align}
\end{subequations}
We see that their \highlight{disagreement} starts from $\order{q^9}$ and are due to derivative interactions. The two extra dimension-9 operators in the $SU(3)$ case are due to baryon combinations of the form
\begin{equation}
\Big[ \epsilon^{ijk} \varphi_i (\pd_\mu \varphi_j) (\pd_\nu \varphi_k) \Big] \Big[ (\pd_\mu \varphi^\dagger) (\pd_\nu \varphi) \Big] + \hc \,.
\label{eqn:OpsN3}
\end{equation}
In summary, the Class I all-order accidental symmetry $SU(3) \implies SU(3)\times U(1)$ is broken by the derivative interactions, but it becomes a finite-order accidental symmetry that holds up to mass dimension eight.

For the case $N=4$, we obtain the following results up to mass dimension ten:
\begin{subequations}\label{eqn:HSISPMN4}
\begin{align}
\HS_{\I,\text{ IBP}}^{\text{Lorentz}\times SU(4)\times U(1),\, \SPM_\Phi} &= 1 + \phi \phi^* 
+ \Big[ (\phi \phi^*)^2 + \phi \phi^* \pd^2 \Big]
+ \Big[ (\phi \phi^*)^3 + 2 (\phi \phi^*)^2 \pd^2 \Big]
\notag\\[5pt]
&\quad
+ \Big[ (\phi \phi^*)^4 + 2 (\phi \phi^*)^3 \pd^2 + 3 (\phi \phi^*)^2 \pd^4 \Big]
\notag\\[5pt]
&\quad
+ \Big[ (\phi \phi^*)^5 + 2 (\phi \phi^*)^4 \pd^2 + 11 (\phi \phi^*)^3 \pd^4 + 4 (\phi \phi^*)^2 \pd^6 \Big]
\notag\\[5pt]
&\quad
+ \order{\text{dim-11}} 
\notag\\[5pt]
&= 1 + q^2 + 2\, q^4 + 3\, q^6 + 6\, q^8 + 18\, q^{10} + \order{q^{11}} \,, \\[12pt]
\HS_{\I,\text{ IBP}}^{\text{Lorentz}\times SU(4),\, \SPM_\Phi} &= 1 + \varphi \varphi^* 
+ \Big[ (\varphi \varphi^*)^2 + \varphi \varphi^* \pd^2 \Big]
+ \Big[ (\varphi \varphi^*)^3 + 2 (\varphi \varphi^*)^2 \pd^2 \Big]
\notag\\[5pt]
&\quad
+ \Big[ (\varphi \varphi^*)^4 + 2 (\varphi \varphi^*)^3 \pd^2 + 3 (\varphi \varphi^*)^2 \pd^4 \Big]
\notag\\[5pt]
&\quad
+ \Big[ (\varphi \varphi^*)^5 + 2 (\varphi \varphi^*)^4 \pd^2 + 11 (\varphi \varphi^*)^3 \pd^4 + 4 (\varphi \varphi^*)^2 \pd^6
\notag\\[0pt]
&\hspace{40pt}
\highlight{ + (\varphi^4 + \varphi^{*4}) (\varphi \varphi^*) \pd^4 + (\varphi^4 + \varphi^{*4}) \pd^6 } \Big]
\notag\\[5pt]
&\quad
+ \order{\text{dim-11}}
\notag\\[5pt]
&= 1 + q^2 + 2\, q^4 + 3\, q^6 + 6\, q^8 + (18 \highlight{+4})\, q^{10} + \order{q^{11}} \,.
\end{align}
\end{subequations}
We see that their \highlight{disagreement} starts from $\order{q^{10}}$ and are due to derivative interactions. The four extra dimension-10 operators in the $SU(4)$ case are due to baryon combinations of the form
\begin{subequations}\label{eqn:OpsN4}
\begin{align}
& \Big[\, \epsilon^{\mu\nu\rho\sigma}\, \epsilon^{ijkl}\, (\pd_\mu \varphi_i)\, (\pd_\nu \varphi_j)\, (\pd_\rho \varphi_k)\, (\pd_\sigma \varphi_l)\, \Big]\, ( \varphi^\dagger \varphi ) + \hc \,, \\[5pt]
& \Big[\, \epsilon^{ijkl}\, \varphi_i\, (\pd_\mu \varphi_j)\, (\pd_\nu \pd_\rho \varphi_k)\, (\pd^\mu \pd^\nu \pd^\rho \varphi_l)\, \Big] + \hc \,.
\end{align}
\end{subequations}
In summary, the Class I all-order accidental symmetry $SU(4) \implies SU(4)\times U(1)$ is broken by the derivative interactions, but it becomes a finite-order accidental symmetry that holds up to mass dimension nine.

\section{Systematic identification of accidental symmetries}
\label{sec:Systematic}

In \cref{sec:AllOrder}, we discussed accidental symmetries that hold to all orders in the EFT potential interactions $V(\Phi)$. In \cref{sec:FiniteOrder}, we explained that when derivative interactions are included (\ie\ when we consider the full Lagrangian $\Lag_\text{\,EFT}$), these all-order accidental symmetries will hold only up to a finite order in the EFT. In either case, we highlighted that computing and comparing the Hilbert series is a reliable approach to checking/verifying an accidental symmetry. Specifically, we provided concrete criteria of using the Hilbert series to verify all-order and finite-order accidental symmetries in 
\eqref{eqn:HSAccidentalV}, \eqref{eqn:HSAccidentalVk} and \eqref{eqn:HSAccidentalLagk}.

Ideally, however, we would like to find a way of verifying accidental symmetries without explicitly computing the Hilbert series, because this calculation is often challenging and time-consuming; see \cref{appsec:ComputingHS} for a variety of examples. Even more ambitiously, a great milestone in the development of model building technologies would be to establish a systematic approach to identifying accidental symmetries for a given field content $\Phi(x)$ and the imposed symmetry group $H$. Note that this would be much harder than verifying a postulated accidental symmetry $H \implies G$, because it requires to determine the group $G$ from $H$ and $\Phi(x)$, as well as its breaking order in the EFT.

In this section, we present some developments towards the above ambitious goals. As a preparation, we first introduce a novel mathematical construct in \cref{subsec:Friendship} --- we define a \emph{friendship} relation among pairs of  subgroups $H_1 \subset G$ and $H_2 \subset G$. In \cref{subsec:VerifyFriends}, we explain how to make use of this definition, together with a theorem by  Brion  \cite{brion1993modules, broer1994hilbert, Grinstein:2023njq}, to derive an approach to verifying all-order accidental symmetries without computing the Hilbert series. It works when the building block fields are bosonic, and we will demonstrate its application to all the three classes of all-order accidental symmetries discussed in \cref{sec:AllOrder}. In \cref{subsec:VerifyFriendsMaNonTroppo}, we further generalize this approach to verifying finite-order accidental symmetries, as well as accommodating derivative interactions. We will show how it works for the examples of finite-order accidental symmetries discussed in \cref{sec:FiniteOrder}. Finally in \cref{subsec:Identify}, we will sketch a possible systematic procedure of identifying the accidental symmetries $G$, from a given set of building block fields $\Phi(x)$ and the imposed symmetry group $H$.

\subsection{\emph{Friendship} between subgroups}
\label{subsec:Friendship}

In this subsection, we introduce a new mathematical construct --- a \emph{friendship} relation between two subgroups $H_1$ and $H_2$ of a group $G$. Before getting to its definition, let us first review some known facts and, along the way, set up our notation.

\subsubsection{Branching Rules}
\label{subsubsec:Decomposition}

Consider a group $G$ and a subgroup $H\subset G$. Any representation of $G$ gives a representation of $H$ (a procedure called \emph{restriction}). In particular, an irrep of $G$ could lead to a reducible rep of $H$, which can then be decomposed into a direct sum of $H$ irreps:
\begin{align}
\Irrep_G=\bigoplus_{\Irrep_H^{}} n_G^{} \big( \Irrep_H^{} \big)\; \Irrep_H^{} \,,
\end{align}
where $n_G^{} \big( \Irrep_H^{} \big)$ denotes the multiplicity. This decomposition is called a \emph{branching rule}; its calculation is often non-trivial.\footnote{They have been worked out and tabulated for many cases as a result of the interest in grand unification theories; see \eg\ Ref.~\cite{Slansky:1981yr} or the \textsc{Mathematica} package LieART \cite{Feger:2019tvk}.}
Rather than reviewing branching rules, we give some explicit examples (without proof) to illustrate this decomposition:
\begin{itemize}
\item \textbf{Example:} The invariant irrep of $G$, denoted $\I_G^{}$, gives the invariant irrep of any of its subgroup $H\subset G$:
    \begin{equation}
    \I_G^{} = \I_H^{} \,,\qquad
    \text{for any}\quad
    H\subset G \,.
    \label{eqn:BranchGInv}
    \end{equation}
\item \textbf{Example:} Let $SO(N-1) \subset SO(N)$ be the regular embedding in the first $(N-1)\times(N-1)$ block, with $N$ sufficiently large.\footnote{The branching rules in \cref{eqn:BranchSONvec,eqn:BranchSONadj,eqn:BranchSONsym} hold as is for $N\ge 6$ (\ie, the right-hand side is $SO(5)$ and above). For lower values of $N$, they still hold technically, but one needs to take into account of the low-rank \emph{specializations} (see \eg\ \cite{Cao:2023psi} for a general \emph{clipping rule}). Specifically, at $N=1$, $SO(1)=\{\mathbf{e}\}$ is the trivial group (\ie\ the identity element $\mathbf{e}$ only), and we have the specializations $\ydiagram{1}_{SO(1)}^{} = \I_{SO(1)}^{}$ and $\adj_{SO(1)}^{} = \ydiagram{2}_{SO(1)}^{} = 0$. At $N=2$, $SO(2)=U(1)$ is abelian, so the two-dimensional reps $\ydiagram{1}_{SO(2)}^{}$ and $\ydiagram{2}_{SO(2)}^{}$ are reducible. They decompose as $\ydiagram{1}_{SO(2)}^{} = (+1) \oplus (-1)$ and $\ydiagram{2}_{SO(2)}^{} = (+2) \oplus (-2)$. In addition, the adjoint irrep specializes to the invariant irrep: $\adj_{SO(2)}^{} = \I_{SO(2)}^{}$. At $N=3$, $SO(3)\simeq SU(2)$, and we have the specialization $\adj_{SO(3)}^{} = \ydiagram{1}_{SO(3)}^{}$. At $N=4$, $SO(4)\simeq SU(2)\times SU(2)$, and the adjoint representation $\adj_{SO(4)}^{}$ is reducible.}
The vector irrep, denoted $\ydiagram{1}_{SO(N)}^{}$, decomposes as
    \begin{equation}
    \ydiagram{1}_{SO(N)}^{} = \ydiagram{1}_{SO(N-1)}^{} \,\oplus\, 
    \I_{SO(N-1)}^{} \,.
    \label{eqn:BranchSONvec}
    \end{equation}
    The adjoint irrep, $\adj_{SO(N)}^{} = \ydiagram{1,1}_{SO(N)}^{}$, which is also the antisymmetric two-index irrep (or the two-form irrep), decomposes as
    \begin{equation}
    \adj_{SO(N)}^{} = \adj_{SO(N-1)}^{} \,\oplus\, \ydiagram{1}_{SO(N-1)}^{} \,.
    \label{eqn:BranchSONadj}
    \end{equation}
    The traceless symmetric two-index irrep $\ydiagram{2}_{SO(N)}^{}$ decomposes as
    \begin{equation}
    \ydiagram{2}_{SO(N)}^{} = \ydiagram{2}_{SO(N-1)}^{} \,\oplus\, \ydiagram{1}_{SO(N-1)}^{}
    \,\oplus\, \I_{SO(N-1)}^{} \,.
    \label{eqn:BranchSONsym}
    \end{equation}
\item \textbf{Example:} Let $SU(N-1) \subset SU(N)$ be the regular embedding in the first $(N-1)\times(N-1)$ block, with $N$ sufficiently large.\footnote{The branching rules in \cref{eqn:BranchSUNffbar,eqn:BranchSUNadj,eqn:BranchSUN2index} hold as is for $N\ge 5$ (\ie, the right-hand side is $SU(4)$ and above). For lower values of $N$, they still hold technically, but receives the low-rank \emph{specializations} much as in the $SO(N)$ case. Specifically, at $N=1$, $SU(1)=\{\mathbf{e}\}$ is the trivial group, and we have the specializations $\ydiagram{1}_{SU(1)}^{} = \overline{\ydiagram{1}}_{SU(1)}^{} = \ydiagram{2}_{SU(1)}^{} = \I_{SU(1)}^{}$ and $\adj_{SU(1)}^{} = \ydiagram{1,1}_{SU(1)}^{} = 0$. At $N=2$, we have the specializations $\overline{\ydiagram{1}}_{SU(2)}^{} = \ydiagram{1}_{SU(2)}^{}$, $\adj_{SU(2)}^{} = \ydiagram{2}_{SU(2)}^{}$, and $\ydiagram{1,1}_{SU(2)}^{} = \I_{SU(2)}^{}$. At $N=3$, we have the specialization $\overline{\ydiagram{1}}_{SU(3)}^{} = \ydiagram{1,1}_{SU(3)}^{}$.}
The fundamental irrep $\ydiagram{1}_{SU(N)}^{}$ and the anti-fundamental irrep $\overline{\ydiagram{1}}_{SU(N)}^{}$ decompose as
    \begin{subequations}\label{eqn:BranchSUNffbar}
    \begin{align}
    \ydiagram{1}_{SU(N)}^{} &= \ydiagram{1}_{SU(N-1)}^{} \,\oplus\, \I_{SU(N-1)}^{} \,, \\[5pt]
    \overline{\ydiagram{1}}_{SU(N)}^{} &= \overline{\ydiagram{1}}_{SU(N-1)}^{} \,\oplus\, \I_{SU(N-1)}^{} \,.
    \end{align}
    \end{subequations}
    The adjoint irrep $\adj_{SU(N)}^{}$ decomposes as
    \begin{equation}
    \adj_{SU(N)}^{} = \adj_{SU(N-1)}^{} \,\oplus\, \ydiagram{1}_{SU(N-1)}^{} \,\oplus\, \overline{\ydiagram{1}}_{SU(N-1)}^{} \,\oplus\, \I_{SU(N-1)}^{} \,.
    \label{eqn:BranchSUNadj}
    \end{equation}
    The antisymmetric two-index irrep $\ydiagram{1,1}_{SU(N)}^{}$ and the symmetric two-index irrep $\ydiagram{2}_{SU(N)}^{}$ decompose as
    \begin{subequations}\label{eqn:BranchSUN2index}
    \begin{align}
    \ydiagram{1,1}_{SU(N)}^{} &= \ydiagram{1,1}_{SU(N-1)}^{} \,\oplus\, \ydiagram{1}_{SU(N-1)}^{} \,, \\[5pt]
    \ydiagram{2}_{SU(N)}^{} &= \ydiagram{2}_{SU(N-1)}^{} \,\oplus\, \ydiagram{1}_{SU(N-1)}^{} 
    \,\oplus\, \I_{SU(N-1)}^{} \,.
    \end{align}
    \end{subequations}
\item \textbf{Example:} Let $G=SU(N)$ and $H=U(1)^{N-1} \subset SU(N)$ be its Cartan subgroup. As $U(1)^{N-1}$ is an abelian group, its irreps are one dimensional, which we denote by $(q_1, q_2, \cdots, q_{N-1})$, with $q_i$ the charge under the $i$-th factor of $U(1)^{N-1}$. The ($N$-dimensional) fundamental and anti-fundamental irreps of $SU(N)$ then decompose respectively as
    \begin{small}
    \begin{subequations}\label{eqn:BranchCartanSUNffbar}
    \begin{align}
    \ydiagram{1}_{SU(N)}^{}
    &= \Bigg[ \bigoplus_{i=1}^{N-1} \left( \cdots,0,\, q_i=+1 \,,0, \cdots \right)_{U(1)^{N-1}}^{} \Bigg]
    \oplus\, \left( -1 \,, \cdots,\, -1 \right)_{U(1)^{N-1}}^{} \,, \\[5pt]
    \overline{\ydiagram{1}}_{SU(N)}^{}
    &= \Bigg[ \bigoplus_{i=1}^{N-1} \left( \cdots,0,\, q_i=-1 \,,0, \cdots \right)_{U(1)^{N-1}}^{} \Bigg]
    \oplus\, \left( +1 \,, \cdots,\, +1 \right)_{U(1)^{N-1}}^{} \,.
    \end{align}
    \end{subequations}
    \end{small}%
    The irreps in the squared brackets have vanishing charges apart from $q_i$. With similar notation, the adjoint irrep of $SU(N)$ decomposes as
    \begin{align}
    \adj_{SU(N)}^{} &= \Bigg[ \bigoplus_{1\le i<j \le N-1} \left( \cdots,\, 0,\, q_i=\pm 1,\, 0,\, \cdots,\, 0,\, q_j=\mp 1,\, 0,\, \cdots \right)_{U(1)^{N-1}}^{} \Bigg]
    \notag\\[5pt]
    &\hspace{20pt}
    \oplus\, \Bigg[ \bigoplus_{i=1}^{N-1} \left( \pm 1 \,, \cdots,\, \pm 1 \,,\, q_i=\pm 2 \,,\, \pm 1 \,, \cdots,\, \pm 1 \right)_{U(1)^{N-1}}^{} \Bigg]
    \notag\\[5pt]
    &\hspace{20pt}
    \oplus\; (N-1)\; \I_{U(1)^{N-1}}^{} \,.
    \label{eqn:BranchCartanSUNadj}
    \end{align}
    The sum in the first line is over $(N-1)(N-2)$ terms that have $q_i=\pm 1, q_j=\mp 1$ with $i<j$ and all the other charges zero. The sum in the second line has $2(N-1)$ terms with $q_i=\pm 2$ and all the other charges $\pm 1$, as indicated. Together with the $(N-1)$ invariants in the last line, they form an $(N^2-1)$-dimensional reducible rep of the Cartan subgroup $U(1)^{N-1}$.
\end{itemize}

\subsubsection{Number of $H$-invariants in a $G$-irrep: $\bm{\tdim \big(\Irrep_G^H\big)}$}
\label{subsubsec:DimIrrepGH}

Above we provided a number of explicit examples of a $G$-irrep decomposing into a direct sum of irreps under a subgroup $H\subset G$. For purposes of studying accidental symmetries, we are in particular interested in the invariant irrep components, $\I_H^{}$, in this kind of decompositions. For convenience, we use ``$\Irrep_G^H$'' to denote the subspace in $\Irrep_G^{}$ that correspond to these components. Consequently, ``$\tdim \big( \Irrep_G^H \big)$'' gives the multiplicity of $\I_H^{}$, \ie, the number of $H$-invariants contained in $\Irrep_G$:
\begin{equation}
\Irrep_G = \tdim \big( \Irrep_G^H \big)\; \I_H \;\oplus\; \text{non-invariant } \Irrep_H \,.
\label{eqn:dimRepGHdef}
\end{equation}
We learned from the various examples above that this multiplicity can be zero in some decompositions, while non-zero in  others.

Specifically, from the example in \cref{eqn:BranchGInv}, we can read off the multiplicity
\begin{equation}
\tdim \big( \I_G^H \big) = 1 \,,\qquad
\text{for any}\quad
H\subset G \,.
\label{eqn:dimInv}
\end{equation}
It says that the invariant $G$-irrep, $\I_G$, contains one, and only one,  $\I_H$ component under any of its subgroups $H\subset G$.

Similarly, from the examples in \cref{eqn:BranchSONvec,eqn:BranchSONadj,eqn:BranchSONsym}, we read off the multiplicities:
\begin{subequations}\label{eqn:dimSON}
\begin{align}
\tdim \Big( \ydiagram{1}_{SO(N)}^{SO(N-1)} \Big) &= 1 \,, \\[5pt]
\tdim \Big( \adj_{SO(N)}^{SO(N-1)} \Big) &= 0 \,, \\[5pt]
\tdim \Big( \ydiagram{2}_{SO(N)}^{SO(N-1)} \Big) &= 1 \,,
\end{align}
\end{subequations}
which generally hold for $N\ge 6$: the $SO(N)$ vector irrep $\ydiagram{1}_{SO(N)}^{}$ and traceless symmetric two-index irrep $\ydiagram{2}_{SO(N)}^{}$ each contains an $\I_{SO(N-1)}^{}$ component, while the adjoint irrep $\adj_{SO(N)}^{}$ does not. At lower values of $N$, \cref{eqn:BranchSONvec,eqn:BranchSONadj,eqn:BranchSONsym} specialize into their low-rank versions (see the footnote around them), which may result in multiplicities different from these general results at $N\ge 6$.

Moving on to the examples in \cref{eqn:BranchSUNffbar,eqn:BranchSUNadj,eqn:BranchSUN2index}, we read off the multiplicities:
\begin{subequations}\label{eqn:dimSUN}
\begin{align}
\tdim\Big( \ydiagram{1}_{SU(N)}^{SU(N-1)} \Big) =
\tdim\Big( \overline{\ydiagram{1}}_{SU(N)}^{SU(N-1)} \Big) &= 1 \,, \\[5pt]
\tdim\Big( \adj_{SU(N)}^{SU(N-1)} \Big) &= 1 \,, \\[5pt]
\tdim\Big( \ydiagram{1,1}_{SU(N)}^{SU(N-1)} \Big) &= 0 \,, \\[5pt]
\tdim\Big( \ydiagram{2}_{SU(N)}^{SU(N-1)} \Big) &= 1 \,,
\end{align}
\end{subequations}
which hold generally for $N\ge5$: the $SU(N)$ fundamental irrep $\ydiagram{1}_{SU(N)}^{}$, anti-fundamental irrep $\overline{\ydiagram{1}}_{SU(N)}^{}$, adjoint irrep $\adj_{SU(N)}^{}$, and symmetric two-index irrep $\ydiagram{2}_{SU(N)}^{}$ each contains one $\I_{SU(N-1)}^{}$ component in the decomposition, while the antisymmetric two-index irrep $\ydiagram{1,1}_{SU(N)}^{}$ does not. At lower values of $N$, \cref{eqn:BranchSUNffbar,eqn:BranchSUNadj,eqn:BranchSUN2index} specialize into their low-rank versions (see the footnote around them), which may result in multiplicities different from these general results at $N\ge 5$.

For the Cartan subgroup examples in \cref{eqn:BranchCartanSUNffbar,eqn:BranchCartanSUNadj}, we read off the multiplicities:
\begin{subequations}\label{eqn:dimSUNCartan}
\begin{align}
\tdim\Big( \ydiagram{1}_{SU(N)}^\text{Cartan} \Big) =
\tdim\Big( \overline{\ydiagram{1}}_{SU(N)}^\text{Cartan} \Big) &= 0 \,, \\[5pt]
\tdim\Big( \adj_{SU(N)}^\text{Cartan} \Big) &= N-1 \,.
\end{align}
\end{subequations}
The $SU(N)$ fundamental irrep $\ydiagram{1}_{SU(N)}^{}$ and anti-fundamental irrep $\overline{\ydiagram{1}}_{SU(N)}^{}$ do not contain an invariant component under the Cartan subgroup. On the other hand, the adjoint irrep $\adj_{SU(N)}^{}$ contains $(N-1)$ invariants under the Cartan subgroup.

Finally, let us mention that when we take the special subgroup $H=G$, it follows from the definition of irreducible representations that
\begin{equation}
\renewcommand\arraystretch{1.3}
\tdim \big( \Irrep_G^G \big) =
\left\{\begin{array}{ll}
1 &\quad \Irrep_G=\I_G \\
0 &\quad \text{otherwise}
\end{array} \right. \,.
\label{eqn:dimH=G}
\end{equation}
When we consider the trivial subgroup $H=\{\mathbf{e}\}$ (\ie\ the identity element $\mathbf{e}$ only), the multiplicity is given by the dimension of the given $G$-irrep:
\begin{equation}
\tdim \big( \Irrep_G^{\{\mathbf{e}\}} \big) = \dim \left( \Irrep_G \right) \,.
\label{eqn:dimH=e}
\end{equation}

We are now ready to introduce the friendship relation.

\subsubsection{Definition of \emph{friends} among subgroups}
\label{subsubsec:Friends}

\vspace{5pt}
\begin{tcolorbox}[colback=light-gray]
\begin{center}
\begin{minipage}{5.5in}\vspace{5pt}
Let $H_1$ and $H_2$ be two subgroups of $G$. We say that $H_1$ and $H_2$ are \emph{friends} of each other when the only $G$-irrep that contains both $\I_{H_1}^{}$ and $\I_{H_2}^{}$ is $\I_G^{}$:
\begin{equation}
\tdim \big( \Irrep_G^{H_1} \big)\, \tdim \big( \Irrep_G^{H_2} \big) = 0
\qquad\forall\qquad
\Irrep_G^{} \ne \I_G^{} \,.
\label{eqn:Friend}
\end{equation}
\end{minipage}\vspace{5pt}
\end{center}
\end{tcolorbox}
\noindent
To visualize this definition, one can imagine listing out all the $G$-irreps. For a Lie group $G$, there are infinitely many of them. They fall into the four types enumerated in \cref{tab:FourTypes}, based on their decompositions under two subgroups $H_1$ and $H_2$. From \cref{eqn:dimInv}, we know that the invariant irrep, $\I_G^{}$, belongs to type-1. When $\I_G^{}$ is the only type-1 $G$-irrep, then the condition in the statement \eqref{eqn:Friend} holds:  $H_1$ and $H_2$ are friends of each other. In other words, for $H_1$ and $H_2$ to be friends, all $\Irrep_G^{} \ne \I_G^{}$ need to fall into the last three types in \cref{tab:FourTypes}.

\begin{table}[t]
\renewcommand{\arraystretch}{1.3}
\setlength{\arrayrulewidth}{.2mm}
\setlength{\tabcolsep}{1.25em}
\centering
\begin{tabular}{ccc}
\toprule
$\Irrep_G^{}$ & $\tdim \big( \Irrep_G^{H_1} \big) \ne 0$ & $\tdim \big( \Irrep_G^{H_2} \big) \ne 0$ \\
\midrule
type-1 & \checkmark & \checkmark \\
\midrule
type-2 & \checkmark & \ding{55} \\
\midrule
type-3  & \ding{55} & \checkmark \\
\midrule
type-4  & \ding{55} & \ding{55} \\
\bottomrule
\end{tabular}
\caption{Four types of $G$-irreps when considering their decompositions under two subgroups $H_1$ and $H_2$, where $\tdim \big( \Irrep_G^{H_i} \big)$ denotes the number of $H_i$-invariants contained in $\Irrep_G$.  When the only type-1 $G$-irrep is the invariant irrep $\I_G^{}$, \cref{eqn:Friend} holds, then $H_1$ and $H_2$ are \emph{friends}. A check mark \checkmark\ means the condition holds, while \ding{55}\ means it does not.}
\label{tab:FourTypes}
\end{table}

The friendship relation defined above is a binary relation, which is manifestly symmetric by the definition \eqref{eqn:Friend}: if $H_1$ and $H_2$ are friends, then $H_2$ and $H_1$ are friends. However, we emphasize that friendship is not an equivalence relation --- it is neither reflexive nor transitive. It is not reflexive, because  no proper subgroup $H\subsetneq G$ is a friend of itself ($H\subset G$ {\em is} its own friend when $H=G$; see the second example below). It is not transitive, because $H_1, H_2$ being friends and $H_2, H_3$ being friends does not imply that $H_1$ and $H_3$ are friends.

To get some intuition about this friendship relation between subgroups, let us give a few simple examples:
\begin{itemize}
\item \textbf{Example:} For $H_1=G$, any subgroup $H_2\subset G$ is a friend. This follows from \cref{eqn:dimH=G}:
    \begin{equation}
    \tdim \big( \Irrep_G^{H_1} \big) = 0
    \qquad\forall\qquad
    \Irrep_G^{} \ne \I_G^{} \,,
    \end{equation}
    which guarantees \cref{eqn:Friend}. In the language of \cref{tab:FourTypes}, all $\Irrep_G^{} \ne \I_G^{}$ fall into type-3 and type-4. Note that this does not imply that all subgroups $H\subset G$ are friends of each other, because the friendship relation is not transitive.
\item \textbf{Example:} For $H_1=\{\mathbf{e}\}$, the only friend is $H_2=G$. This is due to \cref{eqn:dimH=e}, which implies
    \begin{equation}
    \tdim \big( \Irrep_G^{H_1} \big) \ne 0
    \qquad\forall\qquad
    \Irrep_G^{} \,.
    \label{eqn:H1e}
    \end{equation}
    Therefore, for the condition in \cref{eqn:Friend} to hold, its friend $H_2$ needs to satisfy
    \begin{equation}
    \tdim \big( \Irrep_G^{H_2} \big) = 0
    \qquad\forall\qquad
    \Irrep_G^{} \ne \I_G^{} \,.
    \label{eqn:H1eH2}
    \end{equation}
    One can show that the only possibility is $H_2=G$ (using \eg\ Theorem 11.1.13 in Ref.~\cite{goodman2009symmetry}). In the language of \cref{tab:FourTypes}, all $\Irrep_G^{} \ne \I_G^{}$ fall into type-2 in this example.
\item \textbf{Example:} When $G = H_1 \times H_2$, the two subgroups $H_1$ and $H_2$ are friends. This is because all the irreps of $G$ are given by direct products of the irreps of $H_1$ and $H_2$:
    \begin{equation}
    \Irrep_G^{} = \Irrep_{H_1}^{} \;\otimes\; \Irrep_{H_2}^{} \,.
    \end{equation}
    Using \cref{eqn:dimH=G} for both $H_1$ and $H_2$, we see that the only $\Irrep_G^{}$ that contains both $\I_{H_1}^{}$ and $\I_{H_2}^{}$ is $\I_G^{} = \I_{H_1}^{} \otimes \I_{H_2}^{}$, satisfying criterion \eqref{eqn:Friend}.
\item \textbf{Example:} If $F$ is a friend of $H_1$ and $F \subset H_2 \subset G$, then $H_2$ is also a friend of $H_1$. To show this, we divide all $\Irrep_G^{} \ne \I_G^{}$ into two categories: (1) $\tdim \big( \Irrep_G^{H_1} \big) = 0$, and (2) $\tdim \big( \Irrep_G^{H_1} \big) \ne 0$. $G$-irreps in category (1) do not violate the condition \eqref{eqn:Friend}. For $G$-irreps in category (2), we have $\tdim \big( \Irrep_G^F \big) = 0$ because $F$ is a friend of $H_1$. But this implies $\tdim(\Irrep_G^{H_2}) = 0$, as $F$ is a subgroup of $H_2$. So $G$-irreps in category (2) also satisfy the condition \eqref{eqn:Friend}, and therefore $H_1$ and $H_2$ are friends.    
\end{itemize}

Intuitively, when $H_1$ and $H_2$ are friends, they are a kind of ``complementary partners'': the condition $\tdim \big( \Irrep_G^{H_1} \big) = 0$ holds for a certain subset of $\Irrep_G^{} \ne \I_G^{}$, while the condition $\tdim \big( \Irrep_G^{H_2} \big) = 0$ need to cover the rest. Roughly speaking, the bigger the subgroup $H_1$ is, the more $G$-irreps will satisfy $\tdim \big( \Irrep_G^{H_1} \big) = 0$, and the easier it is to find a friend of it. In the extreme case $H_1=G$, $\tdim \big( \Irrep_G^{H_1} \big) = 0$ is satisfied by all $\Irrep_G^{} \ne \I_G^{}$ (see \cref{eqn:dimH=G}), and any subgroup $H_2\subset G$ is a friend of it. In the other extreme $H_1=\{\mathbf{e}\}$, $\tdim \big( \Irrep_G^{H_1} \big) = 0$ holds for none of $\Irrep_G^{} \ne \I_G^{}$ (see \cref{eqn:dimH=e}), and its only friend is $H_2=G$. This ``complementary partners'' nature of the friendship relation are reflected by the examples above.

\subsection{Verifying all-order accidental symmetries: \emph{friends}}
\label{subsec:VerifyFriends}

In this subsection, we make use of the friendship relation introduced in \cref{subsec:Friendship} to derive a criterion for telling when the two Hilbert series $\HS_\I^{G,\, \Phi}$ and $\HS_\I^{H,\, \Phi}$ are equal, without explicitly computing them. We will apply this criterion to verify the three classes of all-order accidental symmetries discussed in \cref{sec:AllOrder}.
\begin{tcolorbox}[colback=light-gray]
\begin{center}
\begin{minipage}{5.5in}\vspace{3pt}
\textbf{Criterion for all-order accidental symmetries:}
Let $H_\Phi$ be the subgroup of $G$ that remains unbroken\footnotemark\ under a generic vacuum expectation value (vev) $\langle\Phi\rangle$ of the bosonic building block fields $\Phi(x)$:
\begin{equation}
G \;\xrightarrow[]{\;\;\langle\Phi\rangle\;\;}\; H_\Phi \subset G \,,
\label{eqn:HPhidef}
\end{equation}
then for any subgroup $H\subset G$, the following two statements are equivalent
\begin{equation}
\HS_\I^{G,\, \Phi}(q) = \HS_\I^{H,\, \Phi}(q)
\qquad\Longleftrightarrow\qquad
\text{$H$ is a friend of $H_\Phi$} \,.
\label{eqn:AllOrderCriterion}
\end{equation}
It means that imposing any friend $H$ of $H_\Phi$ on the $\Phi$ polynomials will be sufficient to guarantee $G$-invariance. We call $H$ a \emph{sufficient subgroup} for $G$-invariance of $\Phi$ polynomials.
\end{minipage}\vspace{3pt}
\end{center}
\end{tcolorbox}
\footnotetext{$H_\Phi$ is also called stabilizer subgroup (of a generic vev), little group or isotropy group.}
\noindent
The proof of this criterion takes a few steps.

First, because $\Phi(x)$ is a $G$-rep, the space of all its polynomials, $\{\Phi^k\}$, forms an infinite-dimensional reducible rep of $G$. Its decomposition into $G$-irreps contains $n\big( \Irrep_G^{} \big)$ copies of $\Irrep_G^{}$:\footnote{The multiplicity $n\big( \Irrep_G^{} \big)$ here is technically infinite. One should think of the decomposition in \cref{eqn:PhikDecomposition} first for each subspace $\Phi^k$ that collects the homogeneous polynomials of degree $k$:
\begin{equation}
\Phi^k = \Big\{ \Phi_1^{k_1} \cdots \Phi_s^{k_s} \,,\quad
k_1  + \cdots + k_s = k \Big\}
\quad\text{with $s$ the total number of components in $\Phi$} \,.
\end{equation}
In each of these subspaces, we have a well-defined multiplicity $n_k\big( \Irrep_G^{} \big)$. The total multiplicity $n\big( \Irrep_G^{} \big)$ above is a ``sum'' of them. A proper grading is typically required to make the sum converge, which would precisely correspond to the Hilbert series (of covariants):
\begin{equation}
\HS_{\Irrep_G^{}}^{G,\,\Phi} (q) = \sum_{k=0}^\infty n_k\big( \Irrep_G^{} \big)\, q^k \,.
\end{equation}
To make the proof easier to follow, we gloss over these technical details below, and simply call these Hilbert series as multiplicities $n\big( \Irrep_G^{} \big)$.}
\begin{equation}
\{\Phi^k\} = n\big( \I_G^{} \big)\; \I_G^{} \oplus
\left[ \bigoplus_{\Irrep_G^{} \ne \I_G^{}} n\big( \Irrep_G^{} \big)\; \Irrep_G^{} \right] \,,
\label{eqn:PhikDecomposition}
\end{equation}
where we have singled out $\I_G^{}$ because of the special role it plays below.

Next, the polynomial space $\{\Phi^k\}$ also forms a rep of any given subgroup $H\subset G$. Therefore, it has a similar decomposition into $H$-irreps, which can be obtained by applying the branching rules to each $G$-irrep on the right-hand side of \eqref{eqn:PhikDecomposition}. Specifically, applying \cref{eqn:dimRepGHdef} (as well as \cref{eqn:BranchGInv}), we get
%
\begin{align}
\{\Phi^k\} &= n\big( \I_G^{} \big)\; \I_H^{} \oplus 
\bigoplus_{\Irrep_G^{} \ne \I_G^{}}
n\big( \Irrep_G^{} \big)\,
\Big[ \tdim \big( \Irrep_G^H \big)\, \I_H^{} \oplus \text{non-invariant } \Irrep_H \Big]
\notag\\[10pt]
&= \Big[ n\big( \I_G^{} \big) + \sum_{\Irrep_G^{} \ne \I_G^{}} n\big( \Irrep_G^{} \big)\; \tdim \big( \Irrep_G^H \big) \Big]\, \I_H^{}
\notag\\[5pt]
&\hspace{40pt}
\oplus  \text{non-invariant } \Irrep_H \,.
\end{align}
Collecting all coefficients of $\I_H^{}$, we see that
\begin{equation}
n\big( \I_H^{} \big) = n\big( \I_G^{} \big) + \sum_{\Irrep_G^{} \ne \I_G^{}} n\big( \Irrep_G^{} \big)\; \tdim \big( \Irrep_G^H \big) \,.
\label{eqn:nInvH}
\end{equation}

It follows that $n\big( \I_H^{} \big) = n\big( \I_G^{} \big)$ (\ie\ $\HS_\I^{H,\, \Phi} = \HS_\I^{G,\, \Phi}$) if and only if the second term in \cref{eqn:nInvH} vanishes. As it is a sum of products of non-negative numbers, this is equivalent to requiring each term to vanish:
\begin{equation}
n\big( \Irrep_G^{} \big)\; \tdim \big( \Irrep_G^H \big) = 0
\qquad\forall\qquad
\Irrep_G^{} \ne \I_G^{} \,.
\label{eqn:nIrrepGdim}
\end{equation}

Finally, we recall that not all $G$-irreps appear in the decomposition in \cref{eqn:PhikDecomposition} --- some have $n\big( \Irrep_G^{} \big)=0$, \ie, they cannot be generated by composing any number of the building blocks. For bosonic building blocks $\Phi$ that generate non-trivial $G$-invariants, a theorem by Brion \cite{brion1993modules, broer1994hilbert, Grinstein:2023njq}\footnote{This theorem says that the rank of the module of $\Irrep_G$ is equal to the number of $H_\Phi$-invariants contained in $\Irrep_G$, \ie\
\begin{equation}
\rank {\left({}_{\mathbb{r}_\I}\mathcal{M}_{\Irrep_G}^{G,\,\Phi} \right) = \tdim \big( \Irrep_G^{H_\Phi} \big)} \,.
\label{eqn:Brion}
\end{equation}
We refer the reader to Ref.~\cite{Grinstein:2023njq} for detailed explanations.
}
implies that those (and only those) $G$-irreps that contain (at least) one $H_\Phi$-invariant will appear in \cref{eqn:PhikDecomposition}:
\begin{equation}
n\big( \Irrep_G^{} \big) \ne 0
\qquad\Longleftrightarrow\qquad
\tdim \big( \Irrep_G^{H_\Phi} \big) \ne 0 \,.
\label{eqn:nIrrepG}
\end{equation}
In these cases, the condition in \cref{eqn:nIrrepGdim} is equivalent to
\begin{equation}
\tdim \big( \Irrep_G^{H_\Phi} \big)\; \tdim \big( \Irrep_G^H \big) = 0
\qquad\forall\qquad
\Irrep_G^{} \ne \I_G^{} \,.
\end{equation}
This is precisely the condition that $H$ and $H_\Phi$ are friends (see \eqref{eqn:Friend}). This completes our proof of the assertion in \eqref{eqn:AllOrderCriterion}.

\subsubsection*{Caveat: fermionic case}

As mentioned just before  \eqref{eqn:HPhidef}, the above criterion is only guaranteed to work for bosonic building block fields. When fermionic building block fields are involved, there are a few technical complications, which could invalidate the proof presented above:
\begin{itemize}
\item When the building block fields $\Psi$ is Grassmann odd, one needs to scrutinize the role that the unbroken subgroup $H_\Psi$ plays in the theorem by Brion in \cref{eqn:Brion}.
\item For purely fermionic building block fields, the polynomial space $\{\Psi^k\}$ is finite dimensional, and consequently the Hilbert series are finite polynomials, as seen in \cref{eqn:HSIFermionDisagree,eqn:HSIIIFermionDistinct}. Because of this, one need to double check if Brion's theorem in \cref{eqn:Brion} still holds, especially regarding the meaning of the rank of the module.
\item The equivalence in \cref{eqn:nIrrepG} no longer holds, whether the Brion's theorem in \cref{eqn:Brion} still holds or not. As the polynomial space $\{\Psi^k\}$ is now finite dimensional, we have $n\big( \Irrep_G^{} \big) \ne 0$ only for a finite number of $G$-irreps, which is not determined by a condition of the type $\tdim \big( \Irrep_G^{H_\Psi} \big) \ne 0$. This failure directly invalidates the proof presented above.
\end{itemize}
We leave an investigation on these complications to future work.

\subsubsection{Application to Class I all-order accidental symmetries}
\label{subsubsec:FriendsClassI}

Let us apply the criterion in \cref{eqn:AllOrderCriterion} to verifying the Class I all-order accidental symmetries discussed in \cref{subsec:ClassI}. In this case, we have $G = SU(N)\times U(1)$ and the building blocks fields $\Phi = (\phi, \phi^*)$ as described in \cref{tab:ClassI}.

Our first step is to determine the unbroken subgroup $H_\Phi\subset G$ when $\Phi$ takes a generic vev $\langle \Phi \rangle$. We begin by writing down all the elements in $G$:
\begin{equation}
g = g_N^{}\, g_1
\qquad\text{with}\qquad
g_N^{} \in SU(N) \,,\quad
g_1 \in U(1) \,.
\end{equation}
Elements in the unbroken subgroup $h_\Phi^{} \in H_\Phi$ are selected from these by requiring that their representation matrices in the $\Phi$ rep, $M_\Phi(h_\Phi^{})$, leave the vev $\langle \Phi \rangle$ unchanged:
\begin{equation}
M_\Phi (h_\Phi^{})\, \langle \Phi \rangle = \langle \Phi \rangle \,.
\end{equation}
We can write $\Phi = (\phi, \phi^*)$ as an $N$-dimensional complex vector, and without loss of generality, put its vev in the last component. In this basis, the $N\times N$ representation matrix $M_\Phi(h_\Phi^{})$ is determined as:\footnote{We have further block-diagonalized this representation matrix, which can be achieved for any rep of a reductive group.}
\begin{equation}
\langle \Phi \rangle = \mqty( 0 \\ \vdots \\ 0 \\ v )
\qquad\Longrightarrow\qquad
M_\Phi (h_\Phi) = M_\Phi(g_N^{})\, M_\Phi(g_1) = \mqty( M_{N-1}^{} & \\ & 1) \,,
\label{eqn:MhPhi}
\end{equation}
where $M_{N-1}^{}$ is an $(N-1)\times(N-1)$ matrix. Note that the above rep is the defining representation for $g\in G=SU(N)\times U(1)$, in which $M_\Phi(g_N^{})$ has to be a special unitary matrix and $M_\Phi(g_1)$ has to be an overall phase factor. Therefore, the form of \cref{eqn:MhPhi} pins down to the following subgroup elements:
\begin{equation}
h_\Phi^{} = g_N^{}\, g_1
\quad\text{with}\quad
g_N^{} = M_\Phi(g_N^{}) = \mqty( u_{N-1}^{} & \\ & e^{-i\theta}) \,,\quad
g_1 = M_\Phi(g_1) = e^{i\theta} \,,
\label{eqn:hPhi1}
\end{equation}
where the matrix $u_{N-1}^{} \in U(N-1)$ is a unitary matrix, and for $M_\Phi(g_N^{})$ to be a special unitary matrix, the phase is constrained to satisfy
\begin{equation}
e^{i\theta} = \det (u_{N-1}^{}) \,.
\label{eqn:theta1}
\end{equation}
For any chosen element $u_{N-1}^{} \in U(N-1)$, the angle $\theta$ is fully determined by the constraint in \cref{eqn:theta1}, and so will be the group element $g_1 \in U(1)$. We see that the subgroup $H_\Phi$ is a $U(N-1)$ group in this case, with the specific elements given in \cref{eqn:hPhi1,eqn:theta1}.

Now the criterion in \eqref{eqn:AllOrderCriterion} implies that
\begin{equation}
\HS_\I^{SU(N)\times U(1),\, \Phi} = \HS_\I^{F,\, \Phi}
\qquad\text{for any $F\subset G$ that is a friend of $H_\Phi$} \,.
\end{equation}
In particular, let us check that the subgroup $H=SU(N)$ in \cref{subsec:ClassI} is indeed a friend of the above $H_\Phi$ given in \cref{eqn:hPhi1,eqn:theta1}. To this end, we check the condition in \eqref{eqn:Friend}. Because $G=H\times U(1)$, all the $G$-irreps can be listed by an $H$-irrep and a charge $Q$ under the $U(1)$ factor:
\begin{equation}
\Irrep_{G=SU(N)\times U(1)}^{} = \left( \Irrep_{H=SU(N)}^{} \;,\; Q \right) \,.
\end{equation}
Among all these $G$-irreps, we only need to concern about those with $\tdim \big( \Irrep_G^H \big) \ne 0$ and $\Irrep_G^{} \ne \I_G^{}$. Applying \cref{eqn:dimH=G}, this means
\begin{equation}
\Irrep_G^{} = \left( \mathbf{1} \;,\; Q\ne 0 \right) \,.
\label{eqn:GirrepsConcern}
\end{equation}
In these one-dimensional irreps of $G$, elements in the subgroup $H_\Phi$ (\ie\ those elements in \cref{eqn:hPhi1}) are represented by the following matrices
\begin{equation}
M ( h_\Phi^{} ) = M( g_N^{} )\, M ( g_1 )
\qquad\text{with}\qquad
M ( g_N^{} ) = 1 \,,\quad
M ( g_1 ) = e^{iQ \theta} \,.
\label{eqn:MhPhiQ}
\end{equation}
Since $Q\ne0$,  these irreps are not $H_\Phi$-invariants, hence
\begin{equation}
\tdim \big[ \left( \mathbf{1} \;,\; Q \ne 0 \right)_G^{H_\Phi} \big] = 0 \,.
\label{eqn:dimGirrepsHPhi}
\end{equation}
Therefore, the condition in \cref{eqn:Friend} holds for $H$ and $H_\Phi$ --- they are friends.

The same argument above can be generalized to the case of multiple flavors summarized in \cref{tab:ClassIk1k2}, where $\Phi = (k_1 \times \phi,\, k_2\times \phi^*)$ with $k = \max \left( k_1, k_2 \right) < N$. Following the same procedure of driving \cref{eqn:hPhi1,eqn:theta1}, we determine the unbroken subgroup elements $h_\Phi^{} \in H_\Phi$ in this case as
\begin{equation}
h_\Phi^{} = g_N^{}\, g_1
\qquad\text{with}\qquad
g_N^{} = \mqty( u_{N-k}^{} & 0 \\ 0 & e^{-i\theta}\, \mathbb{1}_{k\times k} ) \,,\quad
g_1 = e^{i\theta} \,,
\label{eqn:hPhik}
\end{equation}
with a unitary matrix $u_{N-k}^{} \in U(N-k)$ and the phase constrained to satisfy
\begin{equation}
e^{ik\theta} = \det (u_{N-k}^{}) \,.
\label{eqn:thetak}
\end{equation}
We see that \cref{eqn:hPhi1,eqn:theta1} can be recovered from these by taking $k=1$. Now for any chosen element $u_{N-k}^{} \in U(N-k)$, the angle $\theta$ is not fully determined by the constraint in \cref{eqn:thetak} --- one can shift it by any integer multiple of $2\pi/k$. This allows for inequivalent options for the group element $g_1$, corresponding to the shift of $\theta$ by $2\pi/k \times (0, 1, \cdots, k-1)$. Therefore, $H_\Phi$ is a $U(N-k) \times \mathbb{Z}_k$ group, with the elements specifically given in \cref{eqn:hPhik,eqn:thetak}. Similar with the $k=1$ case, \cref{eqn:MhPhiQ,eqn:dimGirrepsHPhi} still hold for the new $H_\Phi$, and therefore it is  a friend of $H$. This explains the multiple flavor case of Class I all-order accidental symmetries. Note that this class stops working when there are too many flavors of fields, \ie, $k = \max \left( k_1, k_2 \right) \ge N$. We will come back to explain this in \cref{subsubsec:VerifyClassIFlavor}.

The above discussions also give us a general intuition about why the all-order accidental symmetries typically stop working at a certain number of flavors: as the number of flavors of irreps increases, the unbroken subgroup $H_\Phi$ will shrink (see \eg\ \cref{eqn:hPhik}), and it becomes harder to find a friend of $H_\Phi$. If $H_\Phi$ shrinks down to the trivial subgroup $\{\mathbf{e}\}$, then its only friend is the full group $H=G$ (see discussions around \cref{eqn:H1e,eqn:H1eH2}). When that is the case, the only sufficient group for $G$-invariance is $G$ itself.

\subsubsection{Application to Class II all-order accidental symmetries}
\label{subsubsec:FriendsClassII}

Now let us make use of the criterion in \cref{eqn:AllOrderCriterion} to verify the Class II all-order accidental symmetries discussed in \cref{subsec:ClassII}. This case was summarized in \cref{tab:ClassII}, where $G = SO(2N)$ and the building blocks fields $\Phi = \phi$ form a vector irrep of $G$. When $\Phi$ takes a generic vev, the unbroken subgroup is $H_\Phi=SO(2N-1)$. In what follows, we verify that the subgroup $H=SU(N)\times U(1) \subset G$ described in \cref{subsec:ClassII} is a friend of this $H_\Phi$, \ie\ the condition in \cref{eqn:Friend} holds for $H$ and $H_\Phi$.

\begin{table}[t]
\renewcommand{\arraystretch}{1.3}
\setlength{\arrayrulewidth}{.2mm}
\setlength{\tabcolsep}{1.25em}
\centering
\ytableausetup{boxsize=0.5em}
\begin{tabular}{lcl}
\toprule
Dynkin labels & $m_{1\le i\le r-1}$ & $m_r$ \\
\midrule
$SU(r+1)$ & \multirow{4}{*}{$w_i-w_{i+1}$} & $w_r$ \\
$SO(2r)$ & & $w_{r-1}+w_r$ \\
$SO(2r+1)$ & & $2w_r$ \\
$Sp(2r)$ & & $w_r$ \\
\bottomrule
\end{tabular}
\caption{Dictionary between the Dynkin labels ($m_i$) and the components of the highest weight vector ($w_i$) for irreps of $SU(r+1)$, $SO(2r)$, $SO(2r+1)$, and $Sp(2r)$.}
\label{tab:DynkinLabels}
\end{table}

We begin by systematically listing out all the irreps of the group $G=SO(2N)$; they can be labeled by the highest weight vector:
\begin{equation}
\Irrep_{SO(2N)}^{} = \vb*{w}_{SO(2N)}^{} = \big( w_1,\, w_2,\, \cdots,\, w_{N-1}^{},\, w_N^{} \big) \,,
\label{eqn:SO2Nirreps}
\end{equation}
where the components satisfy the following constraints
\begin{equation}
w_1 \ge w_2 \ge \cdots \ge w_{N-1}^{} \ge |w_N^{}| \,,\qquad
\text{all integers or all half odd integers}\,.
\label{eqn:wConstraintsSO2N}
\end{equation}
There is an alternative labeling of Lie algebra irreps that is often used --- the \emph{Dynkin labels}. For a given irrep, the relation between its highest weight and its Dynkin labels is simple: the highest weight $\vb*{w}$ is the linear combination of the fundamental weights $\vb*{\mu}_k$, with the Dynkin labels $m_k$ as combination coefficients:
\begin{equation}
\vb*{w} = \sum_{k=1}^r m_k\, \vb*{\mu}_k \,,
\end{equation}
where $r$ is the rank of the Lie algebra. For translation convenience, we provide a dictionary between the Dynkin labels and the highest weight vector in \cref{tab:DynkinLabels} for a few rank $r$ Lie groups: $SU(r+1)$, $SO(2r)$, $SO(2r+1)$, and $Sp(2r)$.

For the purpose of checking the condition in \cref{eqn:Friend}, we concern about non-invariant $G$-irreps in \cref{eqn:SO2Nirreps} that satisfy
\begin{equation}
\tdim \big( \Irrep_G^{H_\Phi} \big) = \tdim \big( \Irrep_{SO(2N)}^{SO(2N-1)} \big) \ne 0 \,,
\end{equation}
\ie, $SO(2N)$-irreps which contain $SO(2N-1)$-invariants. To identify them, we make use of the general branching rule for an $SO(2N)$ irrep decomposing into the $SO(2N-1)$ irreps (see \eg\ section 8 in \cite{goodman2009symmetry}):
\begin{equation}
\vb*{w}_{SO(2N)}^{} = \bigoplus_{w_1 \,\ge\, \lambda_1 \,\ge\, w_2 \,\ge\, \lambda_2 \,\ge\, \cdots \,\ge\, w_{N-1}^{} \,\ge\, \lambda_{N-1}^{} \,\ge\, |w_N^{}|} \vb*{\lambda}_{SO(2N-1)}^{} \,,
\label{eqn:BranchSO2NHPhi}
\end{equation}
where $\vb*{\lambda}_{SO(2N-1)}^{} = (\lambda_1, \lambda_2, \cdots, \lambda_{N-1}^{})$ denotes the highest weight vector of an irrep of $SO(2N-1)$ (which has rank $N-1$). For a given $\vb*{w}_{SO(2N)}^{}$, each $\vb*{\lambda}_{SO(2N-1)}^{}$ that satisfies the constraints above will appear once in this decomposition. The examples given in \cref{eqn:BranchSONvec,eqn:BranchSONadj,eqn:BranchSONsym} are special cases of this general rule. The invariant irrep $\I_{H_\Phi}^{}$ corresponds to the highest weight vector
\begin{equation}
\lambda_1 = \lambda_2 = \cdots = \lambda_{N-1}^{} = 0 \,.
\end{equation}
For this $H_\Phi$-irrep to appear on the right-hand side of \cref{eqn:BranchSO2NHPhi}, we require
\begin{equation}
w_1 \ge 0 \,,\qquad
w_2 = \cdots = w_N^{} = 0 \,.
\end{equation}
Together with the constraints in \cref{eqn:wConstraintsSO2N}, we see that the $G$-irreps that we need to concern about are
\begin{equation}
\Irrep_G^{} = \vb*{w}_{SO(2N)}^{} = (n ,\, 0 ,\, \cdots ,\, 0) \,,\qquad
\text{with $n$ positive integers} \,.
\label{eqn:GirrepConcernII}
\end{equation}

Now our last step is to check if any of the $G$-irreps in \cref{eqn:GirrepConcernII} has an invariant component upon decomposing into irreps of the subgroup $H=SU(N)\times U(1)$. Given the embedding of $H\subset G$ in \cref{eqn:EmbeddingII}, one can work out this decomposition: 
\begin{equation}
(n ,\, 0 ,\, \cdots ,\, 0)_{SO(2N)}^{}
= \bigoplus_{k=0}^n\, \Big[ (n ,\, k ,\, \cdots ,\, k) \;,\; n-2k \Big]_{H=SU(N)\times U(1)}^{} \,.
\label{eqn:BranchSO2NH}
\end{equation}
In the squared brackets, the highest weight vector $(n ,\, k ,\, \cdots ,\, k)$ has $N-1$ components; it is denoting an $SU(N)$ irrep, while $n-2k$ is the charge under the $U(1)$ factor in $H$. As $n>0$, we see that this decomposition does not contain an $\I_H^{}$:
\begin{equation}
\tdim \big( \Irrep_G^H \big) = \tdim \Big( (n ,\, 0 ,\, \cdots ,\, 0)_{SO(2N)}^H \Big) = 0 \,.
\end{equation}
Therefore, \cref{eqn:Friend} holds for $H$ and $H_\Phi$ --- they are friends. We also see from \cref{eqn:BranchSO2NH} that the $U(1)$ factor in $H$ is optional, which reflects a combine of Class I and II all-order accidental symmetries.

To help visualize the friendship relation between $H$ and $H_\Phi$, we generate some example tables in the format of \cref{tab:FourTypes} using the \textsc{Mathematica} package LieART \cite{Feger:2019tvk}: the $N=5$ case ($G=SO(10)$) in \cref{tab:SO10Irreps}, and the $N=6$ case ($G=SO(12)$) in \cref{tab:SO12Irreps}. From these tables, one can see explicitly that there is no $\Irrep_G^{} \ne \I_G^{}$ that contains both $\I_H^{}$ and $\I_{H_\Phi}^{}$, up to the irrep dimension listed. One can also check that the $G$-irreps that contain an $H_\Phi$-invariant are given by the type in \cref{eqn:GirrepConcernII}, whose Dynkin labels are (for $N>2$, see \cref{tab:DynkinLabels})
\begin{equation}
\vb*{w}_{SO(2N)}^{} = (n ,\, 0 ,\, \cdots ,\, 0)
\quad\Longleftrightarrow\quad
\textbf{Dynkin}_{SO(2N)}^{} = (n ,\, 0 ,\, \cdots ,\, 0) \,,
\end{equation}
with $n$ positive integers.

\subsubsection{Application to Class III all-order accidental symmetries}
\label{subsubsec:FriendsClassIII}

Finally, let us apply the criterion in \cref{eqn:AllOrderCriterion} to verify the Class III all-order accidental symmetries discussed in \cref{subsec:ClassIII}. This case was summarized in \cref{tab:ClassIII}, where $G=SU(2N)$ and the building block fields $\Phi=\phi$ form the antisymmetric two-index irrep $\ydiagram{1,1}$ . When $\Phi$ takes a generic vev, the unbroken subgroup is $H_\Phi = Sp(2N)$, as directly follows from the definition of the symplectic group. We would like to verify that $H_1=SU(2N-1)\times U(1)$ and $H=SU(2N-1)$ are both friends of $H_\Phi$. Note that because $H\subset H_1 \subset G$, by our last example in \cref{subsubsec:Friends}, we only need to verify that $H$ is a friend of $H_\Phi$.

We start by listing out all the irreps of $G=SU(2N)$, labeled by the highest weight vector:
\begin{equation}
\Irrep_{SU(2N)}^{} = \vb*{w}_{SU(2N)}^{} = \big( w_1,\, w_2,\, \cdots,\, w_{2N-2}^{},\, w_{2N-1}^{} \big) \,,
\label{eqn:SU2Nirreps}
\end{equation}
where the components satisfy the following constraints
\begin{equation}
w_1 \ge w_2 \ge \cdots \ge w_{2N-2}^{} \ge w_{2N-1}^{} \ge 0 \,,\qquad
\text{all integers} \,.
\label{eqn:wConstraintsSU2N}
\end{equation}
In general, the translation between the highest weight vector $\vb*{w}$ and the Young diagram description of an $SU(r+1)$ irrep is that the component $w_k$ gives the number of boxes in the $k$-th row of the Young diagram. The translation to the description by Dynkin labels is given in \cref{tab:DynkinLabels}.

For the purpose of checking the friendship between $H$ and $H_\Phi$, we concern about non-invariant $G$-irreps in \cref{eqn:SU2Nirreps} that satisfy
\begin{equation}
\tdim \big( \Irrep_G^H \big) = \tdim \big( \Irrep_{SU(2N)}^{SU(2N-1)} \big) \ne 0 \,,
\end{equation}
\ie, $SU(2N)$-irreps which contain $SU(2N-1)$-invariants. To identify them, we make use of the general branching rule for an $SU(2N)$ irrep decomposing into the $SU(2N-1)$ irreps (see \eg\ section 8 in \cite{goodman2009symmetry}):
\begin{equation}
\vb*{w}_{SU(2N)}^{} = \bigoplus_{w_1 \,\ge\, \lambda_1+k \,\ge\, w_2 \,\ge\, \lambda_2+k \,\ge\, \cdots \,\ge\, w_{2N-2}^{} \,\ge\, \lambda_{2N-2}^{}+k \,\ge\, w_{2N-1}^{} \,\ge\, k \,\ge\, 0} \vb*{\lambda}_{SU(2N-1)}^{} \,,
\label{eqn:BranchSU2NH}
\end{equation}
where $\vb*{\lambda}_{SU(2N-1)}^{} = (\lambda_1, \lambda_2, \cdots, \lambda_{2N-2}^{})$ denotes the highest weight vector of an irrep of $SU(2N-1)$ (which has rank $2N-2$), and $k$ is an integer. For a given $\vb*{w}_{SU(2N)}^{}$, each choice of ($\vb*{\lambda}_{SU(2N-1)}^{}$, $k$) that satisfies the constraints above will appear once in this decomposition. The examples given in \cref{eqn:BranchSUNffbar,eqn:BranchSUNadj,eqn:BranchSUN2index} are special cases of this general rule. The invariant irrep $\I_H^{}$ corresponds to the highest weight vector
\begin{equation}
\lambda_1 = \lambda_2 = \cdots = \lambda_{2N-2}^{} = 0 \,.
\end{equation}
For this $H$-irrep to appear on the right-hand side of \cref{eqn:BranchSU2NH}, we need
\begin{equation}
w_1 \ge k \ge 0 \,,\qquad
w_2 = \cdots = w_{2N-1}^{} = k \,.
\end{equation}
Together with the constraints in \cref{eqn:wConstraintsSU2N}, we see that the $G$-irreps that we need to concern about are
\begin{equation}
\Irrep_G^{} = \vb*{w}_{SU(2N)}^{} = (n ,\, k ,\, \cdots ,\, k) \,,\quad
\text{with integers $n\ge k \ge 0$ and $n>0$} \,.
\label{eqn:GirrepConcernIII}
\end{equation}

Now our last step is to check if any of the $G$-irreps in \cref{eqn:GirrepConcernIII} has an invariant component upon decomposing into irreps of its subgroup $H_\Phi=Sp(2N)$. To this end, we work out this decomposition: 
\begin{equation}
(n ,\, k ,\, \cdots ,\, k)_{SU(2N)}^{}
= \bigoplus_{p=0}^{\min(k, n-k)}\, (n-p ,\, p ,\, 0 ,\, \cdots ,\, 0)_{H_\Phi=Sp(2N)}^{} \,.
\label{eqn:BranchSU2NHPhi}
\end{equation}
The highest weight vector $(n-p ,\, p ,\, 0 ,\, \cdots ,\, 0)$ has $N$ components, denoting an $Sp(2N)$ irrep. As $n>0$, we see that this decomposition does not contain an $\I_{H_\Phi}^{}$:
\begin{equation}
\tdim \big( \Irrep_G^{H_\Phi} \big) = \tdim \Big( (n ,\, k ,\, \cdots ,\, k)_{SU(2N)}^{H_\Phi} \Big) = 0 \,.
\end{equation}
Therefore, \cref{eqn:Friend} holds for $H$ and $H_\Phi$ --- they are friends.

To help visualize the friendship relation between $H$ and $H_\Phi$, we generate some example tables in the format of \cref{tab:FourTypes} using the \textsc{Mathematica} package LieART \cite{Feger:2019tvk}: the $N=2$ case ($G=SU(4)$) in \cref{tab:SU4Irreps}, the $N=3$ case ($G=SU(6)$) in \cref{tab:SU6Irreps}, and the $N=4$ case ($G=SU(8)$) in \cref{tab:SU8Irreps}. From these tables, one can see explicitly that there is no $\Irrep_G^{} \ne \I_G^{}$ that contains both $\I_H^{}$ and $\I_{H_\Phi}^{}$, up to the irrep dimension listed. One can also check that the $G$-irreps that contain an $H$-invariant are given by the type in \cref{eqn:GirrepConcernIII}, whose Dynkin labels are (for $N>1$, see \cref{tab:DynkinLabels})
\begin{equation}
\vb*{w}_{SU(2N)}^{} = (n ,\, k ,\, \cdots ,\, k)
\quad\Longleftrightarrow\quad
\textbf{Dynkin}_{SU(2N)}^{} = (n-k ,\, 0 ,\, \cdots ,\, 0 ,\, k) \,,
\end{equation}
with $n\ge k \ge 0$ and $n>0$.

\subsubsection{Summary of advantages}
\label{subsubsec:FriendsAdvantages}

From the discussions above, we see that the criterion in \cref{eqn:AllOrderCriterion} is a very powerful tool. One can exploit group representation theory, in particular, the branching rules, to verify all-order accidental symmetries without explicitly computing the Hilbert series. This provides an alternative approach, which is much easier in some cases, as demonstrated above.

More importantly, this alternative approach enables us to go beyond merely verifying a postulated accidental symmetry. For a given symmetry $G$ and the building block fields $\Phi$, one can work out the unbroken subgroup $H_\Phi$, and then look for all its friends $F\subset G$; these are all sufficient subgroups for the $G$-invariance of $\Phi$ polynomials:
\begin{equation}
F \implies G
\quad\text{for}\quad
V(\Phi) \,.
\end{equation}
We can therefore find more all-order accidental symmetries in a relatively systematic manner. The key to actually achieve this is to develop an efficient algorithm for finding all the friends of a given subgroup $H_\Phi \subset G$.

\subsection{Verifying finite-order accidental symmetries: \emph{friends ma non troppo}}
\label{subsec:VerifyFriendsMaNonTroppo}

In \cref{subsec:VerifyFriends}, we explained how to make use of the friendship relation to verify all-order accidental symmetries. Specifically, we presented a criterion in \cref{eqn:AllOrderCriterion}, which could save us the effort of explicitly computing the Hilbert series. This also enables us to analyze all-order accidental symmetries more systematically by making use of the branching rules. In this subsection, we explain how to generalize this more systematic approach to verifying finite-order accidental symmetries, and demonstrate its application with the examples discussed in \cref{sec:FiniteOrder}.

We learned from \cref{sec:FiniteOrder} that finite-order accidental symmetries can be verified by computing and comparing the Hilbert series. In particular, one can use the criterion in \cref{eqn:HSAccidentalVk} when restricted to non-derivative interactions, and the criterion in \eqref{eqn:HSAccidentalLagk} for general cases.

Let us begin with the relatively easier criterion for non-derivative interactions given in \cref{eqn:HSAccidentalVk}, where the two Hilbert series, $\HS_\I^{G,\, \Phi}(q)$ and $\HS_\I^{H,\, \Phi}(q)$, need not to agree in full, but only up to a certain power in $q$. Our criterion in \cref{eqn:AllOrderCriterion} then implies that $H$ and $H_\Phi$ are not friends --- the condition in \cref{eqn:Friend} is violated by some $G$-irreps:
\begin{equation}
\Irrep_G^{} \ne \I_G^{}
\qquad\text{and}\qquad
\tdim \big( \Irrep_G^H \big)\, \tdim \big( \Irrep_G^{H_\Phi} \big) \ne 0 \,.
\end{equation}
Following the logic of the proof for \cref{eqn:AllOrderCriterion}, in particular, \cref{eqn:nInvH,eqn:nIrrepG}, we know that each such friendship-violating $G$-irrep will make a contribution to the difference between the two Hilbert series. Therefore, they cannot appear ``too soon'' if $\HS_\I^{G,\, \Phi}(q)$ and $\HS_\I^{H,\, \Phi}(q)$ still agree up to a certain order. Putting this idea into concrete language, we obtain the following criterion for finite-order accidental symmetries.
\begin{tcolorbox}[colback=light-gray]
\begin{center}
\begin{minipage}{5.5in}\vspace{3pt}
\textbf{Criterion for finite-order accidental symmetries:}
Let $H_\Phi$ be the subgroup of $G$ that remains unbroken under a generic vev $\langle\Phi\rangle$ of the bosonic building block fields $\Phi(x)$:
\begin{equation}
G \;\xrightarrow[]{\;\;\langle\Phi\rangle\;\;}\; H_\Phi \subset G \,,
\end{equation}
and $H\subset G$ be any subgroup. Then for the two Hilbert series to agree up to order $k$:
\begin{equation}
\HS_\I^{H,\, \Phi}(q) = \HS_\I^{G,\, \Phi}(q) + \order{q^{k+1}} \,,
\label{eqn:HSGHqkplus1}
\end{equation}
a sufficient and necessary condition is that all the $(H, H_\Phi)$ friendship-violating $G$-irreps arise at higher order $\order{q^{k+1}}$:
\begin{equation}
\HS_{\Irrep_G^{}}^{G,\, \Phi}(q) = \order{q^{k+1}}
\qquad\forall\qquad
\left\{ \mqty{\Irrep_G^{} \ne \I_G^{} \\[5pt]
\tdim \big( \Irrep_G^H \big)\, \tdim \big( \Irrep_G^{H_\Phi} \big) \ne 0 } \right. \,.
\label{eqn:FiniteOrderCriterion}
\end{equation}
In such cases, $H$ and $H_\Phi$ are friends effectively up to order $q^k$, and we say that they are \emph{friends ma non troppo}.
\end{minipage}\vspace{3pt}
\end{center}
\end{tcolorbox}
\noindent
To apply the criterion in \cref{eqn:FiniteOrderCriterion}, after identifying the friendship-violating $G$-irreps using branching rules, one needs to compute the Hilbert series $\HS_{\Irrep_G^{}}^{G,\, \Phi}(q)$ for these $G$-irreps, in principle. So admittedly, it is not necessarily easier than directly comparing the two Hilbert series for $G$ and $H$ invariants, \ie\ checking \cref{eqn:HSGHqkplus1}. However, in many cases, the condition in \cref{eqn:FiniteOrderCriterion} can be verified without explicitly computing the Hilbert series, as we will see in a few examples below. This is when our above criterion based on friends ma non troppo is favored.

The criterion in \cref{eqn:FiniteOrderCriterion} was derived specifically for the case of non-derivative interactions in \cref{eqn:HSAccidentalVk}. However, the same logic holds also for the general cases in \eqref{eqn:HSAccidentalLagk}, where derivative interactions are accommodated. Specifically, to tell if two Hilbert series $\HS_{\I,\text{ IBP}}^{\text{Lorentz}\times H,\, \SPM_\Phi} (q)$ and $\HS_{\I,\text{ IBP}}^{\text{Lorentz}\times G,\, \SPM_\Phi} (q)$ agree up to a certain order $k$, one just needs to recast the criterion in \cref{eqn:FiniteOrderCriterion} with a few steps:
\begin{enumerate}
\item Replacing the building blocks fields $\Phi$ with their SPMs $\SPM_\Phi$ to accommodate derivative interactions.
\item Adding Lorentz symmetry into the group $G$ as a factor: $G \;\longrightarrow\; \text{Lorentz}\times G$.
\item Checking the IBP redundancy in the end.
\end{enumerate}
With these recasting, we obtain the following criterion equivalent to \eqref{eqn:HSAccidentalLagk}:
\begin{tcolorbox}[colback=light-gray]
\begin{center}
\begin{minipage}{5.5in}\vspace{-15pt}
\begin{multline}
\HS_{\text{friendship-violating irreps},\text{ IBP}}^{\text{Lorentz}\times G,\, \SPM_\Phi}(q) = \order{q^{k+1}}
\\[10pt]
\Longleftrightarrow\quad
\HS_{\I,\text{ IBP}}^{\text{Lorentz}\times H,\, \SPM_\Phi} (q) = \HS_{\I,\text{ IBP}}^{\text{Lorentz}\times G,\, \SPM_\Phi} (q) + \order{q^{k+1}} \,.
\label{eqn:FiniteOrderCriterionGeneral}
\end{multline}
\end{minipage}\vspace{5pt}
\end{center}
\end{tcolorbox}
\noindent
Here the full group in consideration is $(\text{Lorentz}\times G)$. A generic vev of the building block $\langle \SPM_\Phi \rangle$ breaks it to the subgroup $H_{\SPM_\Phi} \subset (\text{Lorentz}\times G)$. We check the friendship relation between this $H_{\SPM_\Phi}$ and any postulated sufficient subgroup $(\text{Lorentz}\times H) \subset (\text{Lorentz}\times G)$. As $H_{\SPM_\Phi}$ is often the trivial subgroup $\{\mathbf{e}\}$ (although not always), $H$ and $H_{\SPM_\Phi}$ are typically not friends. Using branching rules, we can identify their friendship-violating irreps. The condition in \eqref{eqn:HSAccidentalLagk} is then equivalent to the requirement that these irreps arise at higher order $\order{q^{k+1}}$, once IBP redundancies are also taken into account. We will present some explicit examples below to demonstrate the application of \cref{eqn:FiniteOrderCriterionGeneral}.

\subsubsection{Non-derivative example: Class I with too many flavors}
\label{subsubsec:VerifyClassIFlavor}

In \cref{subsec:ClassI}, we observed that the Class I all-order accidental symmetries hold when the number of flavors is sufficiently small, $k = \max \left( k_1, k_2 \right) < N$, as summarized in \cref{tab:ClassIk1k2}. We showed this by explicitly comparing the Hilbert series in \cref{eqn:HSIk1k2Distinct}. In \cref{subsec:Friendship}, we gave a more systematic and efficient proof making use of our friendship criterion in \cref{eqn:AllOrderCriterion}.

When there are too many flavors, \ie, $k = \max \left( k_1, k_2 \right) \ge N$, we explained in \cref{subsec:Nonderivative} that they are expected to become finite-order accidental symmetries that hold up to the order $q^{N-1}$ (as the baryon combinations arise no earlier than $q^N$):
\begin{equation}
\HS_\I^{H,\, \Phi}(q) = \HS_\I^{G,\, \Phi}(q) + \order{q^N} \,.
\label{eqn:HSIFiniteqN}
\end{equation}
This was verified for the specific cases of $k_1=N>k_2$ and $k_1=k_2=N$ using the explicit results in \cref{eqn:HSINk2Distinct,eqn:HSINNDistinct}. In below, we make use of our systematic approach based on friends ma non troppo, \ie\ the criterion in \cref{eqn:FiniteOrderCriterion}, to prove that \cref{eqn:HSIFiniteqN} holds for all cases of $k = \max \left( k_1, k_2 \right) \ge N$.

To make use of \cref{eqn:FiniteOrderCriterion}, we first identify all the $G$-irreps that violate the friendship relation between $H$ and $H_\Phi$. Revisiting our discussions around \cref{eqn:hPhik,eqn:thetak}, we see that when $k = \max \left( k_1, k_2 \right) \ge N$, the unbroken subgroup elements $h_\Phi^{} \in H_\Phi$ become
\begin{equation}
h_\Phi^{} = g_N^{}\, g_1
\qquad\text{with}\qquad
g_N^{} = e^{-i\theta}\, \mathbb{1}_{N\times N} \,,\quad
g_1 = e^{i\theta} \,,
\label{eqn:hPhikgeN}
\end{equation}
with the angle $\theta$ constrained as $e^{iN\theta} = 1$ (the $k=N$ case of \cref{eqn:hPhik,eqn:thetak}). This is a $\mathbb{Z}_N$ group. For this discrete $H_\Phi$, \cref{eqn:dimGirrepsHPhi} no longer holds for all $Q\ne 0$ (\ie\ all the $G$-irreps in \cref{eqn:GirrepsConcern}), and consequently $H$ and $H_\Phi$ are no longer friends. The following $G$-irreps violate their friendship relation:
\begin{equation}
\Irrep_G^{} = \left( \mathbf{1} \;,\; Q = m N \right) \,,\qquad
\text{with $m$ any nonzero integer} \,.
\label{eqn:GirrepsViolatingI}
\end{equation}
They feed into the difference between $\HS_\I^{H,\, \Phi}(q)$ and $\HS_\I^{G,\, \Phi}(q)$. Nevertheless, $H$ and $H_\Phi$ are friends ma non troppo, as \cref{eqn:GirrepsViolatingI} corresponds to a very small portion of all $G$-irreps.

Now to finish checking the criterion in \cref{eqn:FiniteOrderCriterion}, we see that all the friendship-violating $G$-irreps in \cref{eqn:GirrepsViolatingI} are of order $\order{q^N}$, without computing the Hilbert series for them. Because each component in our building block fields $\Phi=(\phi_i, \phi_j^*)$ has charge $\pm 1$, we need at least $N$ powers of them to make up to a total charge $Q=mN \ne 0$. Therefore, the criterion in \cref{eqn:FiniteOrderCriterion} tells us that $H \implies G$ is a finite-order accidental symmetry up to the order $q^{N-1}$, for \emph{arbitrarily large} $k_1$ and $k_2$. It is worth emphasizing that directly computing the Hilbert series $\HS_\I^{H,\, \Phi}(q)$ and $\HS_\I^{G,\, \Phi}(q)$ at large $k_1$ and $k_2$ would be very difficult, so our friends ma non troppo approach here is much favored. Furthermore, we managed to draw a conclusion for all $k_1$ and $k_2$ with this approach, demonstrating that our criterion in \cref{eqn:FiniteOrderCriterion} is more systematic.

\subsubsection{Non-derivative example: $SU(3)\times SU(2) \implies SU(6)$}
\label{subsubsec:VerifyRectangular}

As another demonstration of the criterion in \cref{eqn:FiniteOrderCriterion}, let us revisit the example discussed in \cref{subsec:Nonderivative} that was summarized in \cref{tab:Rectangular}. In this case, we have $G=SU(6)$ and the building block fields $\Phi=(\phi, \phi^*)$ are a pair of fundamental and anti-fundamental irreps. When $\Phi$ takes a generic vev $\langle\Phi\rangle$, the unbroken subgroup is $H_\Phi = SU(5)$. As this accidental symmetry is not all-order, we know from \cref{eqn:AllOrderCriterion} that the subgroup $H=SU(3) \times SU(2)$ in \cref{tab:Rectangular} is not a friend of $H_\Phi$.

To apply our criterion in \cref{eqn:FiniteOrderCriterion}, we need to identify the $G$-irreps that break the friendship relation between $H$ and $H_\Phi$. To this end, we perform an ``experimental search'' by listing out all the $G$-irreps, in the order of increasing irrep dimension, as shown in \cref{tab:SU6SU3SU2Irreps} generated using the \textsc{Mathematica} package LieART \cite{Feger:2019tvk}. From the table, we see that the lowest dimension $G$-irrep that violates the $(H, H_\Phi)$ friendship has dimension $405$ (given by the Dynkin label $\highlight{(20002)}$, as \highlight{highlighted} in the table). Therefore, any friendship-violating $G$-irrep would require at least four powers of $\phi$ and/or $\phi^*$, because each of them has only $6$ components, and homogeneous polynomials of degree smaller than $4$ do not have enough dimension to contain such an irrep: $6^3=216 < 405$. On the other hand, this irrep is indeed contained in the following tensor product rep:
\begin{equation}
\text{sym}^2(\irrep{6}) \times \text{sym}^2(\irrep{\bar{6}}) = \irrep{405} + \irrep{35} + \irrep{1} \,.
\end{equation}
We therefore conclude using our criterion in \cref{eqn:FiniteOrderCriterion} that this accidental symmetry holds up to order $q^3$, which agrees with our observations in \cref{eqn:HSRectangular,eqn:HSRectangularq,eqn:HSRectangularAgree}. From \cref{tab:SU6SU3SU2Irreps} we also notice that most $G$-irreps do not break the friendship relation between $H$ and $H_\Phi$, so they are friends ma non troppo.

\subsubsection{Derivative example: Custodial symmetry in Higgs sector of SMEFT}
\label{subsubsec:VerifyCustodial}

Now let us move on to some examples of the general case where derivative interactions are included, and demonstrate the application of the criterion \cref{eqn:FiniteOrderCriterion} recast in \cref{eqn:FiniteOrderCriterionGeneral}.

Let us first analyze the custodial violation in the Higgs sector of SMEFT discussed in \cref{subsec:Custodial}. To check the condition in \cref{eqn:FiniteOrderCriterionGeneral}, we list out the relevant groups in this example:
\begin{subequations}
\begin{align}
\text{Full group:}&\qquad
\text{Lorentz}\times G = \text{Lorentz}\times SU(2)_L \times SU(2)_R \,, \\[5pt]
\text{Unbroken subgroup:}&\qquad
H_\Phi = \{\mathbf{e}\} \,, \\[5pt]
\text{Postulated subgroup:}&\qquad
\text{Lorentz}\times H = \text{Lorentz}\times SU(2)_L \times U(1)_Y \,.
\end{align}
\end{subequations}
The subgroups $(\text{Lorentz}\times H)$ and $H_\Phi$ are not friends. As $H_\Phi = \{\mathbf{e}\}$, the friendship-violating irreps are all the non-invariant irreps of $(\text{Lorentz}\times G)$ that contain an invariant under $(\text{Lorentz}\times H)$. These irreps are given by
\begin{equation}
\text{friendship-violating irreps} = \left( \mathbf{1} \;,\; \mathbf{1} \;,\; w_R \in \text{positive even integers} \right) \,,
\label{eqn:GirrepsViolatingCustodial}
\end{equation}
where $w_R = 2j_R$ is the highest weight of the $SU(2)_R$ irrep with spin $j_R$. To see that $w_R$ needs to be an even integer, we recall that $U(1)_Y$ is embedded in the $SU(2)_R$ as the third generator, $Y=t_R^3$, which can be diagonalized as
\begin{equation}
Y = t_R^3 = \text{diag} \left( j_R, j_R-1, \cdots, -j_R \right) \,.
\label{eqn:YtR3}
\end{equation}
We see that for an $SU(2)_R$ irrep to have a $Y=0$ component, $j_R$ needs to be an integer, and hence $w_R = 2j_R$ needs to be an even integer.

With the friendship-violating irreps identified in \cref{eqn:GirrepsViolatingCustodial}, now let us check their appearance, which is encoded in the Hilbert series
\begin{equation}
\HS_{\text{friendship-violating irreps},\text{ IBP}}^{\text{Lorentz}\times SU(2)_L \times SU(2)_R,\, \SPM_{\phi=(H, H^*)}}(q) \,.
\end{equation}
Instead of computing these Hilbert series explicitly, let us try to enumerate the operators that could feed into them.

The Higgs fields $H, H^*$ have hypercharges $Y=\pm 1/2$, so from the highest weight in \cref{eqn:YtR3} we see that an $SU(2)_R$ irrep with $w_R=2j_R$ would require at least $w_R$ powers of the Higgs fields; derivatives would not change the hypercharge. Therefore, irreps with $w_R\ge 6$ would not arise before mass dimension 6.

For the irrep with $w_R=4$, its $Y=j_R=+2$ state would require at least 4 powers of $H$. However, to satisfy the $SU(2)_L$ invariance in \cref{eqn:GirrepsViolatingCustodial}, they need to be contracted. Because $\widetilde{H}^\dagger H=0$ (where $\widetilde{H}\equiv i\sigma^2 H^*$), derivatives need to be involved in such a contraction, and they have to come in pairs to satisfy the Lorentz invariance in \cref{eqn:GirrepsViolatingCustodial}. Therefore, the irrep with $w_R=4$ cannot arise before mass dimension 6 either.

Finally, we are left with the irrep $w_R=2$. Its $Y=j_R=+1$ state would require at least 2 powers of $H$. In addition, at least two powers of derivatives are needed to make an $SU(2)_L$ invariant and Lorentz invariant. Note that $(\pd_\mu \tilde{H})^\dagger (\pd^\mu H) = 0$. So without involving more building blocks, the only option we have is
\begin{equation}
\widetilde{H}^\dagger (\pd^2 H) = \pd_\mu (\widetilde{H}^\dagger \pd^\mu H) \,.
\end{equation}
However, this is a total derivative operator, which is eliminated by the IBP redundancy. Therefore, we need to involve more powers of the Higgs fields or derivatives. They all come in pairs due to the requirement of $SU(2)_L$ and Lorentz invariance in \cref{eqn:GirrepsViolatingCustodial}. So the irrep with $w_R=2$ cannot arise before mass dimension 6.

In summary, we conclude that
\begin{equation}
\HS_{\text{friendship-violating irreps},\text{ IBP}}^{\text{Lorentz}\times SU(2)_L \times SU(2)_R,\, \SPM_{\phi=(H, H^*)}}(q) = \order{q^6} \,.
\end{equation}
Now applying our criterion in \cref{eqn:FiniteOrderCriterionGeneral}, it means that custodial symmetry is an accidental symmetry for the Higgs sector of SMEFT holding up to order $q^5$. This agrees with our previous discussions; see \eg\ \cref{eqn:HSCustodialq}.

\subsubsection{Derivative example: Class I single flavor}
\label{subsubsec:VerifyClassIDerivative}

Finally, we consider the single flavor case of the Class I all-order accidental symmetries summarized in \cref{tab:ClassI}. We included the derivative interactions, and they become finite-order accidental symmetries, as discussed in \cref{subsec:General}. In below, we apply our general friends ma non troppo criterion in \cref{eqn:FiniteOrderCriterionGeneral} to verify the order that they start to break.

We begin with listing out the relevant groups in this example:
\begin{subequations}
\begin{align}
\text{Full group:}&\qquad
\text{Lorentz}\times G = \text{Lorentz}\times SU(N) \times U(1) \,, \\[5pt]
\text{Unbroken subgroup:}&\qquad
H_\Phi = \mathbb{Z}_N \,, \\[5pt]
\text{Postulated subgroup:}&\qquad
\text{Lorentz}\times H = \text{Lorentz}\times SU(N) \,.
\end{align}
\end{subequations}
Note that the unbroken subgroup $H_\Phi$ in this case is not the trivial group, but $\mathbb{Z}_N$, as explained in \cref{subsubsec:VerifyClassIFlavor}, specifically around \cref{eqn:hPhikgeN}. The subgroups $(\text{Lorentz}\times H)$ and $H_\Phi$ are not friends. Their friendship-violating irreps of $(\text{Lorentz}\times G)$ are given by
\begin{equation}
\hspace{-10pt}
\text{friendship-violating irreps} = \left( \mathbf{1} \;,\; \mathbf{1} \;,\; Q = m N \right) \,,\quad
\text{with $m$ non-zero integer} \,.
\label{eqn:GirrepsViolatingIDerivative}
\end{equation}
These irreps are encoded in the Hilbert series
\begin{equation}
\HS_{\text{friendship-violating irreps},\text{ IBP}}^{\text{Lorentz}\times SU(N)\times U(1),\, (\SPM_\phi, \SPM_{\phi^*})}(q) \,.
\end{equation}
Instead of computing these Hilbert series explicitly, let us try to enumerate the operators that could feed into them.

For explicitness, let us specify to the case $N=3$. To make an $SU(3)$ invariant with $Q=3$, we need the factor
\begin{equation}
\Big[ \epsilon^{ijk} \phi_i (\pd_\mu \phi_j) (\pd_\nu \phi_k) \Big] \sim q^5 \,.
\end{equation}
To make a Lorentz invariant, we need to contract it with another factor
\begin{equation}
\Big[ (\pd_\mu \phi^\dagger) (\pd_\nu \phi) \Big] \sim q^4 \,.
\end{equation}
Together, this is precisely the mass dimension 9 operator in \cref{eqn:OpsN3}. One can convince themselves that to make any irrep in \cref{eqn:GirrepsViolatingIDerivative} for the $N=3$ case, this combination gives the lowest mass dimension. We therefore conclude 
\begin{equation}
\HS_{\text{friendship-violating irreps},\text{ IBP}}^{\text{Lorentz}\times SU(3)\times U(1),\, (\SPM_\phi, \SPM_{\phi^*})}(q) = \order{q^9} \,.
\end{equation}
Our criterion in \cref{eqn:FiniteOrderCriterionGeneral} then implies that the accidental symmetry $SU(3) \implies SU(3)\times U(1)$ holds up to order $q^8$, agreeing with the results in \cref{eqn:HSISPMN3}.

The case $N=4$ can be analyzed in a similar manner. One way to make an $SU(4)$ invariant with $Q=4$ is
\begin{equation}
\epsilon^{ijkl}\, \varphi_i\, (\pd_\mu \varphi_j)\, (\pd_\nu \pd_\rho \varphi_k)\, (\pd^\mu \pd^\nu \pd^\rho \varphi_l) \,,
\end{equation}
where different numbers of derivatives have been assigned to each factor $\varphi$. Recall from \cref{eqn:SPMH} that the trace components are not part of the SPM for scalar fields. Therefore, the contraction of Lorentz indices above is the only option. This is precisely the kind of operators in the second line of \cref{eqn:OpsN4}, which have mass dimension 10.

An alternative option to make an $SU(4)$ invariant with $Q=4$ is
\begin{equation}
\epsilon^{ijkl}\, (\pd_\mu \varphi_i)\, (\pd_\nu \varphi_j)\, (\pd_\rho \varphi_k)\, (\pd_\sigma \varphi_l) \,.
\end{equation}
To make a Lorentz invariant, we can contract the Lorentz indices with an $\epsilon$ tensor:
\begin{equation}
\epsilon^{\mu\nu\rho\sigma}\, \epsilon^{ijkl}\, (\pd_\mu \varphi_i)\, (\pd_\nu \varphi_j)\, (\pd_\rho \varphi_k)\, (\pd_\sigma \varphi_l) \,.
\end{equation}
However, this would make it into a total derivative, which is eliminated by the IBP redundancy. To survive the IBP, we need to further multiply it by another factor $\varphi^\dagger \varphi$:
\begin{equation}
\Big[\, \epsilon^{\mu\nu\rho\sigma}\, \epsilon^{ijkl}\, (\pd_\mu \varphi_i)\, (\pd_\nu \varphi_j)\, (\pd_\rho \varphi_k)\, (\pd_\sigma \varphi_l)\, \Big]\, ( \varphi^\dagger \varphi ) \,.
\end{equation}
This gives the kind of operators in the first line of \cref{eqn:OpsN4}, which also have mass dimension 10.

One can convince themselves that to make any irrep in \cref{eqn:GirrepsViolatingIDerivative} for the $N=4$ case, the above two options give the lowest mass dimension. We therefore conclude 
\begin{equation}
\HS_{\text{friendship-violating irreps},\text{ IBP}}^{\text{Lorentz}\times SU(4)\times U(1),\, (\SPM_\phi, \SPM_{\phi^*})}(q) = \order{q^{10}} \,.
\end{equation}
Our criterion in \cref{eqn:FiniteOrderCriterionGeneral} then implies that the accidental symmetry $SU(4) \implies SU(4)\times U(1)$ holds up to order $q^9$, agreeing with the results in \cref{eqn:HSISPMN4}.

\subsection{A process of identifying accidental symmetries}
\label{subsec:Identify}

In \cref{subsec:Friendship}, we introduced a new mathematical construct --- a \emph{friendship} relation between two subgroups (\cref{eqn:Friend}). Making use of this new machinery, we derived more systematic criteria for all-order accidental symmetries in \cref{subsec:VerifyFriends} (\cref{eqn:AllOrderCriterion}) and for finite-order accidental symmetries in \cref{subsec:VerifyFriendsMaNonTroppo} (\cref{eqn:FiniteOrderCriterion,eqn:FiniteOrderCriterionGeneral}). We applied them to verifying the accidental symmetries discussed in \cref{sec:AllOrder,sec:FiniteOrder}, and demonstrated the power of this new approach.

With the success in \cref{subsec:VerifyFriends,subsec:VerifyFriendsMaNonTroppo}, we are encouraged to think about the more ambitious goal beyond simply verifying a postulate accidental symmetry $H \implies G$: is it possible to systematically identify all the accidental symmetries $G$ from a given set of fields $\Phi(x)$ and the imposed symmetry group $H$?

One way of attacking this ambitious task is to carry out a brute-force search. First, we make a list of candidates for $G$. We can collect all the components of $\Phi(x)$ into a big vector, and then consider all the linear transformations of this vector that preserves the kinetic term of the Lagrangian. This gives us the maximal global symmetry group $X$ (within the construction of linear realizations). We list out all the subgroups of $X$ that contains the imposed symmetry group $H$, \ie, $H\subset G_i \subset X$, each of which is then a candidate accidental symmetry. Now for each candidate $G_i$, the task is converted back to that of verifying a postulated accidental symmetry $H \implies G_i$. We can make use of the criteria in \cref{eqn:FiniteOrderCriterion,eqn:FiniteOrderCriterionGeneral} to determine at which order the accidental symmetry $H \implies G_i$ starts to break. If the breaking order is higher than the leading order of the EFT, then it is a valid accidental symmetry. Running this process over all the candidates $G_i$, we can systematically identify all accidental symmetries from a given set of fields $\Phi(x)$ and the imposed symmetry group $H$.

The process described above is feasible in principle. However, as the maximal global symmetry group $X$ is huge in practice, we will have many candidate accidental symmetries to verify. Therefore, a systematic and efficient program for verifying each postulated accidental symmetry $H \implies G_i$ is still the key to enable the above brute-force search. From the examples discussed in \cref{subsec:VerifyFriends,subsec:VerifyFriendsMaNonTroppo}, we see that there are two major tasks in verifying a postulated accidental symmetry:
\begin{enumerate}
\item Identify all the friendship-violating $G_i$ irreps. If there is none, then it is an all-order accidental symmetry.
\item Work out the EFT order at which the friendship-violating $G_i$ irreps start to arise.
\end{enumerate}
We know how to handle the first task in a relatively systematic way. There are known branching rules to make use of, and when analytic analysis gets too complicated, one can also switch to an \emph{experimental} search, by systematically listing out all the $G_i$ irreps, similar to what we did in \cref{tab:SO10Irreps,tab:SO12Irreps,tab:SU4Irreps,tab:SU6Irreps,tab:SU8Irreps,tab:SU6SU3SU2Irreps}. On the other hand, the second task above is more difficult. For the examples discussed in \cref{subsec:VerifyFriendsMaNonTroppo}, we sort of did this part on a case-by-case basis. A more systematic approach would be very interesting, and would enable the whole process of identifying accidental symmetries described above.

\section{Summary and Outlook}
\label{sec:Outlook}

Accidental symmetries are ubiquitous in EFTs and have important phenomenological consequences. A deeper understanding of them, especially a more systematic one, is highly desired and could have far-reaching implications. In this paper, we investigated accidental symmetries in EFTs using the Hilbert series method together with more advanced tools in invariant theory.

In \cref{sec:AllOrder}, we highlighted that when we focus only on the potential interactions $V(\Phi)$ (\ie, polynomials in the fields $\Phi(x)$), there are accidental symmetries that hold to all orders in the EFT. As examples, we showed three classes of such all-order accidental symmetries, each verified by explicitly computing and comparing the Hilbert series. In particular, the famous custodial symmetry in the Higgs sector of SMEFT can be recognized as the $N=2$ case of the Class II discussed in \cref{subsec:ClassII}.

In \cref{sec:FiniteOrder}, we turned to accidental symmetries that hold only up to a finite order in the EFT expansion, and explained how to verify them by computing the Hilbert series. Derivative interactions are accommodated by considering the full EFT Lagrangian $\Lag_\text{\,EFT}$, and we elaborated on their impacts on accidental symmetries that are preserved to all orders by the potential interactions. Several demonstration examples were provided.

In \cref{sec:Systematic}, we presented a more systematic approach of verifying a postulated accidental symmetry. In particular, we introduced a new mathematical construct that we call \emph{friendship} relation between subgroups. Making use of this new definition, we derived more systematic criteria for verifying all-order and finite-order accidental symmetries, without the need to compute the Hilbert series. In \cref{subsec:VerifyFriends}, we applied the new criterion based on \emph{friends} to prove the example all-order accidental symmetries discussed in \cref{sec:AllOrder}. In \cref{subsec:VerifyFriendsMaNonTroppo}, we applied the new criterion based on \emph{friends ma non troppo} to the example finite-order accidental symmetries discussed in \cref{sec:FiniteOrder}. Encouraged by the success of these applications, we also proposed in \cref{subsec:Identify} a feasible approach to the ambitious goal of systematically finding all the accidental symmetries from a given set of fields and the imposed symmetry group.

Our investigations in this paper have already much improved our understanding of the origin, as well as the patterns of accidental symmetries in EFTs, especially with the more systematic approaches developed in \cref{sec:Systematic}. Along this direction, there are several desired technological advancements that, once achieved, would provide us with an even better handling on analyzing accidental symmetries. First, we believe there are many more all-order accidental symmetries than those three classes presented in \cref{sec:AllOrder}. A key to find them, as mentioned in the end of \cref{subsec:VerifyFriends}, is to develop an efficient algorithm for finding all the friends of a given unbroken subgroup $H_\Phi \subset G$. This would allow us to find all the sufficient subgroups $H$ for the $G$ invariance, and hence identify new all-order accidental symmetries $H \implies G \text{ for } V(\Phi)$. Second, regarding the ambitious goal of systematically identifying finite-order accidental symmetries, the key missing technology to enable our proposal in \cref{subsec:Identify} is a systematic way of determining the EFT order at which a concerned friendship-violating $G$-irrep starts to appear. Finally, as emphasized in \cref{sec:Systematic}, our \emph{friends} and \emph{friends ma non troppo} approaches to accidental symmetries only apply to the cases of bosonic building block fields. When Grassmann odd fields are involved, there are several technical complications listed in \cref{subsec:VerifyFriends}. It would be interesting to examine through this list, to identify the sources of failure in the fermionic cases, and then fix them to accommodate fermionic fields into our approaches in \cref{sec:Systematic}.

Our deeper understanding of accidental symmetries, combined with newly developed methodologies for systematically analyzing them, opens the door to numerous interesting phenomenological applications. Accidental symmetries are frequently relied upon in model building to justify the emergence of (approximate) symmetries and to control their breaking. 
Prominent examples include solutions to the hierarchy problem (\eg, within the hyperbolic Higgs scenario \cite{Cohen:2018mgv} an all-order accidental symmetry like the ones studied in \cref{sec:AllOrder} arises); axion models featuring an accidental $U(1)_{PQ}$ symmetry to address the quality problem within composite \cite{Randall:1992ut, Redi:2016esr, Lillard:2018fdt, Gavela:2018paw, Cox:2019rro, Ardu:2020qmo} or weak \cite{Barr:1992qq, Kamionkowski:1992mf, Holman:1992us, Fukuda:2017ylt, Darme:2021cxx} dynamics; \emph{discrete}  Goldstone bosons where an exact discrete symmetry accidentally generates a continuous one \cite{Hook:2018jle, Das:2020arz, Vileta:2022jou, DiLuzio:2021pxd}; and dark matter scenarios where an accidental symmetry ensures dark matter stability \cite{Appelquist:2015yfa, Antipin:2015xia, Harigaya:2016rwr, Mitridate:2017oky, Redi:2018muu, Contino:2018crt, Hertzberg:2019bvt, Contino:2020god}. In all these cases, the new findings of this work could help improve our understanding of these accidental symmetries and guide the development of novel scenarios with enhanced properties.

\acknowledgments
We thank Aneesh Manohar and Hitoshi Murayama for useful discussions.
The work of B.G., X.L., and P.Q. is supported by the U.S. Department of Energy under grant number DE-SC0009919.
X.L.\ is supported in part by Simons Foundation Award 568420.
C.M.\ is funded by \textit{Conselleria de Innovación, Universidades, Ciencia y Sociedad Digital} from \textit{Generalitat Valenciana} and by \textit{Fondo Social Europeo} under grants ACIF/2021/284, and CIBEFP/2023/96. C.M.\ specially thanks UC San Diego Theoretical Physics Department for the hospitality during his visit. 
Likewise, X.L and P.Q. thank the CERN theory group for their warm hospitality during the \emph{Crossroads between Theory and Phenomenology} Program, where part of this work was carried out.

\section*{Appendices}
\addcontentsline{toc}{section}{\protect\numberline{}Appendices}%
\appendix

\section{Computing the Hilbert Series}
\label{appsec:ComputingHS}

In this appendix, we provide some detailed steps in computing the Hilbert series involved in the three classes of all-order accidental symmetries discussed in \cref{sec:AllOrder}. We refer the reader to Ref.~\cite{Grinstein:2023njq} for a review on the basic techniques of calculating the Hilbert series. Here, we focus on the key ingredients and novel steps of the calculations.

\subsection{Haar measures}
\label{appsubsec:Haar}

We first list the Haar measures of the classical Lie groups that are relevant for the calculations in this appendix. In particular, as all the integrands in our calculations will be symmetric under the Weyl group, we use the following reduced version of the Haar measures:
\begin{subequations}\label{eqn:Haar}
\begin{align}
\dd\mu_{U(1)}^{} &= \frac{\dd x}{2\pi i x} \,, \\[8pt]
\dd\mu_{SU(r+1)}^{} &= \prod_{i = 1}^{r} \frac{\dd x_i}{2\pi i x_i} \prod_{1\leq i < j \leq r+1} \left(1-\frac{x_i}{x_j}\right) \,, \\[8pt]
\dd\mu_{SO(2r)}^{} &= \prod_{i = 1}^{r} \frac{\dd x_i}{2\pi i x_i} \prod_{1\leq i < j \leq r} \left(1-\frac{x_i}{x_j}\right)(1 - x_i x_j) \,,
\end{align}
\end{subequations}
where $r$ is the rank of the Lie algebra, $x_i$ are phase variables (\ie, $|x_i|=1$), and for the $SU(r+1)$ case, $x_{r+1} \equiv \prod_{i = 1}^{r} x_i^{-1}$ is defined as an auxiliary variable (\ie, not independent). All independent phase variables are integrated over the unity contour counterclockwise (for one lap). One can check that these Haar measures are properly normalized such that $\int \dd\mu_G^{} = 1$.

The Haar measures in \cref{eqn:Haar} are not the full Haar measure, but only the Weyl integration measure over the Cartan subgroup of the Lie groups. They are written in a specific basis of the Cartan subalgebra, and when use them in the Molien-Weyl formula, the characters in the integrands need to be written in the same basis. We refer the reader to Ref.~\cite{Henning:2017fpj} for detailed derivations and explanations of the above Haar measures.

\begin{allowdisplaybreaks}
\subsection{Hilbert Series in Class I}
\label{appsubsec:ClassI}

In this appendix, we derive the Hilbert series presented in the discussion of Class I all-order accidental symmetries in \cref{subsec:ClassI}.

\subsubsection{Single flavor bosonic}
\label{appsubsubsec:ClassISingleBoson}

In this subsection, we derive the results in \cref{eqn:HSIDistinct}. Let us use the following shorthand in this part to make the expressions more compact:
\begin{subequations}\label{eqn:HSIGHshort}
\begin{align}
\HS_\I^G &\equiv \HS_\I^{SU(N)\times U(1),\, (\phi,\, \phi^*)} (\phi,\, \phi^*) \,, \\[5pt]
\HS_\I^H &\equiv \HS_\I^{SU(N),\, (\varphi,\, \varphi^*)} (\varphi,\, \varphi^*) \,.
\end{align}
\end{subequations}
Using the Molien-Weyl formula \cite{Grinstein:2023njq}, the Hilbert series $\HS_\I^G$ is given by
\begin{align}
\HS_\I^G
&= \oint \frac{\dd x}{2\pi i x} \prod_{i=1}^r \frac{\dd x_i}{2 \pi i x_i} \prod_{1 \le i<j \le N}
\left( 1-\frac{x_i}{x_j} \right)
\prod_{i=1}^N \frac{1}{(1-\phi\, x_i x)(1-\phi^*\, x_i^{-1} x^{-1} )} \,,
\label{eqn:HSIG}
\end{align}
where $r=N-1$ and $x_{r+1} \equiv (x_1 \cdots x_r)^{-1}$. To evaluate this integral, we make the following variable change for the $N=r+1$ integration variables
\begin{equation}
(x_1, \cdots, x_r, x) \quad\longrightarrow\quad
(y_1, \cdots, y_r, y_{r+1}):\qquad
\mqty(y_1 \\ \vdots \\ y_r \\ y_{r+1}) \equiv \mqty(x_1 \\ \vdots \\ x_r \\ \frac{1}{x_1 \cdots x_r}) x \,.
\label{eqn:xtoyVariableChange}
\end{equation}
The integrand becomes less tangled in terms of the new variables $y_i$, and it is easier to factor out the integration over $y_1$:
\begin{align}
\HS_\I^G
&= \oint \prod_{i=1}^N \frac{\dd y_i}{2 \pi i y_i} \prod_{1 \le i<j \le N} \left(1-\frac{y_i}{y_j}\right) \prod_{i=1}^N \frac{1}{(1-\phi\, y_i )(1-\phi^*\, y_i^{-1} )}
\notag\\[10pt]
&= \oint \prod_{i=2}^N \frac{\dd y_i}{2 \pi i y_i} \prod_{2 \le i<j \le N} \left(1-\frac{y_i}{y_j}\right) \prod_{i=2}^N \frac{1}{(1-\phi\, y_i )(1-\phi^*\, y_i^{-1} )}
\notag\\[5pt]
&\hspace{40pt}
\times \oint \frac{\dd y_1}{2\pi i y_1} \prod_{j=2}^N \left( 1-\frac{y_1}{y_j} \right)
\frac{1}{(1-\phi\, y_1 )(1-\phi^*\, y_1^{-1})} \,.
\label{eqn:HSUNy1}
\end{align}
Now the $y_1$ integral can be done straightforwardly by applying Cauchy's residue theorem, as the only \emph{contributing} pole (\ie\ inside the contour $|y_1|=1$) is at $y_1=\phi^*$, since the grading variables $|\phi|,\,|\phi^*|<1$:
\begin{equation}
\oint \frac{\dd y_1}{2\pi i y_1} \prod_{j=2}^N \left( 1-\frac{y_1}{y_j} \right)
\frac{1}{(1-\phi\, y_1 )(1-\phi^*\, y_1^{-1})} = \frac{1}{1-\phi\phi^*} \prod_{j=2}^N \left( 1- \phi^* y_j^{-1} \right) \,.
\end{equation}
With this, we get
\begin{align}
\HS_\I^G
&= \frac{1}{1-\phi\phi^*} \oint \prod_{i=2}^N \frac{\dd y_i}{2 \pi i y_i} \prod_{2 \le i<j \le N} \left(1-\frac{y_i}{y_j}\right) \prod_{i=2}^N \frac{1}{1-\phi\, y_i} \,.
\label{eqn:HSUNy2}
\end{align}
The remaining integrals can be done in the order of the variables $y_2, y_3, \cdots, y_N$. Each time, the only contributing pole is at $y_i=0$, and the integral evaluates to $1$. In the end, we obtain the result in the first line of in \cref{eqn:HSIDistinct}:
\begin{equation}
\HS_\I^G = \frac{1}{1-\phi\phi^*} \,.
\end{equation}

Next, we move on to the second line of \cref{eqn:HSIDistinct}, \ie\ the $\HS_\I^H$ in \cref{eqn:HSIGHshort}. Using the Molien-Weyl formula again, this Hilbert series is given by
\begin{equation}
\HS_\I^H = \oint \prod_{i=1}^r \frac{\dd x_i}{2 \pi i x_i}\prod_{1 \leq i<j \leq N}\left(1-\frac{x_i}{x_j}\right) \prod_{i=1}^N \frac{1}{(1-\varphi x_i)(1-\varphi^* x_i^{-1})} \,,
\label{eqn:HSIH}
\end{equation}
where $r=N-1$ and $x_N \equiv (x_1 \cdots x_r)^{-1}$. To evaluate this integral, we will follow the same strategy used for $\HS_\I^G$ above --- focusing on one integration variable first, say $x_1$. However, note that in \cref{eqn:HSIH}, both $x_1$ and $x_N$ depend on $x_1$. To single out the $x_1$ dependence, we introduce an auxiliary variable
\begin{equation}
w \equiv x_1 x_N = (x_2 \cdots x_r)^{-1} \,,
\end{equation}
with which we can rewrite $\HS_\I^H$ as
\begin{align}
\HS_\I^H
&=\oint \prod_{i=2}^r \frac{\dd x_i}{2 \pi i x_i} 
\prod_{2 \leq i<j \leq r} \left(1-\frac{x_i}{x_j}\right)
\prod_{i=2}^r \frac{1}{(1-\varphi x_i)(1-\varphi^* x_i^{-1})}
\notag\\[5pt]
&\hspace{40pt}
\times \oint \frac{\dd x_1}{2 \pi i x_1}
\left[ \left(1- \frac{x_1^2}{w}\right) \prod_{i=2}^r \left(1- \frac{x_1}{x_i}\right) \left(1- \frac{x_1x_i}{w}\right) \right]
\notag\\[5pt]
&\hspace{80pt}
\times \frac{1}{\left(1-\varphi x_1\right)\left(1-\varphi^* x_1^{-1}\right)} \frac{1}{\left(1-\varphi w x_1^{-1}\right)\left(1-\varphi^* w^{-1} x_1\right)} \,.
\end{align}
We see that there are two contributing poles for the $x_1$ integral, $x_1 = \varphi^*$ and $x_1 = \varphi w$. Performing this integral, we obtain
\begin{align}
\HS_\I^H &= \frac{1}{1-\varphi \varphi^*}
\oint \prod_{i=2}^r \frac{\dd x_i}{2 \pi i x_i}
\prod_{2 \leq i<j \leq r}\left(1-\frac{x_i}{x_j}\right)
\notag\\[5pt]
&\hspace{0pt}
\times \Bigg[ \frac{1}{1-\varphi \varphi^{*-1} w} \prod_{i=2}^r \frac{1-\varphi^* w^{-1} x_i}{1-\varphi x_i}
+ \frac{1}{1-\varphi^{-1} \varphi^* w^{-1}} \prod_{i=2}^r \frac{1-\varphi w x_i^{-1}}{1-\varphi^* x_i^{-1}} \Bigg] \,.
\end{align}
To further simplify this result, we make the following variable change for the second term in the squared brackets:
\begin{equation}
\text{first inverse:}\quad
x_i \longrightarrow x_i^{-1} \,,\quad
\text{and then order flipping:}\quad 
x_i \longrightarrow x_{r+2-i} \,,
\label{eqn:invandflip}
\end{equation}
which leads to
\begin{align}
\HS_\I^H &= \frac{1}{1-\varphi \varphi^*}
\oint \prod_{i=2}^r \frac{\dd x_i}{2 \pi i x_i}
\prod_{2 \leq i<j \leq r}\left(1-\frac{x_i}{x_j}\right)
\notag\\[5pt]
&\hspace{0pt}
\times \Bigg[ \frac{1}{1-\varphi \varphi^{*-1} w} \prod_{i=2}^r \frac{1-\varphi^* w^{-1} x_i}{1-\varphi x_i}
+ \frac{1}{1-\varphi^{-1} \varphi^* w} \prod_{i=2}^r \frac{1-\varphi w^{-1} x_i}{1-\varphi^* x_i} \Bigg] \,.
\label{eqn:HSIHTwoTerms}
\end{align}
Clearly, this two terms in the squared brackets are related simply by the exchange $\varphi \leftrightarrow \varphi^*$. Now consider the integration over the variable $x_2$. Its pole structure is
\begin{subequations}
\begin{align}
x_2 &= (x_3 \cdots x_r)^{-1}\, \varphi \varphi^{*-1}
\,,\qquad\text{for the first term} \,, \\[5pt]
x_2 &= (x_3 \cdots x_r)^{-1}\, \varphi^* \varphi^{-1}
\,,\qquad\text{for the second term} \,.
\end{align}
\end{subequations}
Because \cref{eqn:HSIHTwoTerms} is symmetric between $\varphi$ and $\varphi^*$, without loss of generality, we can proceed its evaluation with the assumption $|\varphi| < |\varphi^*|$. In this case, the second term in the squared brackets does not have a contributing pole of $x_2$, and hence evaluates to zero. Evaluating the first term gives us
\begin{align}
\HS_\I^H &= \frac{1}{1-\varphi\varphi^{*}} \oint \prod_{i=3}^r \frac{\dd x_i}{2 \pi i x_i}
\prod_{3 \leq i<j \leq r}\left(1-\frac{x_i}{x_j}\right)
\prod_{i = 3}^{r} \Big[ 1 - x_i^{-1}\varphi\varphi^{*-1} (x_3 \cdots x_r)^{-1} \Big]
\notag\\[5pt]
&= \frac{1}{1-\varphi\varphi^{*}} \oint \prod_{i=3}^r \frac{\dd x_i}{2 \pi i x_i}
\prod_{3 \leq i<j \leq r}\left(1-\frac{x_i}{x_j}\right)
\prod_{i = 3}^{r} \Big[ 1 - x_i\varphi\varphi^{*-1}(x_3 \cdots x_r) \Big] \,,
\end{align}
where in the second line we have applied the variable change similar to that in \cref{eqn:invandflip} for the set of variables $\{x_3, \cdots, x_r\}$. The remaining integrals can be done in the order of the variables $x_3, \cdots, x_r$. Each time, the only contributing pole is at $x_i=0$, and the integral evaluates to $1$. In the end, we obtain the result in the second line of \cref{eqn:HSIDistinct}:
\begin{equation}
\HS_\I^H = \frac{1}{1-\varphi\varphi^{*}} \,.
\end{equation}

\subsubsection{Multiple flavor}
\label{appsubsubsec:ClassIMultiple}

In this subsection, we derive the results in \cref{eqn:HSIk1k2G,eqn:HSINk2G,eqn:HSINNG}. These results can be summarized as
\begin{align}
\HS_\I^{G_N,\, (k_1,\, k_2)} &\equiv \HS_\I^{SU(N)\times U(1),\, (k_1\times \phi,\, k_2\times \phi^*)} (\phi_a,\, \phi_b^*)
\notag\\[5pt]
&= \oint {\left( {\prod\limits_{i = 1}^N {\frac{{\dd{x_i}}}{{2\pi i{x_i}}}} } \right)\left[ {\prod\limits_{1 \le i < j \le N} {\left( {1 - \frac{{{x_i}}}{{{x_j}}}} \right)} } \right]\left[ {\prod\limits_{a = 1}^{{k_1}} {\prod\limits_{i = 1}^N {\frac{1}{{1 - {\phi _a}{x_i}}}} } } \right]\left[ {\prod\limits_{b = 1}^{{k_2}} {\prod\limits_{i = 1}^N {\frac{1}{{1 - \phi _b^*x_i^{ - 1}}}} } } \right]}
\notag\\[5pt]
&= \prod_{a=1}^{k_1} \prod_{b=1}^{k_2} \frac{1}{ 1 - \phi_a \phi_b^*}
\qquad\text{for}\qquad
N \ge \min(k_1, k_2) \,.
\label{eqn:HSIGgoal}
\end{align}
In below, we prove this by induction over $k_2$. First, we show the base step of the induction, \ie\ the $k_2=1$ and $N\ge 1$ case of \cref{eqn:HSIGgoal}:
\begin{align}
\HS_\I^{G_N,\, (k_1, 1)} &=
\oint {\left( {\prod\limits_{i = 1}^N {\frac{{\dd{x_i}}}{{2\pi i{x_i}}}} } \right)\left[ {\prod\limits_{1 \le i < j \le N} {\left( {1 - \frac{{{x_i}}}{{{x_j}}}} \right)} } \right]\left[ {\prod\limits_{a = 1}^{{k_1}} {\prod\limits_{i = 1}^N {\frac{1}{{1 - {\phi _a}{x_i}}}} } } \right]\left[ {\prod\limits_{i = 1}^N {\frac{1}{{1 - {\phi_1^*}x_i^{ - 1}}}} } \right]}
\notag\\[10pt]
&= \oint {\left( {\prod\limits_{i = 2}^N {\frac{{\dd{x_i}}}{{2\pi i{x_i}}}} } \right)\left[ {\prod\limits_{2 \le i < j \le N} {\left( {1 - \frac{{{x_i}}}{{{x_j}}}} \right)} } \right]\left[ {\prod\limits_{a = 1}^{{k_1}} {\prod\limits_{i = 2}^N {\frac{1}{{1 - {\phi _a}{x_i}}}} } } \right]\left[ {\prod\limits_{i = 2}^N {\frac{1}{{1 - {\phi_1^*}x_i^{ - 1}}}} } \right]}
\notag\\[5pt]
&\hspace{40pt}
\times \oint {\frac{{\dd{x_1}}}{{2\pi i{x_1}}}\left[ {\prod\limits_{i = 2}^N {\left( {1 - \frac{{{x_1}}}{{{x_i}}}} \right)} } \right]\left[ {\prod\limits_{a = 1}^{{k_1}} {\frac{1}{{1 - {\phi _a}{x_1}}}} } \right]\frac{1}{{1 - {\phi_1^*}x_1^{ - 1}}}}
\notag\\[10pt]
&= \left[ {\prod\limits_{a = 1}^{{k_1}} {\frac{1}{{1 - {\phi _a}{\phi_1^*}}}} } \right]\oint {\left( {\prod\limits_{i = 2}^N {\frac{{\dd{x_i}}}{{2\pi i{x_i}}}} } \right)\left[ {\prod\limits_{2 \le i < j \le N} {\left( {1 - \frac{{{x_i}}}{{{x_j}}}} \right)} } \right]\left[ {\prod\limits_{a = 1}^{{k_1}} {\prod\limits_{i = 2}^N {\frac{1}{{1 - {\phi _a}{x_i}}}} } } \right]}
\notag\\[10pt]
&= \prod\limits_{a = 1}^{{k_1}} {\frac{1}{{1 - {\phi _a}{\phi_1^*}}}}
\qquad\text{for}\qquad
N\ge 1 \,.
\end{align}
Next, we prove the induction step: \emph{if} \cref{eqn:HSIGgoal} holds up to $k_2-1$, for all $N-1\ge k_2-1$:
\begin{equation}
\HS_\I^{G_{N-1},\, (k_1,\, k_2-1)} = \prod\limits_{a = 1}^{{k_1}} \prod\limits_{b = 1}^{{k_2-1}} \frac{1}{1 - \phi_a \phi_b^*}
\qquad\text{for}\qquad
N-1 \ge k_2-1 \,,
\label{eqn:HSIGgoalAssumption}
\end{equation}
\emph{then} it also holds at $k_2$, for all $N\ge k_2$. This can be seen by a brute-force calculation:
\begin{align}
\HS_\I^{G_N,\, (k_1,\, k_2)}
&= \oint {\left( {\prod\limits_{i = 1}^N {\frac{{\dd{x_i}}}{{2\pi i{x_i}}}} } \right)\left[ {\prod\limits_{1 \le i < j \le N} {\left( {1 - \frac{{{x_i}}}{{{x_j}}}} \right)} } \right]\left[ {\prod\limits_{a = 1}^{{k_1}} {\prod\limits_{i = 1}^N {\frac{1}{{1 - {\phi _a}{x_i}}}} } } \right]\left[ {\prod\limits_{b = 1}^{{k_2}} {\prod\limits_{i = 1}^N {\frac{1}{{1 - \phi _b^*x_i^{ - 1}}}} } } \right]}
\notag\\[10pt]
&= \oint \left( {\prod\limits_{i = 2}^N {\frac{{\dd{x_i}}}{{2\pi i{x_i}}}} } \right)\left[ {\prod\limits_{2 \le i < j \le N} {\left( {1 - \frac{{{x_i}}}{{{x_j}}}} \right)} } \right]\left[ {\prod\limits_{a = 1}^{{k_1}} {\prod\limits_{i = 2}^N {\frac{1}{{1 - {\phi _a}{x_i}}}} } } \right]\left[ {\prod\limits_{b = 1}^{{k_2}} {\prod\limits_{i = 2}^N {\frac{1}{{1 - \phi _b^*x_i^{ - 1}}}} } } \right]
\notag\\[5pt]
&\hspace{40pt}
\times \oint {\frac{{\dd{x_1}}}{{2\pi i{x_1}}}\left[ {\prod\limits_{i = 2}^N {\left( {1 - \frac{{{x_1}}}{{{x_i}}}} \right)} } \right]\left[ {\prod\limits_{a = 1}^{{k_1}} {\frac{1}{{1 - {\phi _a}{x_1}}}} } \right]\left[ {\prod\limits_{b = 1}^{{k_2}} {\frac{1}{{1 - \phi _b^*x_1^{ - 1}}}} } \right]}
\notag\\[10pt]
&= \oint \left( {\prod\limits_{i = 2}^N {\frac{{\dd{x_i}}}{{2\pi i{x_i}}}} } \right)\left[ {\prod\limits_{2 \le i < j \le N} {\left( {1 - \frac{{{x_i}}}{{{x_j}}}} \right)} } \right]\left[ {\prod\limits_{a = 1}^{{k_1}} {\prod\limits_{i = 2}^N {\frac{1}{{1 - {\phi _a}{x_i}}}} } } \right]\left[ {\prod\limits_{b = 1}^{{k_2}} {\prod\limits_{i = 2}^N {\frac{1}{{1 - \phi _b^*x_i^{ - 1}}}} } } \right]
\notag\\[5pt]
&\hspace{40pt}
\times \sum\limits_{m = 1}^{{k_2}} {\left[ {\prod\limits_{i = 2}^N {\left( {1 - \phi _m^*x_i^{ - 1}} \right)} } \right]\left[ {\prod\limits_{a = 1}^{{k_1}} {\frac{1}{{1 - {\phi _a}\phi _m^*}}} } \right]\left[ {\prod\limits_{1 \le b \le {k_2}}^{b \ne m} {\frac{1}{{1 - \phi _b^*\phi _m^{* - 1}}}} } \right]}
\notag\\[10pt]
&= \sum\limits_{m = 1}^{{k_2}} \left[ {\prod\limits_{a = 1}^{{k_1}} {\frac{1}{{1 - {\phi _a}\phi _m^*}}} } \right]\left[ {\prod\limits_{1 \le b \le {k_2}}^{b \ne m} {\frac{1}{{1 - \phi _b^*\phi _m^{* - 1}}}} } \right]
\HS_\I^{G_{N-1},\, (k_1,\, k_2-1)} \left( \phi_a, \phi_{b \ne m}^* \right)
\notag\\[10pt]
&= \sum\limits_{m = 1}^{{k_2}} {\left[ {\prod\limits_{a = 1}^{{k_1}} {\frac{1}{{1 - {\phi _a}\phi _m^*}}} } \right]\left[ {\prod\limits_{1 \le b \le {k_2}}^{b \ne m} {\frac{1}{{1 - \phi _b^*\phi _m^{* - 1}}}} } \right]\left[ {\prod\limits_{a = 1}^{{k_1}} {\prod\limits_{1 \le b \le {k_2}}^{b \ne m} {\frac{1}{{1 - {\phi _a}\phi _b^*}}} } } \right]}
\notag\\[5pt]
&= \left[ {\prod\limits_{a = 1}^{{k_1}} {\prod\limits_{b = 1}^{{k_2}} {\frac{1}{{1 - {\phi _a}\phi _b^*}}} } } \right]\left[ {\sum\limits_{m = 1}^{{k_2}} {\prod\limits_{1 \le b \le {k_2}}^{b \ne m} {\frac{1}{{1 - \phi _b^*\phi _m^{* - 1}}}} } } \right]
\notag\\[5pt]
&= \prod\limits_{a = 1}^{{k_1}} {\prod\limits_{b = 1}^{{k_2}} {\frac{1}{{1 - {\phi _a}\phi _b^*}}} }
\qquad\text{for}\qquad
N \ge k_2 \,.
\label{eqn:InductionStep1}
\end{align}
To get the third-to-last line above, we have used the assumption of the induction step in \cref{eqn:HSIGgoalAssumption}, \ie,
\begin{align}
\HS_\I^{G_{N-1},\, (k_1,\, k_2-1)} \left( \phi_a, \phi_{b \ne m}^* \right)
= \prod\limits_{a = 1}^{{k_1}} {\prod\limits_{1 \le b \le {k_2}}^{b \ne m} {\frac{1}{{1 - {\phi _a}\phi _b^*}}} } \quad \;\;\;{\mkern 1mu} ,\quad \;\;\;{\mkern 1mu} N-1 \ge k_2-1 \,.
\end{align}
This completes the proof of \cref{eqn:HSIGgoal}.

\subsection{Hilbert Series in Class II}
\label{appsubsec:ClassII}

In this appendix, we derive the Hilbert series involved in the discussion of Class II all-order accidental symmetries in \cref{subsec:ClassII}.

\subsubsection{Single flavor}
\label{appsubsubsec:ClassIISingle}

In this subsection, we provide detailed steps in computing the results in \cref{eqn:HSIIDistinct}. More precisely, we exclusively focus on \cref{eqn:HSIIG}, since \cref{eqn:HSIIH} has already been derived in \cref{appsubsubsec:ClassISingleBoson}. 

Introducing the shorthand notation
\begin{equation}
\HS_\I^{G} \equiv \HS_\I^{SO(2N),\, \phi}(\phi)\, ,
\end{equation}
and using the Molien-Weyl formula, the Hilbert series for $G = SO(2N)$ invariants reads
\begin{equation}
\HS_\I^{G} = \oint \prod_{i=1}^N \frac{\dd x_i}{2 \pi i x_i} \left[\prod_{1 \leq i<j \leq N}\left(1-\frac{x_i}{x_j}\right)\left(1-x_i x_j\right)\right] \left[\prod_{i=1}^N \frac{1}{\left(1-\phi x_i\right)\left(1-\phi x_i^{-1}\right)}\right] \, .
\end{equation}
To evaluate the previous integral, we can first focus on the variable $x_1$:
\begin{align}
\HS_\I^{G} &= \oint \prod_{i=2}^N \frac{\dd x_i}{2 \pi i x_i} \left[\prod_{2 \leq i<j \leq N}\left(1-\frac{x_i}{x_j}\right)\left(1-x_i x_j\right)\right] \left[\prod_{i=2}^N \frac{1}{\left(1-\phi x_i\right)\left(1-\phi x_i^{-1}\right)}\right] \notag \\[5pt]
&\hspace{40pt} \times \oint \frac{\dd x_1}{2 \pi i x_1} \left[ \prod_{i = 2}^{N} \left(1-\frac{x_1}{x_i}\right)\left(1-x_1 x_i\right) \right] \frac{1}{\left(1-\phi x_1\right)\left(1-\phi x_1^{-1}\right)}\, .
\label{eqn:HSSO2Nstep2}
\end{align}
The $x_1$ integral can be done straightforwardly since the only contributing pole is at $x_1 = \phi$, yielding
\begin{align}
&\oint \frac{\dd x_1}{2 \pi i x_1} \left[ \prod_{i = 2}^{N} \left(1-\frac{x_1}{x_i}\right)\left(1-x_1 x_i\right) \right] \frac{1}{\left(1-\phi x_1\right)\left(1-\phi x_1^{-1}\right)} \notag \\[5pt]
&\hspace{80pt}
=\frac{1}{1-\phi^2} \left[\prod_{i = 2}^{N} \left(1-\phi x_i^{-1}\right)\left(1-\phi x_i\right)\right] \, .
\end{align}
Plugging this result into \cref{eqn:HSSO2Nstep2} leads to
\begin{equation}
\HS_\I^{G} = \frac{1}{1-\phi^2} \oint \prod_{i=2}^N \frac{\dd x_i}{2 \pi i x_i} \left[\prod_{2 \leq i<j \leq N}\left(1-\frac{x_i}{x_j}\right)\left(1-x_i x_j\right)\right] \, .
\end{equation}
The remaining integrals can be computed in the order of the variables $x_2, x_3, \cdots, x_N$. Each time, the only contributing pole is at $x_i=0$, and the integral evaluates to $1$. Therefore, our final result is the one quoted in \cref{eqn:HSIIG}:
\begin{equation}
\HS_\I^{G} = \frac{1}{1-\phi^2} \,.
\end{equation}

\subsubsection{Multiple flavor}
\label{appsubsubsec:ClassIIMultiple}

In this subsection, we provide detailed steps in computing the results in \cref{eqn:HSIIkDistinct}, exclusively focusing on deriving \cref{eqn:HSIIkG}, since \cref{eqn:HSIIkH} has already been derived in \cref{appsubsubsec:ClassIMultiple}.

We introduce the following shorthand for this part
\begin{equation}
\HS_\I^{G_N,\, k} \equiv \HS_\I^{SO(2N),\, k\times \phi} (\phi_a) = \prod_{1\le a\le b \le k} \frac{1}{1 - \phi_a \phi_b}
\qquad\text{for}\qquad
N > k \,,
\label{eqn:HSIIGgoal}
\end{equation}
and we prove this by induction over $k$. First, the base case of \cref{eqn:HSIIGgoal}, $k=1$ for all $N>1$, is nothing but the single flavor result in \cref{eqn:HSIIG}, which have just been proved in \cref{appsubsubsec:ClassIISingle}. Now, we prove the induction step: \emph{if} \cref{eqn:HSIIGgoal} holds up to $k-1$, for all $N-1>k-1$:
\begin{equation}
\HS_\I^{G_{N-1},\, k-1} = \prod_{1\le a\le b \le k-1} \frac{1}{1 - \phi_a \phi_b}
\qquad\text{for}\qquad
N-1 > k-1 \,,
\label{eqn:HSIIGgoalAssumption}
\end{equation}
\emph{then} it also holds at $k$, for all $N>k$. This can be seen by a brute-force calculation:
\begin{align}
\hspace{-5pt}
\HS_\I^{G_N,\, k}
&= \oint {\left( {\prod\limits_{i = 1}^N {\frac{{\dd{x_i}}}{{2\pi i{x_i}}}} } \right)\left[ {\prod\limits_{1 \le i < j \le N} {\left( {1 - \frac{{{x_i}}}{{{x_j}}}} \right)\left( {1 - {x_i}{x_j}} \right)} } \right]\left[ {\prod\limits_{a = 1}^k {\prod\limits_{i = 1}^N {\frac{1}{{\left( {1 - {\phi _a}{x_i}} \right)\left( {1 - {\phi _a}x_i^{ - 1}} \right)}}} } } \right]}
\notag\\[10pt]
&= \oint {\left( {\prod\limits_{i = 2}^N {\frac{{\dd{x_i}}}{{2\pi i{x_i}}}} } \right)\left[ {\prod\limits_{2 \le i < j \le N} {\left( {1 - \frac{{{x_i}}}{{{x_j}}}} \right)\left( {1 - {x_i}{x_j}} \right)} } \right]\left[ {\prod\limits_{a = 1}^k {\prod\limits_{i = 2}^N {\frac{1}{{\left( {1 - {\phi _a}{x_i}} \right)\left( {1 - {\phi _a}x_i^{ - 1}} \right)}}} } } \right]}
\notag\\[5pt]
&\hspace{0pt}\quad
\times \oint {\frac{{\dd{x_1}}}{{2\pi i{x_1}}}\left[ {\prod\limits_{i = 2}^N {\left( {1 - {x_1}x_i^{ - 1}} \right)\left( {1 - {x_1}{x_i}} \right)} } \right]\left[ {\prod\limits_{a = 1}^k {\frac{1}{{\left( {1 - {\phi _a}{x_1}} \right)\left( {1 - {\phi _a}x_1^{ - 1}} \right)}}} } \right]}
\notag\\[10pt]
&= \sum\limits_{m = 1}^k {\frac{1}{{1 - \phi _m^2}}\left[ {\prod\limits_{1 \le a \le k}^{a \ne m} {\frac{1}{{\left( {1 - {\phi _a}{\phi _m}} \right)\left( {1 - {\phi _a}\phi _m^{ - 1}} \right)}}} } \right]}
\notag\\[5pt]
&\hspace{0pt}\quad
\times \oint {\left( {\prod\limits_{i = 2}^N {\frac{{\dd{x_i}}}{{2\pi i{x_i}}}} } \right)\left[ {\prod\limits_{2 \le i < j \le N} {\left( {1 - \frac{{{x_i}}}{{{x_j}}}} \right)\left( {1 - {x_i}{x_j}} \right)} } \right]\left[ {\prod\limits_{1 \le a \le k}^{a \ne m} {\prod\limits_{i = 2}^N {\frac{1}{{\left( {1 - {\phi _a}{x_i}} \right)\left( {1 - {\phi _a}x_i^{ - 1}} \right)}}} } } \right]}
\notag\\[10pt]
&= \sum\limits_{m = 1}^k {\frac{1}{{1 - \phi _m^2}}\left[ {\prod\limits_{1 \le a \le k}^{a \ne m} {\frac{1}{{\left( {1 - {\phi _a}{\phi _m}} \right)\left( {1 - {\phi _a}\phi _m^{ - 1}} \right)}}} } \right] \HS_\I^{G_{N-1},\, k-1} \left( {{\phi _{a \ne m}}} \right)}
\notag\\[10pt]
&= \sum\limits_{m = 1}^k {\frac{1}{{1 - \phi _m^2}}\left[ {\prod\limits_{1 \le a \le k}^{a \ne m} {\frac{1}{{\left( {1 - {\phi _a}{\phi _m}} \right)\left( {1 - {\phi _a}\phi _m^{ - 1}} \right)}}} } \right]\prod\limits_{1 \le a \le b \le k}^{a \ne m, b \ne m} {\frac{1}{{1 - {\phi _a}{\phi _b}}}} }
\notag\\[10pt]
&= \left( {\prod\limits_{1 \le a \le b \le k} {\frac{1}{{1 - {\phi _a}{\phi _b}}}} } \right)\left( {\sum\limits_{m = 1}^k {\prod\limits_{1 \le a \le k}^{a \ne m} {\frac{1}{{1 - {\phi _a}\phi _m^{ - 1}}}} } } \right) = \prod\limits_{1 \le a \le b \le k} {\frac{1}{{1 - {\phi _a}{\phi _b}}}} \,,
\end{align}
where to get the second-to-last line, we have used the assumption of the induction step in \cref{eqn:HSIIGgoalAssumption}. This completes the proof of \cref{eqn:HSIIGgoal}.

\subsection{Hilbert Series in Class III}
\label{appsubsec:ClassIII}

In this appendix, we compute the Hilbert series involved in the discussion of Class III all-order accidental symmetries in \cref{subsec:ClassIII}. In particular, we derive the results in \cref{eqn:HSIIIDistinct,eqn:HSIIIFermionDistinct}. For convenience, we introduce the following shorthand for the bosonic case in \cref{eqn:HSIIIDistinct}:
\begin{subequations}\label{eqn:HSIIIShort}
\begin{align}
\HS_\I^{G, \Phi} &\equiv \HS_\I^{SU(2N),\, \phi} (\phi) \,, \label{eqn:HSIIIGShort}  \\[5pt]
\HS_\I^{H_1, \Phi} &\equiv \HS_\I^{SU(2N-1) \times U(1),\, (\varphi_1,\, \varphi_2)} (\varphi_1, \varphi_2) \,, \label{eqn:HSIIIH1Short} \\[5pt]
\HS_\I^{H_2, \Phi} &\equiv \HS_\I^{SU(2N-1),\, (\varphi_1,\, \varphi_2)} (\varphi_1, \varphi_2) \,, \label{eqn:HSIIIH2Short} 
\end{align}
\end{subequations}
and for the fermionic case in \cref{eqn:HSIIIFermionDistinct}:
\begin{subequations}\label{eqn:HSIIIFermionShort}
\begin{align}
\HS_\I^{G, \Psi} &\equiv \HS_\I^{SU(2N),\, \psi} (\psi) \,, \label{eqn:HSIIIFermionGShort} \\[5pt]
\HS_\I^{H, \Psi} &\equiv \HS_\I^{SU(2N-1) \times U(1),\, (\chi_1,\, \chi_2)} (\chi_1, \chi_2) \,. \label{eqn:HSIIIFermionHShort}
\end{align}
\end{subequations}

\subsubsection{The Hilbert series $\HS_\I^{G,\Phi}$}
\label{appsubsubsec:ClassIIIG}

In this subsection, we compute the Hilbert series in \cref{eqn:HSIIIGShort} to derive the result in \cref{eqn:HSIIIG}.

Using the Molien-Weyl formula, the Hilbert series $\HS_\I^{G,\Phi}$ can be computed from
\begin{equation}
\HS_\I^{G,\Phi} = \oint \prod_{i=1}^{2 N-1} \frac{\dd x_i}{2 \pi i x_i} \prod_{1 \leq i<j \leq 2 N}\left(1-\frac{x_i}{x_j}\right) \prod_{1 \leq i<j \leq 2 N} \frac{1}{1-\phi x_i x_j} \,,
\end{equation}
where $x_{2N} = \left(x_1 \cdots x_{2N-1}\right)^{-1}$. First, we focus on the integration over the variable $x_1$. In this sense, it is important to notice that the auxiliary $x_{2N}$ also depends on $x_1$, so that in order to single out the $x_1$ dependence of the integrand we define
\begin{equation}
w \equiv x_1 x_{2N} = \left(x_2 \cdots x_{2N-1}\right)^{-1} \,.
\end{equation}
Then, we can factor out the integration over $x_1$ as
\begin{align}
\HS_\I^{G, \Phi} &= \oint \prod_{i=2}^{2 N-1} \frac{\dd x_i}{2 \pi i x_i} \left[ \prod_{2 \leq i<j \leq 2 N-1}\left(1-\frac{x_i}{x_j}\right) \frac{1}{1-\phi x_i x_j} \right] \frac{1}{1-\phi w} \notag \\[5pt]
&\times \oint \frac{\dd x_1}{2 \pi i x_1} \left(1-\frac{x_1^2}{w}\right) \left[\prod_{i=2}^{2N-1} \left(1-\frac{x_1}{x_i}\right) \left(1-\frac{x_1 x_i}{w} \right) \frac{1}{1-\phi x_1 x_i} \frac{1}{1-\phi x_1^{-1} x_i w} \right] \,.
\end{align}
To evaluate the integral over $x_1$, we have to take into account the contributions from the poles at $x_1 = \phi x_m w$, with $m = 2,\cdots,2N-1$, yielding
\begin{align}
\HS_\I^{G, \Phi} &= \oint \prod_{i=2}^{2 N-1} \frac{\dd x_i}{2 \pi i x_i} \left[ \prod_{2 \leq i<j \leq 2 N-1}\left(1-\frac{x_i}{x_j}\right) \frac{1}{1-\phi x_i x_j} \right] \frac{1}{1-\phi w} \notag \\[5pt]
&\times \sum_{m=2}^{2N-1} \left(1-\phi^2 x_m^2 w\right) \left[ \prod_{i=2}^{2N-1} \frac{\left(1-\phi x_m x_i^{-1} w\right)\left(1-\phi x_m x_i\right)}{1-\phi^2 x_m x_i w}\right] \left[\prod_{2 \le i \le 2N-1}^{i \ne m} \frac{1}{1-x_i x_m^{-1}}\right]\, .
\end{align}
It is convenient to separate the $i = m$ contribution in the second line of the expression above as
\begin{multline}
\left[\prod_{i=2}^{2N-1} \frac{\left(1-\phi x_m x_i^{-1} w\right)\left(1-\phi x_m x_i\right)}{1-\phi^2 x_m x_i w}\right] \\
= \frac{\left(1-\phi w\right)\left(1-\phi x_m^2\right)}{1-\phi^2 x_m^2 w} \left[\prod_{2 \le i \le 2N-1}^{i \ne m} \frac{\left(1-\phi x_m x_i^{-1} w\right)\left(1-\phi x_m x_i\right)}{1-\phi^2 x_m x_i w}\right] \,,
\end{multline}
such that some factors cancel out, and the integrand reduces to
\begin{align}\label{eqn:HSSU2Ninterstep1}
\HS_\I^{G, \Phi} &= \oint \prod_{i=2}^{2N-1} \frac{\dd x_i}{2 \pi i x_i} \left[ \prod_{2 \leq i<j \leq 2 N-1}\left(1-\frac{x_i}{x_j}\right) \frac{1}{1-\phi x_i x_j} \right]
\notag\\[5pt]
&\hspace{40pt}
\times \sum_{m=2}^{2N-1} \left(1-\phi x_m^2\right) \left[\prod_{2 \le i \le 2N-1}^{i \ne m} \frac{1-\phi x_m x_i^{-1} w}{1-\phi^2 x_m x_i w}\  \frac{1-\phi x_m x_i}{1-x_i x_m^{-1}} \right] \,.
\end{align}
Next, we address the integration over $x_m$. As in the previous case, we have to single out all $x_m$-dependent terms of the integrand. In particular, the $i = m$ and $j = m$ contributions from the squared brackets in the first line of \cref{eqn:HSSU2Ninterstep1} can be factored out as
\begin{equation}\label{eqn:split-ij=m}
\prod_{2 \leq i<j \leq 2N-1} \frac{1-x_i x_j^{-1}}{1-\phi x_i x_j} = \left[\prod_{2 \leq i<j \leq 2 N-1}^{i \ne m, j \ne m} \frac{1-x_i x_j^{-1}}{1-\phi x_i x_j} \right] \left[\prod_{i = 2}^{m-1} \frac{1-x_i x_m^{-1}}{1-\phi x_i x_m}\right] \left[\prod_{i = m+1}^{2N-1} \frac{1-x_m x_i^{-1}}{1-\phi x_i x_m}\right]\, .
\end{equation}
On the other hand, we can rewrite the last factor in the second line of \cref{eqn:HSSU2Ninterstep1} as
\begin{equation}\label{eqn:split-i}
\prod_{2 \le i \le 2N-1}^{i \ne m} \frac{1-\phi x_m x_i}{1-x_i x_m^{-1}} = \left[\prod_{i = 2}^{m-1} \frac{1-\phi x_i x_m}{1-x_i x_m^{-1}}\right] \left[\prod_{i=m+1}^{2N-1} \frac{1-\phi x_i x_m}{1-x_i x_m^{-1}}\right] \,.
\end{equation}
A great simplification can be carried out by canceling repeated factors in \cref{eqn:split-ij=m,eqn:split-i}. It is also convenient to define the $x_m$-independent auxiliary variable
\begin{equation}
w_m \equiv w x_m = \prod_{2 \le i \le 2N-1}^{i \ne m} \frac{1}{x_i} \,.
\end{equation}
Then, it is easy to factor out the integration over $x_m$ in \cref{eqn:HSSU2Ninterstep1}:
\begin{align}\label{eqn:HSSU2Ninterstep2}
\HS_\I^{G, \Phi} = \sum_{m=2}^{2N-1} &\oint \prod_{2 \le i \le 2N-1}^{i \ne m} \frac{\dd x_i}{2 \pi i x_i} \left[\prod_{2 \leq i<j \leq 2 N-1}^{i \ne m, j \ne m} \frac{1-x_i x_j^{-1}}{1-\phi x_i x_j}\right] \left[\prod_{2 \le i \le 2N-1}^{i \ne m} \frac{1-\phi w_m x_i^{-1}}{1-\phi^2 w_m x_i}\right] \notag \\[5pt]
\times &\oint \frac{\dd x_m}{2 \pi i x_m} \left(1-\phi x_m^2\right) \left[ \prod_{i=m+1}^{2N-1} \left(-\frac{x_m}{x_i}\right) \right] \,.
\end{align}
To evaluate the $x_m$ integral, it is important to notice that the product factor in the second line of \cref{eqn:HSSU2Ninterstep2} is absent when $m = 2N-1$. Only in that case, there exists a contributing pole at $x_m = 0$, and thus the integral does not vanish. We make evident this condition by introducing $\delta_{2N-1}^m$, which selects the single non-vanishing contribution. Consequently, we obtain
\begin{align}
\HS_\I^{G, \Phi} &= \sum_{m=2}^{2N-1} \oint \prod_{2 \le i \le 2N-1}^{i \ne m} \frac{\dd x_i}{2 \pi i x_i} \left[\prod_{2 \leq i<j \leq 2 N-1}^{i \ne m, j \ne m} \frac{1-x_i x_j^{-1}}{1-\phi x_i x_j}\right] \left[\prod_{2 \le i \le 2N-1}^{i \ne m} \frac{1-\phi w_m x_i^{-1}}{1-\phi^2 w_m x_i}\right] \delta_{2N-1}^m \notag \\[8pt]
&= \oint \prod_{i=2}^{2N-2} \frac{\dd x_i}{2 \pi i x_i} \left[\prod_{2 \leq i<j \leq 2N-2} \left(1-\frac{x_i}{x_j}\right) \frac{1}{1-\phi x_i x_j} \right] \left[\prod_{i=2}^{2N-2} \frac{1-\phi w_{2N-1} x_i^{-1}}{1-\phi^2 w_{2N-1} x_i}\right]\, .
\end{align}
We can relabel the remaining variables in the set such that $i$ ranges from 1 to $2N-3$, \ie, $\{x_2,\cdots,x_{2N-2}\} \rightarrow \{x_1,\cdots,x_{2N-3}\}$. If we further flip their order, similarly as explained in \cref{eqn:invandflip}, we are left with
\begin{equation}
\HS_\I^{G, \Phi} = \oint \prod_{i=1}^{2N-3} \frac{\dd x_i}{2 \pi i x_i} \left[\prod_{1 \leq i<j \leq 2N-3} \frac{1-x_i^{-1}x_j}{1-\phi x_i x_j}\right] \left[\prod_{i=1}^{2N-3} \frac{1-\phi w_{2N-2} x_i^{-1}}{1-\phi^2 w_{2N-2} x_i}\right] \, ,
\label{eqn:HSSU2Ninterstep3}
\end{equation}
with $w_{2N-2} = \left(x_1 \cdots x_{2N-3}\right)^{-1}$. This integral falls into the form of the auxiliary function $I_k (\phi_1, \phi_2)$ generally defined in \cref{eqn:recursion}, which we will compute by an induction in below. With \cref{eqn:recursion}, we readily obtain the result quoted in \cref{eqn:HSIIIG}:
\begin{equation}
\HS_\I^{G, \Phi} = I_{N-1}(\phi, \phi) = \frac{1}{1-\phi^N} \,.
\end{equation}

\subsubsection*{Computing the auxiliary function $I_k\left(\phi_1, \phi_2\right)$ by induction}

We show that the auxiliary function $I_k(\phi_1,\phi_2)$ defined in below, for any integer $k\ge 1$ and the grading variables $|\phi_1|<1$, $|\phi_2|<1$, evaluates to
\begin{align}
I_k\left(\phi_1, \phi_2\right) &\equiv \oint \prod_{i=1}^{2k-1} \frac{\dd x_i}{2 \pi i x_i} \left[\prod_{1 \leq i<j \leq 2k-1} \frac{1-x_i^{-1}x_j}{1-\phi_1 x_i x_j}\right] \left[\prod_{i=1}^{2k-1} \frac{1-\phi_2 w_{2k} x_i^{-1}}{1-\phi_1\phi_2 w_{2k} x_i}\right]
\notag\\[5pt]
&= \frac{1}{1-\phi_1^k\phi_2} \,,
\label{eqn:recursion}
\end{align}
where $w_{2k} \equiv \left(x_1 \cdots x_{2k-1}\right)^{-1}$, and hereafter we will be generally using the notation
\begin{equation}
w_{n+1} \equiv \left(x_1 \cdots x_n\right)^{-1} \,.
\label{eqn:w}
\end{equation}
In order to prove \cref{eqn:recursion} by induction, we first verify its validity for lower values of $k$; then, assuming it is true for any $k$, we demonstrate it also holds for $k+1$. 

Starting with $k = 1$, the integral $I_1\left(\phi_1, \phi_2\right)$ can be done straightforwardly since it only involves one variable, $x_1$, with a single contributing pole at $x_1 = 0$:
\begin{equation}
I_1\left(\phi_1, \phi_2\right) \equiv \oint \frac{\dd x_1}{2\pi i x_1} \frac{1-\phi_2 x_1^{-1} w_2}{1-\phi_1\phi_2 x_1 w_2} = \oint \frac{\dd x_1}{2\pi i x_1} \frac{1-\phi_2 x_1^{-2}}{1-\phi_1\phi_2} = \frac{1}{1-\phi_1\phi_2}\, .
\end{equation}
The first non-trivial case, that is $k = 2$, is given by
\begin{align}
I_2\left(\phi_1, \phi_2\right) &\equiv \oint \frac{\dd x_1}{2\pi i x_1}\frac{\dd x_2}{2\pi i x_2}\frac{\dd x_3}{2\pi i x_3} \frac{1-x_1^{-1}x_2}{1-\phi_1 x_1 x_2} \frac{1-x_1^{-1}x_3}{1-\phi_1 x_1 x_3} \frac{1-x_2^{-1}x_3}{1-\phi_1 x_2 x_3} \notag \\[5pt]
& \hspace{40pt}\times \frac{1-\phi_2 x_1^{-1} w_4}{1-\phi_1\phi_2 x_1 w_4} \frac{1-\phi_2 x_2^{-1} w_4}{1-\phi_1\phi_2 x_2 w_4} \frac{1-\phi_2 x_3^{-1} w_4}{1-\phi_1\phi_2 x_3 w_4}\, .
\end{align}
We focus on the integration over $x_3$ first. Aiming to simplify the pole structure in the expression above, we perform the variable change and flip the order of the variable set analogously to \cref{eqn:invandflip}. Factoring out the $x_3$ integral, we can write down
\begin{align}\label{eqn:I2interstep1}
I_2\left(\phi_1, \phi_2\right) &\equiv \oint \frac{\dd x_1}{2\pi i x_1}\frac{\dd x_2}{2\pi i x_2} \frac{1-x_1^{-1}x_2}{1-\phi_1 x_1^{-1} x_2^{-1}} \notag \\[5pt]
&\hspace{40pt} \times \oint \frac{\dd x_3}{2\pi i x_3} \frac{1-x_1^{-1}x_3}{1-\phi_1 x_1^{-1} x_3^{-1}} \frac{1-x_2^{-1}x_3}{1-\phi_1 x_2^{-1} x_3^{-1}} \notag \\[5pt]
& \hspace{80pt} \times \frac{1-\phi_2 x_1^2 x_2 x_3}{1-\phi_1\phi_2 x_2 x_3} \frac{1-\phi_2 x_1 x_2^2 x_3}{1-\phi_1\phi_2 x_1 x_3} \frac{1-\phi_2 x_1 x_2 x_3^2}{1-\phi_1\phi_2 x_1 x_2}\, .
\end{align}
The only contributing poles are placed at $x_3 = \phi_1 x_1^{-1}$ and $x_3 = \phi_1 x_2^{-1}$, such that after integration we obtain
\begin{align}
&\oint \frac{\dd x_3}{2\pi i x_3} \frac{1-x_1^{-1}x_3}{1-\phi_1 x_1^{-1} x_3^{-1}} \frac{1-x_2^{-1}x_3}{1-\phi_1 x_2^{-1} x_3^{-1}} \notag \frac{1-\phi_2 x_1^2 x_2 x_3}{1-\phi_1\phi_2 x_2 x_3} \frac{1-\phi_2 x_1 x_2^2 x_3}{1-\phi_1\phi_2 x_1 x_3} \frac{1-\phi_2 x_1 x_2 x_3^2}{1-\phi_1\phi_2 x_1 x_2}\\[5pt]
&=\left(1-\phi_1 x_1^{-2}\right) \frac{1-\phi_1 x_1^{-1} x_2^{-1}}{1-x_1 x_2^{-1}} \frac{1-\phi_1\phi_2 x_2^2}{1-\phi_1^2\phi_2} + \left(1-\phi_1 x_2^{-2}\right) \frac{1-\phi_1 x_1^{-1} x_2^{-1}}{1-x_1^{-1} x_2} \frac{1-\phi_1\phi_2 x_1^2}{1-\phi_1^2\phi_2}\, .
\end{align}
Plugging the previous result into \cref{eqn:I2interstep1} leads to
\begin{align}
I_2\left(\phi_1, \phi_2\right) \equiv \frac{1}{1-\phi_1^2\phi_2} \oint \frac{\dd x_1}{2\pi i x_1}\frac{\dd x_2}{2\pi i x_2} &\Bigg[\left(1-\phi_1 x_1^{-2}\right)\left(1-\phi_1\phi_2 x_2^2\right)\left(-\frac{x_2}{x_1}\right) \notag \\[5pt]
&+ \left(1-\phi_1 x_2^{-2}\right)\left(1-\phi_1\phi_2 x_1^2\right)\Bigg]\, .
\end{align}
The integration over $x_2$ follows straightforwardly as the first term in the squared brackets has no contributing poles, and thus vanishes, while the second term presents a single pole at $x_2=0$. Once the $x_2$ integral is performed, it is easy to check that the remaining integral over $x_1$ evaluates to 1 when computing the residue at the pole $x_1 = 0$. More precisely, we have
\begin{equation}
I_2\left(\phi_1, \phi_2\right) \equiv \frac{1}{1-\phi_1^2\phi_2} \oint \frac{\dd x_1}{2\pi i x_1} \left(1-\phi_1\phi_2 x_1^2\right) = \frac{1}{1-\phi_1^2\phi_2}\, .
\end{equation}
At this point, we assume that the result in \cref{eqn:recursion} is true for any $k$, and prove it also holds for $k+1$. The latter case is given by
\begin{equation}
I_{k+1}\left(\phi_1, \phi_2\right) \equiv \oint \prod_{i=1}^{2k+1} \frac{\dd x_i}{2 \pi i x_i} \left[\prod_{1 \leq i<j \leq 2k+1} \frac{1-x_i^{-1}x_j}{1-\phi_1 x_i x_j}\right] \left[\prod_{i=1}^{2k+1} \frac{1-\phi_2 w_{2k+2} x_i^{-1}}{1-\phi_1\phi_2 w_{2k+2} x_i}\right]\, ,
\end{equation}
with $w_{2k+2} \equiv \left(x_1 \cdots x_{2k+1}\right)^{-1}$. As done for lower values of $k$, we start integrating the last variable of the set, namely $x_{2k+1}$. To this aim, it is convenient to apply the variable change and order flipping as explained in \cref{eqn:invandflip}, so that the pole structure of the integrand gets simplified: 
\begin{equation}
I_{k+1}\left(\phi_1, \phi_2\right) \equiv \oint \prod_{i=1}^{2k+1} \frac{\dd x_i}{2 \pi i x_i} \left[\prod_{1 \leq i<j \leq 2k+1} \frac{1-x_i^{-1} x_j}{1-\phi_1 x_i^{-1} x_j^{-1}}\right] \left[\prod_{i=1}^{2k+1} \frac{1-\phi_2 x_i w_{2k+2}^{-1}}{1-\phi_1\phi_2 x_i^{-1} w_{2k+2}^{-1}}\right] \, .
\end{equation}
Taking into account that $w_{2k+2}$ depends on $x_{2k+1}$, while the combination $w_{2k+1} \equiv x_{2k+1} w_{2k+2}$ is independent of the target variable, it is easy to factor out the $x_{2k+1}$ integral as
\begin{align}
I_{k+1}\left(\phi_1, \phi_2\right) &\equiv \oint \prod_{i=1}^{2k} \frac{\dd x_i}{2 \pi i x_i} \left[ \prod_{1 \leq i<j \leq 2k} \frac{1-x_i^{-1} x_j}{1-\phi_1 x_i^{-1} x_j^{-1}} \right]
\oint \frac{\dd x_{2k+1}}{2 \pi i x_{2k+1}} \left[\prod_{i=1}^{2k} \frac{1-x_i^{-1} x_{2k+1}}{1-\phi_1 x_i^{-1} x_{2k+1}^{-1}}\right] \notag\\[5pt]
&\hspace{40pt}
\times \left[\frac{1-\phi_2 x_{2k+1}^2 w_{2k+1}^{-1}}{1-\phi_1\phi_2 w_{2k+1}^{-1}} \prod_{i=1}^{2k} \frac{1-\phi_2 x_{2k+1} x_i w_{2k+1}^{-1}}{1-\phi_1\phi_2 x_{2k+1} x_i^{-1} w_{2k+1}^{-1}}\right] \,.
\end{align}
One can readily check that the contributing poles are placed at $x_{2k+1} = \phi_1 x_m^{-1}$, with $m=1,\cdots,2k$. Evaluating the corresponding integral yields
\begin{align}\label{eqn:recursioninterstep1}
I_{k+1}\left(\phi_1, \phi_2\right) &\equiv \oint \prod_{i=1}^{2k} \frac{\dd x_i}{2 \pi i x_i} \left[\prod_{1 \leq i<j \leq 2k} \frac{1-x_i^{-1} x_j}{1-\phi_1 x_i^{-1} x_j^{-1}}\right] \notag\\[5pt]
&\hspace{-20pt}
\times\sum_{m=1}^{2k} \left(1-\phi_1 x_m^{-2}\right) \left[\prod_{1 \le i \le 2k}^{i \ne m} \frac{1-\phi_1 x_i^{-1} x_m^{-1}}{1-x_i^{-1} x_m} \frac{1-\phi_1 \phi_2 x_m^{-1} x_i w_{2k+1}^{-1}}{1-\phi_1^2 \phi_2 x_m^{-1} x_i^{-1} w_{2k+1}^{-1}} \right] \,.
\end{align}
Next, we proceed with the integration over $x_m$. Following our previous strategy, we should first apply the change of variables and order flipping as done in \cref{eqn:invandflip}, that is,
\begin{align}\label{eqn:recursioninterstep2}
I_{k+1}\left(\phi_1, \phi_2\right) &\equiv \oint \prod_{i=1}^{2k} \frac{\dd x_i}{2 \pi i x_i} \left[\prod_{1 \leq i<j \leq 2k} \frac{1-x_i^{-1} x_j}{1-\phi_1 x_i x_j}\right] \notag \\[5pt]
&\times \sum_{m=1}^{2k} \left(1-\phi_1 x_m\right) \left[\prod_{1 \le i \le 2k}^{i \ne m} \frac{1-\phi_1 x_i x_m}{1-x_i x_m^{-1}} \frac{1-\phi_1 \phi_2 x_m x_i^{-1} w_{2k+1}}{1-\phi_1^2 \phi_2 x_m x_i w_{2k+1}} \right] \, .
\end{align}
We also define the $x_m$-independent auxiliary variable
\begin{equation}
u_m \equiv x_m w_{2k+1} = \prod_{1 \le i \le 2k}^{i\ne m} \frac{1}{x_i} \,.
\label{eqn:auxiliary-um}
\end{equation}
The structure of the integrand can be further simplified once we separate the $i=m$ and $j=m$ contributions from the squared brackets in the first line of \cref{eqn:recursioninterstep2} as
\begin{equation}
\prod_{1 \leq i<j \leq 2k} \frac{1-x_i^{-1} x_j}{1-\phi_1 x_i x_j} = \left[\prod_{1 \leq i<j \leq 2k}^{i\ne m, j\ne m} \frac{1-x_i^{-1} x_j}{1-\phi_1 x_i x_j}\right] \left[\prod_{i=1}^{m-1} \frac{1-x_i^{-1} x_m}{1-\phi_1 x_i x_m}\right] \left[\prod_{i=m+1}^{2k} \frac{1-x_i x_m^{-1}}{1-\phi_1 x_i x_m}\right] \, ,
\label{eqn:splitfactor}
\end{equation}
which partially cancel out the corresponding factors arising from the first term within the square brackets in the second line, namely
\begin{equation}
\prod_{1 \le i \le 2k}^{i \ne m} \frac{1-\phi_1 x_i x_m}{1-x_i x_m^{-1}} = \left[\prod_{i=1}^{m-1} \frac{1-\phi_1 x_i x_m}{1-x_i x_m^{-1}}\right] \left[\prod_{i=m+1}^{2k} \frac{1-\phi_1 x_i x_m}{1-x_i x_m^{-1}}\right]\, .
\end{equation}
After simplification, it is easy to factor out the $x_m$ integral, in such a way that the integrand in \cref{eqn:recursioninterstep2} reads
\begin{align}
I_{k+1}\left(\phi_1, \phi_2\right) \equiv \sum_{m=1}^{2k} &\oint \prod_{1 \le i \le 2k}^{i\ne m} \frac{\dd x_i}{2 \pi i x_i} \left[\prod_{1 \leq i<j \leq 2k}^{i\ne m, j\ne m} \frac{1-x_i^{-1} x_j}{1-\phi_1 x_i x_j}\right] \left[\prod_{1 \le i \le 2k}^{i\ne m} \frac{1-\phi_1\phi_2 x_i^{-1} u_m}{1-\phi_1^2\phi_2 x_i u_m}\right] \notag \\[5pt] 
\times &\oint \frac{\dd x_m}{2 \pi i x_m} \left(1-\phi_1 x_m^2\right) \left[\prod_{i=1}^{m-1} \left(-\frac{x_m}{x_i}\right)\right] \,,
\end{align}
where we have introduced the auxiliary variable defined in \cref{eqn:auxiliary-um}. From the second line, it is clear that there exists a pole at $x_m = 0$ when the factor within the squared brackets is absent, that is, for $m = 1$; otherwise, there are no poles and the integral vanishes. For that particular case, the integral evaluates to 1, leading to
\begin{align}
I_{k+1}\left(\phi_1, \phi_2\right) &\equiv \sum_{m=1}^{2k} \oint \prod_{1 \le i \le 2k}^{i\ne m} \frac{\dd x_i}{2 \pi i x_i} \left[\prod_{1 \leq i<j \leq 2k}^{i\ne m, j\ne m} \frac{1-x_i^{-1} x_j}{1-\phi_1 x_i x_j}\right] \left[\prod_{1 \le i \le 2k}^{i\ne m} \frac{1-\phi_1\phi_2 x_i^{-1} u_m}{1-\phi_1^2\phi_2 x_i u_m}\right] \delta_1^m \notag \\[8pt]
&= \oint \prod_{i=2}^{2k} \frac{\dd x_i}{2 \pi i x_i} \left[\prod_{2\le i < j \le 2k} \frac{1-x_i^{-1} x_j}{1-\phi_1 x_i x_j}\right] \left[\prod_{i=2}^{2k} \frac{1-\phi_1\phi_2 x_i^{-1} u_1}{1-\phi_1^2\phi_2 x_i u_1}\right] \, .
\end{align}
Relabeling the variables in the set such that $i$ ranges from 1 to $2k-1$, this means changing $x_i \rightarrow x_{i-1}$ and thus $u_1 \rightarrow w_{2k} = \left(x_1 \cdots x_{2k-1} \right)^{-1}$, yields
\begin{equation}
\hspace{-10pt}
I_{k+1}\left(\phi_1, \phi_2\right) \equiv \oint \prod_{i=1}^{2k-1} \frac{\dd x_i}{2 \pi i x_i} \left[\prod_{1\le i < j \le 2k-1} \frac{1-x_i^{-1} x_j}{1-\phi_1 x_i x_j}\right] \left[\prod_{i=1}^{2k-1} \frac{1-\phi_1\phi_2 x_i^{-1} w_{2k}}{1-\phi_1^2\phi_2 x_i w_{2k}}\right] \, .
\end{equation}
In this form, it is easy to check that
\begin{equation}
I_{k+1}\left(\phi_1, \phi_2\right) = I_k\left(\phi_1, \phi_1\phi_2\right) = \frac{1}{1-\phi_1^{k+1}\phi_2}\, .
\end{equation}
Therefore, we can induce that \cref{eqn:recursion} is valid for any $k$.

\subsubsection{The Hilbert series $\HS_\I^{H_1,\Phi}$}
\label{appsubsubsec:ClassIIIH1}

In this subsection, we turn into the computation of the Hilbert series in \cref{eqn:HSIIIH1Short} to derive the result quoted in \cref{eqn:HSIIIH1}. 

Using the Molien-Weyl formula, the Hilbert series $\HS_\I^{H_1,\Phi}$ is given by
\begin{equation}
\HS_\I^{H_1,\Phi} = \oint \frac{\dd x}{2 \pi i x} \prod_{i=1}^{r} \frac{\dd x_i}{2 \pi i x_i} \left[\prod_{1 \le i < j \le r+1} \frac{1-x_i x_j^{-1}}{1-\varphi_1 x_i x_j x^2}\right] \left[\prod_{i=1}^{r+1} \frac{1}{1-\varphi_2 x_i x^{-2(N-1)}}\right] \, ,
\end{equation}
with $r = 2N-2$ and $x_{r+1} = \left(x_1 \cdots x_r\right)^{-1}$. To evaluate this integral, we first perform the same change of variables introduced in \cref{eqn:xtoyVariableChange}, followed by the order flipping explained in \cref{eqn:invandflip}:
\begin{equation}
\HS_\I^{H_1,\Phi} = \oint \prod_{i=1}^{2N-1}\frac{\dd y_i}{2 \pi i y_i} \left[\prod_{1 \le i < j \le 2N-1} \frac{1-y_i^{-1} y_j}{1-\varphi_1 y_i y_j}\right] \left[\prod_{i=1}^{2N-1} \frac{1}{1-\varphi_2 y_i w_{2N}}\right]\, ,
\label{eqn:HSH1interstep1}
\end{equation}
where $w_{2N} \equiv \left(y_1 \cdots y_{2N-1}\right)^{-1} = x^{-(2N-1)}$. The integral becomes less tangled in terms of the new variables $y_i$. Then, it is easier to follow our previous strategy and integrate one variable first, say $y_{2N-1}$. In this sense, we can introduce $w_{2N-1} \equiv y_{2N-1} w_{2N}$ in order to make all $y_{2N-1}$ dependence transparent. Once we factor out the $y_{2N-1}$ integral, \cref{eqn:HSH1interstep1} reads
\begin{align}
\HS_\I^{H_1,\Phi} &= \oint \prod_{i=1}^{2N-2}\frac{\dd y_i}{2 \pi i y_i} \left[ \prod_{1\le i < j \le 2N-2} \frac{1-y_i^{-1}y_j}{1-\varphi_1 y_i y_j} \right] \frac{1}{1-\varphi_2 w_{2N-1}} \notag \\[5pt]
&\times \oint \frac{\dd y_{2N-1}}{2 \pi i y_{2N-1}} \left[\prod_{i=1}^{2N-2} \frac{1-y_i^{-1}y_{2N-1}}{1-\varphi_1 y_i y_{2N-1}} \frac{1}{1-\varphi_2 y_{2N-1}^{-1} y_i w_{2N-1}}\right] \, .
\end{align}
The contributing poles are placed at $y_{2N-1} = \varphi_2 y_m w_{2N-1}$, with $m=1,\cdots,2N-2$. We can further define the auxiliary variable
\begin{equation}
v_m \equiv y_m w_{2N-1} = \prod_{1\le i \le 2N-2}^{i\ne m} \frac{1}{y_i}\, ,
\end{equation}
which does not depend $y_m$. Then, the previous poles are simply $y_{2N-1} \equiv \varphi_2 v_m$. After integration, we obtain the following result:
\begin{align}\label{eqn:HSH1interstep2}
\HS_\I^{H_1,\Phi} = \oint &\prod_{i=1}^{2N-2}\frac{\dd y_i}{2 \pi i y_i} \left[ \prod_{1\le i < j \le 2N-2} \frac{1-y_i^{-1}y_j}{1-\varphi_1 y_i y_j} \right] \notag\\[5pt]
\times &\sum_{m=1}^{2N-2} \frac{1}{1-\varphi_1\varphi_2 y_m v_m} \left[ \prod_{1\le i \le 2N-2}^{i\ne m} \frac{1}{1-y_i y_m^{-1}} \frac{1-\varphi_2 y_i^{-1} v_m}{1-\varphi_1\varphi_2 y_i v_m}\right] \, .
\end{align}
Next, we continue with the integration over $y_m$. In order to make clear the $x_m$ dependence of the integrand, it is convenient to factor out the $i=m$ and $j=m$ contributions within the squared brackets in the first line of \cref{eqn:HSH1interstep2} as 
\begin{equation}
\prod_{1\le i < j \le 2N-2} \frac{1-y_i^{-1}y_j}{1-\varphi_1 y_i y_j} = \prod_{1\le i < j \le 2N-2}^{i\ne m, j\ne m} \frac{1-y_i^{-1}y_j}{1-\varphi_1 y_i y_j} \prod_{i=1}^{m-1} \frac{1-y_i^{-1}y_m}{1-\varphi_1 y_i y_m} \prod_{i=m+1}^{2N-2} \frac{1-y_i y_m^{-1}}{1-\varphi_1 y_i y_m} \, ,
\label{eqn:split-ij=m-2}
\end{equation}
while the first term within the squared brackets in the second line of \cref{eqn:HSH1interstep2} can be separate as
\begin{equation}
\prod_{1\le i \le 2N-2}^{i\ne m} \frac{1}{1-y_i y_m^{-1}} = \prod_{i=1}^{m-1} \frac{1}{1-y_i y_m^{-1}} \prod_{i=m+1}^{2N-2} \frac{1}{1-y_i y_m^{-1}}\, .
\label{eqn:split-i-2}
\end{equation}
After simplification of repeated factors in \cref{eqn:split-ij=m-2,eqn:split-i-2}, we can write down
\begin{align}
\HS_\I^{H_1,\Phi} = \sum_{m=1}^{2N-2} &\oint \prod_{1\le i \le 2N-2}^{i\ne m} \frac{\dd y_i}{2 \pi i y_i} \left[\prod_{1\le i < j \le 2N-2}^{i\ne m, j\ne m} \frac{1-y_i^{-1}y_j}{1-\varphi_1 y_i y_j}\right] \left[\prod_{1\le i \le 2N-2}^{i\ne m} \frac{1-\varphi_2 y_i^{-1} v_m}{1-\varphi_1\varphi_2 y_i v_m}\right] \notag \\[5pt]
\times &\oint \frac{\dd y_m}{2 \pi i y_m} \frac{1}{1-\varphi_1\varphi_2 y_m v_m} \left[\prod_{i=1}^{m-1} \left(-\frac{y_m}{y_i}\right) \right] \left[\prod_{1\le i \le 2N-2}^{i\ne m} \frac{1}{1-\varphi_1 y_i y_m}\right] \, .
\end{align}
Similarly to previous computations, it is clear that there exists a single contributing pole at $y_m = 0$ when the first product factor in squared brackets is absent, that is, for $m = 1$; otherwise, there are no poles and the integral vanishes. For that particular case, the integration over $y_m$ evaluates to 1, yielding
\begin{align}
\HS_\I^{H_1,\Phi} &= \sum_{m=1}^{2N-2} \oint \prod_{1\le i \le 2N-2}^{i\ne m} \frac{\dd y_i}{2 \pi i y_i} \left[\prod_{1\le i < j \le 2N-2}^{i\ne m, j\ne m} \frac{1-y_i^{-1}y_j}{1-\varphi_1 y_i y_j}\right] \left[\prod_{1\le i \le 2N-2}^{i\ne m} \frac{1-\varphi_2 y_i^{-1} v_m}{1-\varphi_1\varphi_2 y_i v_m}\right] \delta_1^m \notag \\[8pt]
&= \oint \prod_{2\le i \le 2N-2} \frac{\dd y_i}{2 \pi i y_i} \left[\prod_{2\le i < j \le 2N-2} \frac{1-y_i^{-1}y_j}{1-\varphi_1 y_i y_j}\right] \left[\prod_{2\le i \le 2N-2} \frac{1-\varphi_2 y_i^{-1} v_1}{1-\varphi_1\varphi_2 y_i v_1}\right] \, .
\end{align}
Changing variables as $y_i \rightarrow y_{i-1}$ in such a way that $i$ ranges from 1 to $2N-3$ and $v_1 \rightarrow w_{2N-2} = \left(y_1 \cdots y_{2N-3}\right)^{-1}$, we are left with
\begin{equation}
\HS_\I^{H_1,\Phi} = \oint \prod_{i=1}^{2N-3} \frac{\dd y_i}{2 \pi i y_i} \left[\prod_{1\le i < j \le 2N-3} \frac{1-y_i^{-1}y_j}{1-\varphi_1 y_i y_j}\right] \left[\prod_{i=1}^{2N-3} \frac{1-\varphi_2 y_i^{-1} w_{2N-2}}{1-\varphi_1\varphi_2 y_i w_{2N-2}}\right] \,.
\end{equation}
Finally, this falls to the form of the auxiliary function defined in \cref{eqn:recursion}, with which we readily obtain the result quoted in \cref{eqn:HSIIIH1}:
\begin{equation}
\HS_\I^{H_1,\Phi} = I_{N-1}\left(\varphi_1, \varphi_2\right)
= \frac{1}{1-\varphi_1^{N-1}\varphi_2} \,.
\end{equation}

\subsubsection{The Hilbert series $\HS_\I^{H_2,\Phi}$}
\label{appsubsubsec:ClassIIIH2}

In this subsection, we compute the Hilbert series in \cref{eqn:HSIIIH2Short} to derive the result in \cref{eqn:HSIIIH2}.

Using the Molien-Weyl formula, the Hilbert series for $\HS_\I^{H_2,\Phi}$ invariants reads
\begin{equation}
\HS_\I^{H_2,\Phi} \equiv \oint \prod_{i=1}^{r} \frac{\dd x_i}{2 \pi i x_i} \left[\prod_{1 \le i<j \le r+1} \frac{1-x_i x_j^{-1}}{1-\varphi_1 x_i x_j}\right] \left[\prod_{i=1}^{r+1} \frac{1}{1-\varphi_2 x_i}\right] \, ,
\label{eqn:HSIIIH2ShortMolien-Weyl}
\end{equation}
being $r = 2N-2$, and $x_{r+1} \equiv \left(x_1 \cdots x_{r}\right)^{-1}$. Taking into account our definition in \cref{eqn:w}, we can rewrite the previous expression as
\begin{equation}
\HS_\I^{H_2,\Phi} \equiv \oint \prod_{i=1}^{r} \frac{\dd x_i}{2 \pi i x_i} \left[\prod_{1 \le i<j \le r} \frac{1-x_i x_j^{-1}}{1-\varphi_1 x_i x_j}\right] \left[\prod_{i=1}^{r} \frac{1}{1-\varphi_2 x_i} \frac{1-x_i w_{2N-1}^{-1}}{1-\varphi_1 x_i w_{2N-1}} \right] \frac{1}{1-\varphi_2 w_{2N-1}}\, .
\end{equation}
To evaluate this integral, we first focus on the integration variable $x_{2N-2}$. The corresponding pole structure can be simplified by flipping the order of the variables in the set $\{x_1,\cdots,x_{2N-2}\}$, as explained in \cref{eqn:invandflip}. The dependence on $x_{2N-2}$ is made explicit once we use $w_{2N-2} \equiv x_{2N-2} w_{2N-1}$. This yields
\begin{align}
\HS_\I^{H_2,\Phi} &\equiv \oint \prod_{i=1}^{2N-3} \frac{\dd x_i}{2 \pi i x_i} \left[ \prod_{1 \le i<j \le 2N-3} \frac{1-x_i^{-1} x_j}{1-\varphi_1 x_i x_j} \right] \left[ \prod_{i=1}^{2N-3} \frac{1}{1-\varphi_2 x_i} \right] \frac{1}{1-\varphi_1 w_{2N-2}} \notag \\[5pt]
&\hspace{40pt} \times \oint \frac{\dd x_{2N-2}}{2 \pi i x_{2N-2}} \frac{1}{1-\varphi_2 x_{2N-2}^{-1} w_{2N-2}} \frac{1-x_{2N-2}^2 w_{2N-2}^{-1}}{1-\varphi_2 x_{2N-2}} \notag \\[5pt]
&\hspace{80pt} \times \left[ \prod_{i=1}^{2N-3} \frac{1-x_i^{-1} x_{2N-2}}{1-\varphi_1 x_i x_{2N-2}} \frac{1-x_i x_{2N-2} w_{2N-2}^{-1}}{1-\varphi_1 x_i x_{2N-2}^{-1} w_{2N-2}} \right]\, .
\end{align}
We can identify two contributing poles, namely $x_{2N-2} = \varphi_1 x_m w_{2N-2}$, with $m = 1,\cdots,2N-3$, and $x_{2N-2} = \varphi_2 w_{2N-2}$. At this point, aiming to keep track of the subsequent computations more easily, we split the result after integration over $x_{2N-2}$ as
\begin{equation}
\HS_\I^{H_2,\Phi} \equiv \HS_1 + \HS_2\, ,
\end{equation}
where
\begin{align}\label{eqn:H1}
\HS_1 &\equiv \oint \prod_{i=1}^{2N-3} \frac{\dd x_i}{2 \pi i x_i} \left[ \prod_{1 \le i<j \le 2N-3} \frac{1-x_i^{-1} x_j}{1-\varphi_1 x_i x_j} \right] \left[\prod_{i=1}^{2N-3} \frac{1}{1-\varphi_2 x_i} \right] \notag \\[5pt]
&\times \sum_{m=1}^{2N-3} \frac{1-\varphi_1 x_m^2}{1-\varphi_1^{-1}\varphi_2 x_m^{-1}}\frac{1}{1-\varphi_1\varphi_2 x_m w_{2N-2}} \left[\prod_{1\le i \le 2N-3}^{i\ne m} \frac{1-\varphi_1 x_i x_m}{1-x_i x_m^{-1}} \frac{1-\varphi_1 x_i^{-1} x_m w_{2N-2}}{1-\varphi_1^2 x_i x_m w_{2N-2}}\right] \, ,
\end{align}
and
\begin{align}\label{eqn:H2}
\HS_2 &\equiv \oint \prod_{i=1}^{2N-3} \frac{\dd x_i}{2 \pi i x_i} \left[ \prod_{1 \le i<j \le 2N-3} \frac{1-x_i^{-1} x_j}{1-\varphi_1 x_i x_j} \right] \frac{1}{1-\varphi_1 w_{2N-2}} \notag \\[5pt]
&\hspace{40pt} \times \left[\prod_{i=1}^{2N-3} \frac{1}{1-\varphi_1\varphi_2^{-1} x_i} \frac{1-\varphi_2 x_i^{-1} w_{2N-2}}{1-\varphi_1\varphi_2 x_i w_{2N-2}} \right] \, .
\end{align}
Next, we proceed with the $x_{2N-3}$ integral in both $\HS_1$ and $\HS_2$ contributions. On the one hand, regarding $\HS_1$ in \cref{eqn:H1}, we can significantly simplify the sum over $m$ by applying, term by term, the following variable change:
\begin{equation}
\mqty(x_m \\ x_{m+1} \\ \vdots \\ x_{2N-3}) \longrightarrow \mqty(x_{2N-3} \\ x_m \\ \vdots \\ x_{2N-4})\, ,
\label{eqn:xmtox2N-3VariableChange}
\end{equation}
that is, for each value of $m$, we change $x_m \rightarrow x_{2N-3}$, and then relabel the remaining variables so that they range from 1 to $2N-4$. This variable change implies that 
\begin{equation}
\prod_{1\le i < j \le 2N-3} \left(1-x_i^{-1}x_j\right) \longrightarrow \left[\prod_{i=m}^{2N-4} \left(-x_{2N-3}^{-1} x_i\right)\right] \left[\prod_{1\le i < j \le 2N-3} \left(1-x_i^{-1}x_j\right)\right] \, ,
\label{eqn:factorafterVariableChange}
\end{equation}
in the first line of \cref{eqn:H1}. Further introducing $w_{2N-3} \equiv x_{2N-3} w_{2N-2}$, we are left with
\begin{align}\label{eqn:H1interstep1}
\HS_1 &\equiv \oint \prod_{i=1}^{2N-3} \frac{\dd x_i}{2 \pi i x_i} \left[ \prod_{1 \le i<j \le 2N-3} \frac{1-x_i^{-1} x_j}{1-\varphi_1 x_i x_j} \right] \left[\prod_{i=1}^{2N-3} \frac{1}{1-\varphi_2 x_i} \right] \notag \\[5pt]
&\hspace{40pt} \times \left[\prod_{i=1}^{2N-4} \frac{1-\varphi_1 x_i x_{2N-3}}{1-x_i x_{2N-3}^{-1}} \frac{1-\varphi_1 x_i^{-1} w_{2N-3}}{1-\varphi_1^2 x_i w_{2N-3}}\right] \frac{1}{1-\varphi_1 \varphi_2 w_{2N-3}} \frac{1-\varphi_1 x_{2N-3}^2}{1-\varphi_1^{-1}\varphi_2 x_{2N-3}^{-1}} \notag \\[5pt]
&\hspace{80pt} \times \left[\sum_{m=1}^{2N-3} \prod_{i=m}^{2N-4} \left(-x_{2N-3}^{-1} x_i\right)\right]\, .
\end{align}
The first factor within squared brackets in the first line of the expression above can be expanded as
\begin{equation}
\prod_{1 \le i<j \le 2N-3} \frac{1-x_i^{-1} x_j}{1-\varphi_1 x_i x_j} = \left[\prod_{1 \le i<j \le 2N-4} \frac{1-x_i^{-1} x_j}{1-\varphi_1 x_i x_j}\right] \left[\prod_{i=1}^{2N-4} \frac{1-x_i^{-1} x_{2N-3}}{1-\varphi_1 x_1 x_{2N-3}} \right]\, ,
\end{equation}
and thus we can cancel out repeated factors appearing in the second and third lines of \cref{eqn:H1interstep1}. After simplification, it is straightforward to factor out the $x_{2N-3}$ integral as
\begin{align}
\HS_1 &\equiv \oint \prod_{i=1}^{2N-4} \frac{\dd x_i}{2 \pi i x_i} \left[ \prod_{1 \le i<j \le 2N-4} \frac{1-x_i^{-1} x_j}{1-\varphi_1 x_i x_j} \right] \notag \\[5pt]
&\hspace{20pt} \times \left[\prod_{i=1}^{2N-4} \frac{1}{1-\varphi_2 x_i} \frac{1-\varphi_1 x_i^{-1} w_{2N-3}}{1-\varphi_1^2 x_i w_{2N-3}} \right] \frac{1}{1-\varphi_1\varphi_2 w_{2N-3}}\notag \\[5pt]
&\hspace{40pt} \times \oint \frac{\dd x_{2N-3}}{2 \pi i x_{2N-3}} \frac{1}{1-\varphi_2 x_{2N-3}} \frac{1-\varphi_1 x_{2N-3}^2}{1-\varphi_1^{-1}\varphi_2 x_{2N-3}^{-1}} \left[\sum_{m=1}^{2N-3} \prod_{i=1}^{m-1} \left(-x_i^{-1} x_{2N-3}\right)\right]\, .
\end{align}
The only contributing pole is at $x_{2N-3} = \varphi_1^{-1}\varphi_2$ provided that the grading variables satisfy $|\varphi_1| > |\varphi_2|$; otherwise, there are no poles inside the contour $|x_{2N-3}| = 1$ and the integral vanishes. We make evident this condition by introducing the notation $\delta_{|\varphi_1| > |\varphi_2|}$ in the following way:
\begin{align}
\HS_1 \equiv \delta_{|\varphi_1| > |\varphi_2|} &\oint \prod_{i=1}^{2N-4} \frac{\dd x_i}{2 \pi i x_i} \left[ \prod_{1 \le i<j \le 2N-4} \frac{1-x_i^{-1} x_j}{1-\varphi_1 x_i x_j} \right] \frac{1}{1-\varphi_1\varphi_2 w_{2N-3}} \notag \\[5pt]
&\times \left[\prod_{i=1}^{2N-4} \frac{1}{1-\varphi_2 x_i} \frac{1-\varphi_1 x_i^{-1} w_{2N-3}}{1-\varphi_1^2 x_i w_{2N-3}} \right]  \sum_{m=1}^{2N-3} \prod_{i=1}^{m-1} \left(-\varphi_1^{-1}\varphi_2 x_i^{-1}\right)\, .
\end{align}
Finally, we perform the variable change and order flipping analogously to \cref{eqn:invandflip}, yielding
\begin{align}\label{eqn:HS1result}
\HS_1 \equiv \delta_{|\varphi_1| > |\varphi_2|} &\oint \prod_{i=1}^{2N-4} \frac{\dd x_i}{2 \pi i x_i} \left[ \prod_{1 \le i<j \le 2N-4} \frac{1-x_i^{-1} x_j}{1-\varphi_1 x_i^{-1} x_j^{-1}} \right] \frac{1}{1-\varphi_1\varphi_2 w_{2N-3}^{-1}} \notag \\[5pt]
&\times \left[\prod_{i=1}^{2N-4} \frac{1}{1-\varphi_2 x_i^{-1}} \frac{1-\varphi_1 x_i w_{2N-3}^{-1}}{1-\varphi_1^2 x_i^{-1} w_{2N-3}^{-1}} \right]  \sum_{m=1}^{2N-3} \prod_{i=1}^{m-1} \left(-\varphi_1^{-1}\varphi_2 x_i\right)\, .
\end{align}
On the other hand, let us now turn into the computation of $\HS_2$ in \cref{eqn:H2}. As mentioned before, we first focus on the integration over $x_{2N-3}$. In order to simplify the pole structure of the integrand, we make use of the change of variables and order reversal quoted in \cref{eqn:invandflip}. Further introducing $w_{2N-2} = x_{2N-3}^{-1} w_{2N-3}$, the $x_{2N-3}$ integral can be factor out as
\begin{align}
\HS_2 &\equiv \oint \prod_{i=1}^{2N-4} \frac{\dd x_i}{2 \pi i x_i} \left[\prod_{1 \le i<j \le 2N-4} \frac{1-x_i^{-1} x_j}{1-\varphi_1 x_i^{-1} x_j^{-1}}\right] \left[\prod_{i=1}^{2N-4} \frac{1}{1-\varphi_1\varphi_2^{-1}x_i^{-1}}\right] \frac{1}{1-\varphi_1\varphi_2 w_{2N-3}^{-1}} \notag \\[5pt]
&\hspace{40pt} \times \oint \frac{\dd x_{2N-3}}{2 \pi i x_{2N-3}} \frac{1}{1-\varphi_1\varphi_2^{-1} x_{2N-3}^{-1}} \frac{1-\varphi_2 x_{2N-3}^2 w_{2N-3}^{-1}}{1-\varphi_1 x_{2N-3} w_{2N-3}^{-1}} \notag \\[5pt]
&\hspace{80pt} \times \left[\prod_{i=1}^{2N-4} \frac{1-x_i^{-1} x_{2N-3}}{1-\varphi_1 x_i^{-1} x_{2N-3}^{-1}} \frac{1-\varphi_2 x_i x_{2N-3} w_{2N-3}^{-1}}{1-\varphi_1\varphi_2 x_i^{-1} x_{2N-3} w_{2N-3}^{-1}}\right]\, .
\end{align}
The contributing poles are placed at $x_{2N-3} = \varphi_1 x_m^{-1}$, with $m=1,\cdots,2N-4$, and $x_{2N-3} = \varphi_1 \varphi_2^{-1}$, provided that $|\varphi_1| < |\varphi_2|$. Similarly to what has been done before, we deal separately with these two contributions for the sake of clarity. Hence, after integration, we obtain
\begin{equation}
\HS_2 \equiv \HS_{2,1} + \HS_{2,2}\, ,
\end{equation}
where
\begin{align}\label{eqn:HS21}
&\HS_{2,1} \equiv \oint \prod_{i=1}^{2N-4} \frac{\dd x_i}{2 \pi i x_i} \left[\prod_{1 \le i<j \le 2N-4} \frac{1-x_i^{-1} x_j}{1-\varphi_1 x_i^{-1} x_j^{-1}}\right] \left[\prod_{i=1}^{2N-4} \frac{1}{1-\varphi_1\varphi_2^{-1}x_i^{-1}}\right] \notag \\[5pt]
&\times \sum_{m=1}^{2N-4} \frac{1-\varphi_1 x_m^{-2}}{1-\varphi_2^{-1} x_m} \frac{1}{1-\varphi_1^2 x_m^{-1} w_{2N-3}^{-1}} \left[\prod_{1\le i\le 2N-4}^{i\ne m} \frac{1-\varphi_1 x_i^{-1} x_m^{-1}}{1-x_i^{-1}x_m} \frac{1-\varphi_1\varphi_2 x_i x_m^{-1} w_{2N-3}^{-1}}{1-\varphi_1^2 \varphi_2 x_i^{-1} x_m^{-1} w_{2N-3}^{-1}}\right] \, ,
\end{align}
and
\begin{align}\label{eqn:HS22}
\HS_{2,2} &\equiv \delta_{|\varphi_1| < |\varphi_2|} \oint \prod_{i=1}^{2N-4} \frac{\dd x_i}{2 \pi i x_i} \left[\prod_{1 \le i<j \le 2N-4} \frac{1-x_i^{-1} x_j}{1-\varphi_1 x_i^{-1} x_j^{-1}}\right] \frac{1}{1-\varphi_1\varphi_2 w_{2N-3}^{-1}} \notag \\[5pt]
&\hspace{80pt} \times \left[\prod_{i=1}^{2N-4} \frac{1}{1-\varphi_2 x_i^{-1}} \frac{1-\varphi_1 x_i w_{2N-3}^{-1}}{1-\varphi_1^2 x_i^{-1} w_{2N-3}^{-1}} \right]\, .
\end{align}
Now, we carry on with the integration over $x_{2N-4}$ in both $\HS_{2,1}$ and $\HS_{2,2}$. Starting with $\HS_{2,1}$ in \cref{eqn:HS21}, it is convenient to apply the variable change introduced in \cref{eqn:xmtox2N-3VariableChange}, which, in this case, induces 
\begin{equation}
\prod_{1\le i < j \le 2N-4} \left(1-x_i^{-1}x_j\right) \longrightarrow \left[\prod_{i=m}^{2N-5} \left(-x_{2N-4}^{-1} x_i\right)\right] \left[\prod_{1\le i < j \le 2N-4} \left(1-x_i^{-1}x_j\right)\right] \, .
\end{equation}
After variable change, one should notice that
\begin{equation}\label{eqn:splitfactorafterVariableChange}
\prod_{1\le i < j \le 2N-4} \frac{1-x_i^{-1}x_j}{1-\varphi_1 x_i^{-1}x_j^{-1}} = \left[\prod_{1\le i < j \le 2N-5} \frac{1-x_i^{-1}x_j}{1-\varphi_1 x_i^{-1}x_j^{-1}}\right] \left[\prod_{i=1}^{2N-5} \frac{1-x_i^{-1} x_{2N-4}}{1-\varphi_1 x_i^{-1} x_{2N-4}^{-1}}\right] \, .
\end{equation}
With all these ingredients, it is easy to check that the integrand in \cref{eqn:HS21} simplifies to
\begin{align}\label{eqn:HS21interstep1}
\HS_{2,1} &\equiv \oint \prod_{i=1}^{2N-4} \frac{\dd x_i}{2 \pi i x_i} \left[\prod_{1 \le i<j \le 2N-5} \frac{1-x_i^{-1} x_j}{1-\varphi_1 x_i^{-1} x_j^{-1}}\right] \notag \\[5pt]
&\hspace{40pt} \times \left[\prod_{i=1}^{2N-5} \frac{1}{1-\varphi_1\varphi_2^{-1} x_i^{-1}} \frac{1-\varphi_1\varphi_2 x_i w_{2N-4}^{-1}}{1-\varphi_1^2\varphi_2 x_i^{-1} w_{2N-4}^{-1}} \right] \frac{1}{1-\varphi_1^2 w_{2N-4}^{-1}}  \notag \\[5pt] 
&\hspace{80pt} \times \frac{1}{1-\varphi_1\varphi_2^{-1} x_{2N-4}^{-1}} \frac{1-\varphi_1 x_{2N-4}^{-2}}{1-\varphi_2^{-1} x_{2N-4}} \sum_{m=1}^{2N-4}\prod_{i=m}^{2N-5}\left(-x_{2N-4}^{-1} x_i\right)\, ,
\end{align}
where we have introduced $w_{2N-4} \equiv x_{2N-4} w_{2N-3}$. Note that all $x_{2N-4}$ dependence is contained in the last line of \cref{eqn:HS21interstep1}. The pole structure of the corresponding integrand can be simplified by changing variables as $x_{2N-4} \rightarrow x_{2N-4}^{-1}$, so that there exists a single contributing pole at $x_{2N-4} = \varphi_1^{-1}\varphi_2$ if $|\varphi_1| > |\varphi_2|$. This gives:
\begin{align}
&\oint \frac{\dd x_{2N-4}}{2 \pi i x_{2N-4}} \frac{1}{1-\varphi_1\varphi_2^{-1} x_{2N-4}^{-1}} \frac{1-\varphi_1 x_{2N-4}^{-2}}{1-\varphi_2^{-1} x_{2N-4}} \sum_{m=1}^{2N-4}\prod_{i=m}^{2N-5}\left(-x_{2N-4}^{-1} x_i\right) \notag \\[8pt]
&= \varphi_1^{-1} \varphi_2^2 \sum_{m=1}^{2N-4}\prod_{i=m}^{2N-5}\left(-\varphi^{-1}\varphi_2 x_i\right)\, ,
\end{align}
Plugging the previous result in \cref{eqn:HS21interstep1} yields
\begin{align}\label{eqn:HS21result}
\HS_{2,1} \equiv \delta_{|\varphi_1| > |\varphi_2|}\ \varphi_1^{-1}\varphi_2^2\ &\oint \prod_{i=1}^{2N-5} \frac{\dd x_i}{2 \pi i x_i} \left[\prod_{1 \le i<j \le 2N-5} \frac{1-x_i^{-1} x_j}{1-\varphi_1 x_i^{-1} x_j^{-1}}\right] \notag \\[5pt]
&\times \left[\prod_{i=1}^{2N-5} \frac{1}{1-\varphi_1\varphi_2^{-1} x_i^{-1}} \frac{1-\varphi_1\varphi_2 x_i w_{2N-4}^{-1}}{1-\varphi_1^2\varphi_2 x_i^{-1} w_{2N-4}^{-1}} \right] \frac{1}{1-\varphi_1^2 w_{2N-4}^{-1}}  \notag \\[5pt] 
&\times \left[\sum_{m=1}^{2N-4}\prod_{i=m}^{2N-5} \left(-\varphi_1^{-1}\varphi_2 x_i\right)\right]\, .
\end{align}
Let us now focus on $\HS_{2,2}$ in \cref{eqn:HS22}. Also in this case, we proceed with the integration over $x_{2N-4}$. Taking into account the definition $w_{2N-4} \equiv x_{2N-4} w_{2N-3}$, we can factor out the $x_{2N-4}$ integral as
\begin{equation}
\HS_{2,2} \equiv \delta_{|\varphi_1| < |\varphi_2|} \oint \prod_{i=1}^{2N-5} \frac{\dd x_i}{2 \pi i x_i} \left[\prod_{1 \le i<j \le 2N-5} \frac{1-x_i^{-1} x_j}{1-\varphi_1 x_i^{-1} x_j^{-1}} \prod_{i=1}^{2N-5} \frac{1}{1-\varphi_2 x_i^{-1}}\right] \frac{\widetilde{I}}{1-\varphi_1^2 w_{2N-4}^{-1}} \, ,
\end{equation}
with
\begin{align}
\widetilde{I} &\equiv \oint \frac{\dd x_{2N-4}}{2 \pi i x_{2N-4}} \left[\prod_{i=1}^{2N-5} \frac{1-x_i^{-1} x_{2N-4}}{1-\varphi_1 x_i^{-1} x_{2N-4}^{-1}}\right] \frac{1}{1-\varphi_2 x_{2N-4}^{-1}} \frac{1-\varphi_1 x_{2N-4}^2 w_{2N-4}^{-1}}{1-\varphi_1\varphi_2 x_{2N-4} w_{2N-4}^{-1}} \notag \\[5pt]
&\hspace{40pt} \times \left[\prod_{i=1}^{2N-5} \frac{1-\varphi_1 x_i x_{2N-4} w_{2N-4}^{-1}}{1-\varphi_1^2 x_i^{-1} x_{2N-4} w_{2N-4}^{-1}}\right] \, .
\end{align}
The contributing poles are placed at $x_{2N-4} = \varphi_1 x_m^{-1}$, with $m=1,\cdots,2N-5$, and $x_{2N-4} = \varphi_2$. Once again, we treat separately these two contributions, so that, after integration, we can write down
\begin{equation}
\HS_{2,2} \equiv \HS_{2,2,1} + \HS_{2,2,2}\, ,
\end{equation}
where
\begin{align}\label{eqn:HS221}
\HS_{2,2,1} &\equiv \delta_{|\varphi_1| < |\varphi_2|} \oint \prod_{i=1}^{2N-5} \frac{\dd x_i}{2 \pi i x_i} \left[\prod_{1 \le i<j \le 2N-5} \frac{1-x_i^{-1} x_j}{1-\varphi_1 x_i^{-1} x_j^{-1}}\right] \left[\prod_{i=1}^{2N-5} \frac{1}{1-\varphi_2 x_i^{-1}}\right] \notag \\[5pt]
&\hspace{60pt} \times \sum_{m=1}^{2N-5} \frac{1-\varphi_1 x_m^{-2}}{1-\varphi_1^{-1}\varphi_2 x_m} \frac{1}{1-\varphi_1^2\varphi_2 x_m^{-1} w_{2N-4}^{-1}} \notag \\[5pt]
&\hspace{80pt} \times \left[\prod_{1\le i\le 2N-5}^{i\ne m} \frac{1-\varphi_1 x_i^{-1} x_m^{-1}}{1-x_i^{-1} x_m} \frac{1-\varphi_1^2 x_i x_m^{-1} w_{2N-4}^{-1}}{1-\varphi_1^3 x_i^{-1} x_m^{-1} w_{2N-4}^{-1}}\right]\, ,
\end{align}
and
\begin{align}\label{eqn:HS222}
\HS_{2,2,2} &\equiv \delta_{|\varphi_1| < |\varphi_2|} \oint \prod_{i=1}^{2N-5} \frac{\dd x_i}{2 \pi i x_i} \left[\prod_{1 \le i<j \le 2N-5} \frac{1-x_i^{-1} x_j}{1-\varphi_1 x_i^{-1} x_j^{-1}}\right] \frac{1}{1-\varphi_1^2 w_{2N-4}^{-1}} \notag \\[5pt]
&\hspace{60pt} \times \left[\prod_{i=1}^{2N-5} \frac{1}{1-\varphi_1\varphi_2^{-1}x_i^{-1}} \frac{1-\varphi_1\varphi_2 x_i w_{2N-4}^{-1}}{1-\varphi_1^2\varphi_2 x_i^{-1} w_{2N-4}^{-1}}\right] \, .
\end{align}
For the $\HS_{2,2,1}$ piece given in \cref{eqn:HS221}, we compute the $x_{2N-5}$ integral. To this aim, we follow our previous strategy, that is, we first apply the change of variables presented in \cref{eqn:xmtox2N-3VariableChange}, which implies the changes in \cref{eqn:factorafterVariableChange}, and then separate factors as in \cref{eqn:splitfactorafterVariableChange} in order to simplify our integrand. This leads us to
\begin{align}
\HS_{2,2,1} \equiv \delta_{|\varphi_1| < |\varphi_2|} &\oint \prod_{i=1}^{2N-5} \frac{\dd x_i}{2 \pi i x_i} \left[\prod_{1 \le i<j \le 2N-6} \frac{1-x_i^{-1} x_j}{1-\varphi_1 x_i^{-1} x_j^{-1}}\right] \notag \\[5pt]
&\times \left[\prod_{i=1}^{2N-6} \frac{1}{1-\varphi_2 x_i^{-1}} \frac{1-\varphi_1^2 x_i w_{2N-5}^{-1}}{1-\varphi_1^3 x_i^{-1} w_{2N-5}^{-1}}\right] \frac{1}{1-\varphi_1^2\varphi_2 w_{2N-5}^{-1}} \notag \\[5pt]
&\times \frac{1}{1-\varphi_1^{-1}\varphi_2 x_{2N-5}} \frac{1-\varphi_1 x_{2N-5}^{-2}}{1-\varphi_2 x_{2N-5}^{-1}} \sum_{m=1}^{2N-5} \prod_{i=m}^{2N-6} \left(-x_{2N-5}^{-1} x_i\right) \,,
\end{align}
where we have introduced $w_{2N-5} = x_{2N-5} w_{2N-4}$. Notice that the last line in the expression above contains all $x_{2N-5}$ dependence. If we change variables as $x_{2N-5} \rightarrow x_{2N-5}^{-1}$, it is easy to check that there are no contributing poles, and thus the integral vanishes:
\begin{equation}
\HS_{2,2,1} = 0 \,.
\end{equation}

All in all, we have decomposed the Hilbert series $\HS_\I^{H_2,\Phi}$ in \cref{eqn:HSIIIH2ShortMolien-Weyl} into three different contributions:
\begin{equation}
\HS_\I^{H_2,\Phi} \equiv \HS_1 + \bigg[ \HS_{2,1} + \big( \underbrace{\HS_{2,2,1}}_0 + \HS_{2,2,2} \big) \bigg] = \HS_1 + \HS_{2,1} + \HS_{2,2,2}\, ,
\end{equation}
with $\HS_1$, $\HS_{2,1}$, and $\HS_{2,2,2}$, given in \cref{eqn:HS1result,eqn:HS21result,eqn:HS222}, respectively. Now using the three auxiliary functions $A_{k,s}(\varphi_1, \varphi_2)$, $B_{k,s}(\varphi_1, \varphi_2)$, $C_{k,s}(\varphi_1, \varphi_2)$ defined respectively in \cref{eqn:Aks,eqn:Bks,eqn:Cks}, and their general evaluation results (shown by induction later) in \cref{eqn:AksResult,eqn:BksResult,eqn:CksResult}, we obtain
\begin{subequations}
\begin{align}
\HS_1 &= C_{N-2,0} =\delta_{|\varphi_1| > |\varphi_2|}\frac{1}{1-\varphi_1^{N-1}\varphi_2} \,, \\[8pt]
\HS_{2,1} &= \varphi_1^{-1} \varphi_2^2\ B_{N-3,1} = 0 \,, \\[8pt]
\HS_{2,2,2} &= A_{N-3,1} = \delta_{|\varphi_1| < |\varphi_2|}\frac{1}{1-\varphi_1^{N-1}\varphi_2} \,.
\end{align}
\end{subequations}
They add up to give the result quoted in \cref{eqn:HSIIIH2}:
\begin{equation}
\HS_\I^{H_2,\Phi} = \frac{1}{1-\varphi_1^{N-1}\varphi_2} \,.
\end{equation}
In the following, we provide detailed steps regarding the computation of these auxiliary functions $A_{k,s}(\varphi_1, \varphi_2)$, $B_{k,s}(\varphi_1, \varphi_2)$, and $C_{k,s}(\varphi_1, \varphi_2)$.

\subsubsection*{Computing $A_{k,s}(\varphi_1, \varphi_2)$}

We define the auxiliary function
\begin{align}\label{eqn:Aks}
A_{k,s}(\varphi_1, \varphi_2) \equiv \delta_{|\varphi_1| < |\varphi_2|} &\oint \prod_{i=1}^{2k+1} \frac{\dd x_i}{2 \pi i x_i} \left[\prod_{1 \le i<j \le 2k+1} \frac{1-x_i^{-1} x_j}{1-\varphi_1 x_i^{-1} x_j^{-1}}\right] \frac{1}{1-\varphi_1^{s+1} w_{2k+2}^{-1}} \notag \\[5pt]
&\times \left[\prod_{i=1}^{2k+1} \frac{1}{1-\varphi_1\varphi_2^{-1}x_i^{-1}} \frac{1-\varphi_1^s \varphi_2 x_i w_{2k+2}^{-1}}{1-\varphi_1^{s+1}\varphi_2 x_i^{-1} w_{2k+2}^{-1}}\right] \, .
\end{align}
It is clear that \cref{eqn:HS222} is reproduced by setting $k=N-3$ and $s=1$ in \cref{eqn:Aks}. To evaluate this expression, we first focus on the integration over $x_{2k+1}$. Introducing $w_{2k+1} \equiv x_{2k+1} w_{2k+2}$, the $x_{2k+1}$ integral can be factor out as
\begin{align}
A_{k,s} \equiv \delta_{|\varphi_1| < |\varphi_2|} &\oint \prod_{i=1}^{2k} \frac{\dd x_i}{2 \pi i x_i} \left[\prod_{1 \le i<j \le 2k} \frac{1-x_i^{-1} x_j}{1-\varphi_1 x_i^{-1} x_j^{-1}} \prod_{i=1}^{2k} \frac{1}{1-\varphi_1\varphi_2^{-1}x_i^{-1}} \right] \frac{1}{1-\varphi_1^{s+1} \varphi_2 w_{2k+1}^{-1}} \notag \\[5pt]
\times &\oint \frac{\dd x_{2k+1}}{2 \pi i x_{2k+1}} \frac{1}{1-\varphi_1\varphi_2^{-1}x_{2k+1}^{-1}} \left[\prod_{i=1}^{2k} \frac{1-x_i^{-1} x_{2k+1}}{1-\varphi_1 x_i^{-1} x_{2k+1}^{-1}} \right] \notag \\[5pt]
&\hspace{30pt} \times \frac{1-\varphi_1^s \varphi_2 x_{2k+1}^2 w_{2k+1}^{-1}}{1-\varphi_1^{s+1} x_{2k+1} w_{2k+1}^{-1}} \left[\prod_{i=1}^{2k} \frac{1-\varphi_1^s \varphi_2 x_i x_{2k+1} w_{2k+1}^{-1}}{1-\varphi_1^{s+1} \varphi_2 x_i^{-1} x_{2k+1} w_{2k+1}^{-1}} \right] \, .
\end{align}
The contributing poles are placed at $x_{2k+1} = \varphi_1 x_m^{-1}$, with $m=1,\cdots,2k$, and $x_{2k+1} = \varphi_1\varphi_2^{-1}$, since we are in the scenario where the grading variables satisfy $|\varphi_1| < |\varphi_2|$. It is convenient to treat separately these two contributions in order to keep track of the computations to follow more easily. Then, we can write down
\begin{equation}
A_{k,s} \equiv A_{k,s}^{(1)} + A_{k,s}^{(2)} \,,
\end{equation}
where
\begin{align}\label{eqn:Aks1}
A_{k,s}^{(1)} \equiv \delta_{|\varphi_1| < |\varphi_2|} &\oint \prod_{i=1}^{2k} \frac{\dd x_i}{2 \pi i x_i} \left[ \prod_{1 \le i<j \le 2k} \frac{1-x_i^{-1} x_j}{1-\varphi_1 x_i^{-1} x_j^{-1}}\right] \left[\prod_{i=1}^{2k} \frac{1}{1-\varphi_1\varphi_2^{-1}x_i^{-1}}\right] \notag \\[5pt]
&\times \sum_{m=1}^{2k} \frac{1-\varphi_1 x_m^{-2}}{1-\varphi_2^{-1} x_m} \frac{1}{1-\varphi_1^{s+2} x_m^{-1} w_{2k+1}^{-1}} \notag \\[5pt] 
&\hspace{30pt} \times \left[\prod_{1 \le i \le 2k}^{i\ne m} \frac{1-\varphi_1 x_i^{-1} x_m^{-1}}{1-x_i^{-1} x_m} \frac{1-\varphi_1^{s+1}\varphi_2 x_i x_m^{-1} w_{2k+1}^{-1}}{1-\varphi_1^{s+2} \varphi_2 x_i^{-1} x_m^{-1} w_{2k+1}^{-1}}\right] \,,
\end{align}
and
\begin{align}\label{eqn:Aks2}
A_{k,s}^{(2)} \equiv \delta_{|\varphi_1| < |\varphi_2|} &\oint \prod_{i=1}^{2k} \frac{\dd x_i}{2 \pi i x_i} \left[\prod_{1 \le i<j \le 2k} \frac{1-x_i^{-1} x_j}{1-\varphi_1 x_i^{-1} x_j^{-1}} \right] \frac{1}{1-\varphi_1^{s+1} \varphi_2 w_{2k+1}^{-1}} \notag \\[5pt]
&\times \left[ \prod_{i=1}^{2k} \frac{1}{1-\varphi_2 x_i^{-1}} \frac{1-\varphi_1^{s+1} x_i w_{2k+1}^{-1}}{1-\varphi_1^{s+2} x_i^{-1} w_{2k+1}^{-1}} \right]\, .
\end{align}
Next, we proceed with the integration over $x_{2k}$ in both $A_{k,s}^{(1)}$ and $A_{k,s}^{(2)}$. On the one hand, the sum in $A_{k,s}^{(1)}$ can be reduced by applying the variable change explained in \cref{eqn:xmtox2N-3VariableChange}, which, similarly to \cref{eqn:factorafterVariableChange}, induces
\begin{equation}
\prod_{1\le i < j \le 2k} \left(1-x_i^{-1}x_j\right) \longrightarrow \left[\prod_{i=m}^{2k} \left(-x_{2k}^{-1} x_i\right)\right] \left[\prod_{1\le i < j \le 2k} \left(1-x_i^{-1}x_j\right)\right] \, .
\label{eqn:inducedchanges}
\end{equation}
The last product factor in the right-hand side of the previous line can be decomposed as
\begin{equation}
\left[\prod_{1\le i < j \le 2k} \left(1-x_i^{-1}x_j\right)\right] = \left[\prod_{1\le i < j \le 2k-1} \left(1-x_i^{-1}x_j\right)\right] \left[\prod_{i=1}^{2k-1} \left(1-x_i^{-1} x_{2k} \right)\right]\, .
\end{equation}
Further introducing $w_{2k} \equiv x_{2k} w_{2k+1}$, the integral $A_{k,s}^{(1)}$ in \cref{eqn:Aks1} becomes
\begin{align}\label{eqn:Aks1interstep1}
A_{k,s}^{(1)} \equiv \delta_{|\varphi_1| < |\varphi_2|} &\oint \prod_{i=1}^{2k} \frac{\dd x_i}{2 \pi i x_i} \left[ \prod_{1 \le i<j \le 2k-1} \frac{1-x_i^{-1} x_j}{1-\varphi_1 x_i^{-1} x_j^{-1}}\right] \left[\prod_{i=1}^{2k-1} \frac{1}{1-\varphi_1\varphi_2^{-1}x_i^{-1}} \right] \notag \\[5pt]
&\times \left[\prod_{i=1}^{2k-1} \frac{1-\varphi_1^{s+1}\varphi_2 x_i w_{2k}^{-1}}{1-\varphi_1^{s+2}\varphi_2 x_i^{-1} w_{2k}^{-1}}\right] \frac{1}{1-\varphi_1^{s+2} w_{2k}^{-1}} \notag \\[5pt]
&\times \frac{1}{1-\varphi_1\varphi_2^{-1}x_{2k}^{-1}} \frac{1-\varphi_1 x_{2k}^{-2}}{1-\varphi_2^{-1} x_{2k}} \sum_{m=1}^{2k} \left[ \prod_{i=m}^{2k-1} \left(-x_{2k}^{-1} x_i \right) \right]\, .
\end{align}
Written in this form, all $x_{2k}$ dependence of the integrand is contained in the last line of \cref{eqn:Aks1interstep1}. Changing variables as $x_{2k} \rightarrow x_{2k}^{-1}$, it is easy to check that there are no contributing poles to the $x_{2k}$ integral (recall that $|\varphi_1| < |\varphi_2|$), and thus $A_{k,s}^{(1)}$ vanishes:
\begin{equation}
A_{k,s}^{(1)} = 0\, .
\end{equation}

On the other hand, we can factor out the $x_{2k}$ integral in $A_{k,s}^{(2)}$ as
\begin{align}
A_{k,s}^{(2)} \equiv \delta_{|\varphi_1| < |\varphi_2|} &\oint \prod_{i=1}^{2k-1} \frac{\dd x_i}{2 \pi i x_i} \left[\prod_{1 \le i<j \le 2k-1} \frac{1-x_i^{-1} x_j}{1-\varphi_1 x_i^{-1} x_j^{-1}} \prod_{i=1}^{2k-1} \frac{1}{1-\varphi_2 x_i^{-1}} \right] \frac{1}{1-\varphi_1^{s+2} w_{2k}^{-1}} \notag \\[5pt]
\times &\oint \frac{\dd x_{2k}}{2 \pi i x_{2k}}  \frac{1}{1-\varphi_2 x_{2k}^{-1}} \frac{1-\varphi_1^{s+1} x_{2k}^2 w_{2k}^{-1}}{1-\varphi_1^{s+1} \varphi_2 x_{2k} w_{2k}^{-1}} \notag \\[5pt]
&\hspace{30pt} \times \left[ \prod_{i=1}^{2k-1} \frac{1-x_i^{-1} x_{2k}}{1-\varphi_1 x_i^{-1} x_{2k}^{-1}} \frac{1-\varphi_1^{s+1} x_i x_{2k} w_{2k}^{-1}}{1-\varphi_1^{s+2} x_i^{-1} x_{2k} w_{2k}^{-1}} \right]\, ,
\end{align}
using again $w_{2k} = x_{2k} w_{2k+1}$. In this case, the contributing poles are at $x_{2k} = \varphi_1 x_m^{-1}$, with $m=1,\cdots,2k-1$, and $x_{2k} = \varphi_2$. Similarly to what has been done before, we can separate these two contributions as
\begin{equation}
A_{k,s}^{(2)} \equiv A_{k,s}^{(2,1)} + A_{k,s}^{(2,2)} \,,
\end{equation}
where
\begin{align}\label{eqn:Aks21}
A_{k,s}^{(2,1)} \equiv \delta_{|\varphi_1| < |\varphi_2|} &\oint \prod_{i=1}^{2k-1} \frac{\dd x_i}{2 \pi i x_i} \left[\prod_{1 \le i<j \le 2k-1} \frac{1-x_i^{-1} x_j}{1-\varphi_1 x_i^{-1} x_j^{-1}}\right] \left[\prod_{i=1}^{2k-1} \frac{1}{1-\varphi_2 x_i^{-1}}\right] \notag \\[5pt]
&\times \sum_{m=1}^{2k-1} \frac{1-\varphi_1 x_m^{-2}}{1-\varphi_1^{-1}\varphi_2 x_m} \frac{1}{1-\varphi_1^{s+2}\varphi_2 x_m^{-1} w_{2k}^{-1}} \notag \\[5pt]
&\hspace{30pt} \times \left[ \prod_{1\le i\le 2k-1}^{i\ne m} \frac{1-\varphi_1 x_i^{-1} x_m^{-1}}{1-x_i^{-1}x_m} \frac{1-\varphi_1^{s+2}x_i x_m^{-1} w_{2k}^{-1}}{1-\varphi_1^{s+3} x_i^{-1} x_m^{-1} w_{2k}^{-1}}\right] \,,
\end{align}
and
\begin{align}\label{eqn:Aks22}
A_{k,s}^{(2,2)} \equiv \delta_{|\varphi_1| < |\varphi_2|} &\oint \prod_{i=1}^{2k-1} \frac{\dd x_i}{2 \pi i x_i} \left[\prod_{1 \le i<j \le 2k-1} \frac{1-x_i^{-1} x_j}{1-\varphi_1 x_i^{-1} x_j^{-1}} \right] \frac{1}{1-\varphi_1^{s+2} w_{2k}^{-1}} \notag \\[5pt]
&\times \left[ \prod_{i=1}^{2k-1} \frac{1}{1-\varphi_1\varphi_2^{-1} x_i^{-1}} \frac{1-\varphi_1^{s+1}\varphi_2 x_i w_{2k}^{-1}}{1-\varphi_1^{s+2}\varphi_2 x_i^{-1} w_{2k}^{-1}} \right] \,.
\end{align}
We can then focus on the integration over $x_{2k-1}$ in $A_{k,s}^{(2,1)}$. Following our previous strategy, we change variables as explained in \cref{eqn:xmtox2N-3VariableChange}. This implies
\begin{equation}
\prod_{1\le i < j \le 2k-1} \left(1-x_i^{-1}x_j\right) \longrightarrow \left[\prod_{i=m}^{2k-2} \left(-x_{2k}^{-1} x_i\right)\right] \left[\prod_{1\le i < j \le 2k-1} \left(1-x_i^{-1}x_j\right)\right] \,.
\label{eqn:inducedchange2}
\end{equation}
After the change of variables, we can conveniently decompose the factor
\begin{equation}
\prod_{1\le i < j \le 2k-1} \frac{1-x_i^{-1}x_j}{1-\varphi_1 x_i^{-1}x_j^{-1}} = \left[\prod_{1\le i < j \le 2k-2} \frac{1-x_i^{-1}x_j}{1-\varphi_1 x_i^{-1}x_j^{-1}}\right] \left[\prod_{i=1}^{2k-2} \frac{1-x_i^{-1} x_{2k-1}}{1-\varphi_1 x_i^{-1} x_{2k-1}^{-1}}\right] \, .
\label{eqn:splitfactors2}
\end{equation}
Taking this into account, and introducing $w_{2k-1} = x_{2k-1} w_{2k-2}$, $A_{k,s}^{(2,1)}$ in \cref{eqn:Aks21} is reduced to
\begin{align}\label{eqn:Aks21interstep1}
A_{k,s}^{(2,1)} \equiv \delta_{|\varphi_1| < |\varphi_2|} &\oint \prod_{i=1}^{2k-1} \frac{\dd x_i}{2 \pi i x_i} \left[\prod_{1 \le i<j \le 2k-2} \frac{1-x_i^{-1} x_j}{1-\varphi_1 x_i^{-1} x_j^{-1}}\right] \left[\prod_{i=1}^{2k-2} \frac{1}{1-\varphi_2 x_i^{-1}}\right] \notag \\[5pt]
&\times \left[\prod_{i=1}^{2k-2} \frac{1-\varphi_1^{s+2} x_i w_{2k-1}^{-1}}{1-\varphi_1^{s+3} x_i^{-1} w_{2k-1}^{-1}} \right] \frac{1}{1-\varphi_1^{s+2} \varphi_2 w_{2k-1}^{-1}} \notag \\[5pt]
&\times \frac{1}{1-\varphi_2 x_{2k-1}^{-1}} \frac{1-\varphi_1 x_{2k-1}^{-2}}{1-\varphi_1^{-1}\varphi_2 x_{2k-1}} \sum_{m=1}^{2k-1}\left[\prod_{i=m}^{2k-2} \left(-x_{2k-1}^{-1} x_i \right) \right]\, .
\end{align}
All the dependence on $x_{2k-1}$ is transferred to the last line of \cref{eqn:Aks21interstep1}. Once we apply the variable change $x_{2k-1} \rightarrow x_{2k-1}^{-1}$, it is straightforward to check that there are no contributing poles to this integral. Hence:
\begin{equation}
A_{k,s}^{(2,1)} = 0 \,.
\end{equation}
Therefore, we are only left with $A_{k,s}^{(2,2)}$. Importantly, one may realize it precisely corresponds to:
\begin{equation}
A_{k,s}^{(2,2)} = A_{k-1,s+1}\, .
\end{equation}
Then, we end up with:
\begin{equation}
A_{k,s} \equiv A_{k,s}^{(1)} + A_{k,s}^{(2)} = \underbrace{A_{k,s}^{(1)}}_0 + \bigg[\underbrace{A_{k,s}^{(2,1)}}_0 + A_{k,s}^{(2,2)} \bigg] = A_{k,s}^{(2,2)} = A_{k-1,s+1} \,.
\label{eqn:Aksrecursion}
\end{equation}
In light of \cref{eqn:Aksrecursion}, it is then clear that $A_{k,s} = A_{0,s+k}$. The latter, in turn, can be computed from $A_{0,s}$ by simply substituting $s \rightarrow s+k$, that is:
\begin{align}
A_{0,s} &\equiv \delta_{|\varphi_1| < |\varphi_2|} \oint \frac{\dd x_1}{2 \pi i x_1} \frac{1}{1-\varphi_1\varphi_2^{-1}x_1^{-1}} \frac{1-\varphi_1^s\varphi_2 x_1 w_2^{-1}}{1-\varphi_1^{s+1}\varphi_2 x_1^{-1} w_2^{-1}} \frac{1}{1-\varphi_1^{s+1} w_2^{-1}} \notag \\[8pt]
&= \delta_{|\varphi_1| < |\varphi_2|} \frac{1}{1-\varphi_1^{s+1}\varphi_2} \oint \frac{\dd x_1}{2 \pi i x_1} \frac{1}{1-\varphi_1\varphi_2^{-1}x_1^{-1}} \frac{1-\varphi_1^s\varphi_2 x_1^2}{1-\varphi_1^{s+1}x_1} \notag \\[8pt]
&= \delta_{|\varphi_1| < |\varphi_2|} \frac{1}{1-\varphi_1^{s+1}\varphi_2} \,,
\end{align}
where we have used $w_2 = x_1^{-1}$. The integration over $x_1$ can be done straightforwardly since there is a single contributing pole at $x_1 = \varphi_1 \varphi_2^{-1}$. Substituting $s \rightarrow s+k$, we obtain
\begin{equation}
A_{k,s} \equiv \delta_{|\varphi_1| < |\varphi_2|} \frac{1}{1-\varphi_1^{s+k+1}\varphi_2} \,.
\label{eqn:AksResult}
\end{equation}

\subsubsection*{Computing $B_{k,s}(\varphi_1, \varphi_2)$}

We define the auxiliary function
\begin{align}\label{eqn:Bks}
B_{k,s}(\varphi_1, \varphi_2) &\equiv \delta_{|\varphi_1| > |\varphi_2|} \oint \prod_{i=1}^{2k+1} \frac{\dd x_i}{2 \pi i x_i} \left[\prod_{1 \le i<j \le 2k+1} \frac{1-x_i^{-1} x_j}{1-\varphi_1 x_i^{-1} x_j^{-1}}\right] \notag \\[5pt]
&\hspace{60pt} \times \left[\prod_{i=1}^{2k+1} \frac{1}{1-\varphi_1\varphi_2^{-1} x_i^{-1}} \frac{1-\varphi_1^s\varphi_2 x_i w_{2k+2}^{-1}}{1-\varphi_1^{s+1}\varphi_2 x_i^{-1} w_{2k+2}^{-1}} \right] \frac{1}{1-\varphi_1^{s+1} w_{2k+2}^{-1}}  \notag \\[5pt] 
&\hspace{60pt} \times \left[\sum_{n=1}^{2k+2}\prod_{i=n}^{2k+1} \left(-\varphi_1^{-1}\varphi_2 x_i\right)\right]\, ,
\end{align}
in such a way that \cref{eqn:HS21result} corresponds to $k=N-3$ and $s=1$. As in previous cases, our first step consists of integrating the variable $x_{2k+1}$. Introducing $w_{2k+1} \equiv x_{2k+1} w_{2k+2}$, we can factor out the $x_{2k+1}$ integral, that we call $J_{2k+1}$, as
\begin{align}
B_{k,s} \equiv \delta_{|\varphi_1| > |\varphi_2|} &\oint \prod_{i=1}^{2k} \frac{\dd x_i}{2 \pi i x_i} \left[\prod_{1 \le i<j \le 2k} \frac{1-x_i^{-1} x_j}{1-\varphi_1 x_i^{-1} x_j^{-1}} \right] \left[ \prod_{i=1}^{2k} \frac{1}{1-\varphi_1\varphi_2^{-1} x_i^{-1}} \right] \notag \\[5pt]
&\times \frac{1}{1-\varphi_1^{s+1}\varphi_2 w_{2k+1}^{-1}} J_{2k+1}\, ,
\end{align}
with 
\begin{align}\label{eqn:J2k+1}
J_{2k+1} \equiv &\oint \frac{\dd x_{2k+1}}{2 \pi i x_{2k+1}} \frac{1}{1-\varphi_1\varphi_2^{-1} x_{2k+1}^{-1}} \left[ \prod_{i=1}^{2k} \frac{1-x_i^{-1}x_{2k+1}}{1-\varphi_1 x_i^{-1} x_{2k+1}^{-1}} \right] \notag \\[5pt]
&\times \frac{1-\varphi_1^s\varphi_2 x_{2k+1}^2 w_{2k+1}^{-1}}{1-\varphi_1^{s+1}x_{2k+1} w_{2k+1}^{-1}} \left[ \prod_{i=1}^{2k} \frac{1-\varphi_1^s \varphi_2 x_i x_{2k+1} w_{2k+1}^{-1}}{1-\varphi_1^{s+1} \varphi_2 x_i^{-1} x_{2k+1} w_{2k+1}^{-1}} \right] \notag \\[5pt]
&\times \left[1-\varphi_1^{-1}\varphi_2 x_{2k+1} \sum_{n=1}^{2k+1} \prod_{i=n}^{2k} \left(-\varphi_1^{-1}\varphi_2 x_i\right)\right]\, .
\end{align}
The contributing poles to the integral $J_{2k+1}$ in \cref{eqn:J2k+1} are at $x_{2k+1} = \varphi_1 x_m^{-1}$, with $m=1,\cdots,2k$. After integration, we obtain
\begin{align}\label{eqn:J2k+1Result}
J_{2k+1} &\equiv \sum_{m=1}^{2k} \frac{1-\varphi_1 x_m^{-2}}{1-\varphi_2^{-1} x_m} \frac{1-\varphi_1^{s+1}\varphi_2 w_{2k+1}^{-1}}{1-\varphi_1^{s+2}x_m^{-1}w_{2k+1}^{-1}} \left[1-\varphi_2 x_m^{-1} \sum_{n=1}^{2k+1} \prod_{i=n}^{2k} \left(-\varphi_1^{-1}\varphi_2 x_i \right)\right] \notag \\[5pt]
&\hspace{30pt} \times \left[\prod_{1\le i\le 2k}^{i\ne m} \frac{1-\varphi_1 x_i^{-1} x_m^{-1}}{1-x_i^{-1} x_m} \frac{1-\varphi_1^{s+1} \varphi_2 x_i x_m^{-1} w_{2k+1}^{-1}}{1-\varphi_1^{s+2}\varphi_2 x_i^{-1} x_m^{-1} w_{2k+1}^{-1}}\right] \, .
\end{align}
In order to simplify the sum over $m$ in \cref{eqn:J2k+1}, we apply the transformation introduced in \cref{eqn:xmtox2N-3VariableChange}. The latter induces the same changes quoted in \cref{eqn:inducedchanges}, together with 
\begin{equation}\label{eqn:factorafterVariableChange2}
x_m^{-1} \sum_{n=1}^{2k+1} \prod_{i=n}^{2k} \left(-\varphi_1^{-1}\varphi_2 x_i \right) \rightarrow -\varphi_1^{-1}\varphi_2 \sum_{n=1}^m \prod_{i=n}^{2k-1} \left(-\varphi_1^{-1}\varphi_2 x_i\right) + x_{2k}^{-1} \sum_{n=m}^{2k} \prod_{i=n}^{2k-1} \left(-\varphi_1^{-1}\varphi_2 x_i\right)\, .
\end{equation}
The cancellation of repeated factors can be made more transparent once we separate
\begin{equation}
\prod_{1\le i < j \le 2k} \frac{1-x_i^{-1}x_j}{1-\varphi_1 x_i^{-1}x_j^{-1}} = \left[\prod_{1\le i < j \le 2k-1} \frac{1-x_i^{-1}x_j}{1-\varphi_1 x_i^{-1}x_j^{-1}}\right] \left[\prod_{i=1}^{2k-1} \frac{1-x_i^{-1} x_{2k}}{1-\varphi_1 x_i^{-1} x_{2k}^{-1}}\right] \, .
\end{equation}
Further introducing $w_{2k} \equiv x_{2k} w_{2k-1}$, and after variable change, $B_{k,s}$ reduces to
\begin{align}
B_{k,s} \equiv \delta_{|\varphi_1| > |\varphi_2|} &\oint \prod_{i=1}^{2k-1} \frac{\dd x_i}{2 \pi i x_i} \left[ \prod_{1 \le i<j \le 2k-1} \frac{1-x_i^{-1} x_j}{1-\varphi_1 x_i^{-1} x_j^{-1}} \right] \notag \\[5pt]
&\times \left[\prod_{i=1}^{2k-1} \frac{1}{1-\varphi_1\varphi_2^{-1} x_i^{-1}} \frac{1-\varphi_1^{s+1}\varphi_2 x_i w_{2k}^{-1}}{1-\varphi_1^{s+2}\varphi_2 x_i^{-1} w_{2k}^{-1}}\right] \frac{1}{1-\varphi_1^{s+2} w_{2k}^{-1}} J_{2k}\, ,
\end{align}
where we have factored out $J_{2k}$
\begin{align}\label{eqn:J2k}
J_{2k} \equiv &\oint \frac{\dd x_{2k}}{2 \pi i x_{2k}} \frac{1}{1-\varphi_1\varphi_2^{-1}x_{2k}^{-1}} \frac{1-\varphi_1 x_{2k}^{-2}}{1-\varphi_2^{-1} x_{2k}} \sum_{m=1}^{2k} \left[ \prod_{i=m}^{2k-1} \left(-x_{2k}^{-1} x_i \right) \right] \notag \\[5pt]
&\times \left[1 + \varphi_1^{-1}\varphi_2^2 \sum_{n=1}^{m}\prod_{i=n}^{2k-1} \left(-\varphi_1^{-1}\varphi_2 x_i\right) - \varphi_2 x_{2k}^{-1} \sum_{n=m}^{2k} \prod_{i=n}^{2k-1} \left(-\varphi_1^{-1}\varphi_2 x_i \right) \right]\, ,
\end{align}
that is, the integral over $x_{2k}$. In order to simplify the pole structure of the integrand in \cref{eqn:J2k}, we perform the variable change $x_{2k} \rightarrow x_{2k}^{-1}$, such that $J_{2k}$ becomes
\begin{align}
J_{2k} \equiv &\oint \frac{\dd x_{2k}}{2 \pi i x_{2k}} \frac{1}{1-\varphi_1\varphi_2^{-1}x_{2k}} \frac{1-\varphi_1 x_{2k}^2}{1-\varphi_2^{-1} x_{2k}^{-1}} \sum_{m=1}^{2k} \left[ \prod_{i=m}^{2k-1} \left(-x_{2k} x_i \right) \right] \notag \\[5pt]
&\times \left[1 + \varphi_1^{-1}\varphi_2^2 \sum_{n=1}^{m}\prod_{i=n}^{2k-1} \left(-\varphi_1^{-1}\varphi_2 x_i\right) - \varphi_2 x_{2k} \sum_{n=m}^{2k} \prod_{i=n}^{2k-1} \left(-\varphi_1^{-1}\varphi_2 x_i \right) \right]\, .
\end{align}
In this way, there is a single contributing pole at $x_{2k} = \varphi_1^{-1}\varphi_2$, since we have previously assumed that $|\varphi_1| > |\varphi_2|$. Integration over $x_{2k}$ then yields
\begin{align}\label{eqn:J2kResult}
J_{2k} &= \varphi_1^{-1}\varphi_2^2 \sum_{m=1}^{2k} \left[\prod_{i=m}^{2k-1} \left(-\varphi_1^{-1}\varphi_2 x_i\right) \right] \notag \\[5pt] 
&\times \left[1 + \varphi_1^{-1}\varphi_2^2 \sum_{n=1}^{m} \prod_{i=n}^{2k-1} \left(-\varphi_1^{-1}\varphi_2 x_i\right) - \varphi_1^{-1}\varphi_2^2 \sum_{n=m}^{2k} \prod_{i=n}^{2k-1} \left(-\varphi_1^{-1}\varphi_2 x_i\right) \right]\, .
\end{align}
The result in \cref{eqn:J2kResult} can be further simplified as follows. Let us define
\begin{equation}
a_n \equiv \prod_{i=n}^{2k-1} \left(-\varphi_1^{-1} \varphi_2 x_i \right)\, .
\label{eqn:andefinition}
\end{equation}
The last two terms in the second line of \cref{eqn:J2kResult} cancel out once we take into account that
\begin{equation}
\sum_{m=1}^{2k} a_m \left( \sum_{n=1}^{m} a_n - \sum_{n=m}^{2k} a_n \right) = \sum_{1\le n\le m\le 2k} a_m a_n - \sum_{1\le m\le n\le 2k} a_m a_n = 0\, .
\label{eqn:ansimplify}
\end{equation}
As a result, $B_{k,s}$ is given by
\begin{align}
B_{k,s} \equiv \varphi_1^{-1}\varphi_2^2\ \delta_{|\varphi_1| > |\varphi_2|} &\oint \prod_{i=1}^{2k-1} \frac{\dd x_i}{2 \pi i x_i} \left[ \prod_{1 \le i<j \le 2k-1} \frac{1-x_i^{-1} x_j}{1-\varphi_1 x_i^{-1} x_j^{-1}} \right] \notag \\[5pt]
&\times \left[\prod_{i=1}^{2k-1} \frac{1}{1-\varphi_1\varphi_2^{-1} x_i^{-1}} \frac{1-\varphi_1^{s+1}\varphi_2 x_i w_{2k}^{-1}}{1-\varphi_1^{s+2}\varphi_2 x_i^{-1} w_{2k}^{-1}}\right] \frac{1}{1-\varphi_1^{s+2} w_{2k}^{-1}} \notag \\[5pt]
&\times \left[\sum_{m=1}^{2k}\prod_{i=m}^{2k-1} \left(-\varphi_1^{-1}\varphi_2 x_i \right) \right]\, .
\end{align}
In this form, it is easy to see that
\begin{equation}
B_{k,s} \equiv \varphi_1^{-1}\varphi_2^2\ B_{k-1,s+1}\, .
\end{equation}
The previous relation implies that $B_{k,s} = \left(\varphi_1^{-1}\varphi_2^2\right)^k B_{0,s+k}$. In turn, $B_{0,s+k}$ can be computed from $B_{0,s}$ by replacing $s \rightarrow s+k$. Then, we have:
\begin{align}
B_{0,s} &\equiv \delta_{|\varphi_1| > |\varphi_2|} \oint \frac{\dd x_1}{2 \pi i x_1} \frac{1}{1-\varphi_1\varphi_2^{-1} x_1^{-1}} \frac{1-\varphi_1^s \varphi_2 x_1^2}{1-\varphi_1^{s+1}\varphi_2} \frac{1}{1-\varphi_1^{s+1}x_1}\left(1-\varphi_1^{-1}\varphi_2 x_1\right) \notag \\[8pt]
&= \delta_{|\varphi_1| > |\varphi_2|} \frac{1}{1-\varphi_1^{s+1}\varphi_2} \oint \frac{\dd x_1}{2 \pi i x_1} \frac{1-\varphi_1^s\varphi_2 x_1^2}{1-\varphi_1\varphi_2^{-1} y_1^{-1}} \frac{1-\varphi_1^{-1}\varphi_2 x_1}{1-\varphi_1^{s+1} x_1} = 0\, ,
\end{align}
as the integrand has no contributing poles. Therefore, it is clear that
\begin{equation}
B_{k,s} = 0 \,.
\label{eqn:BksResult}
\end{equation}

\subsubsection*{Computing $C_{k,s}(\varphi_1, \varphi_2)$}

We define the auxiliary function
\begin{align}\label{eqn:Cks}
C_{k,s}(\varphi_1, \varphi_2) \equiv \delta_{|\varphi_1| > |\varphi_2|} &\oint \prod_{i=1}^{2k} \frac{\dd x_i}{2 \pi i x_i} \left[ \prod_{1 \le i<j \le 2k} \frac{1-x_i^{-1} x_j}{1-\varphi_1 x_i^{-1} x_j^{-1}} \right] \notag \\[5pt]
&\times \left[\prod_{i=1}^{2k} \frac{1}{1-\varphi_2 x_i^{-1}} \frac{1-\varphi_1^{s+1}x_i w_{2k+1}^{-1}}{1-\varphi_1^{s+2}x_i^{-1} w_{2k+1}^{-1}}\right] \frac{1}{1-\varphi_1^{s+1}\varphi_2 w_{2k+1}^{-1}} \notag \\[5pt]
&\times \left[\sum_{n=1}^{2k+1} \prod_{i=n}^{2k} \left(-\varphi_1^{-1}\varphi_2 x_i \right)\right] \,.
\end{align}
Then \cref{eqn:HS1result} is reproduced by setting $k=N-2$ and $s=0$. To evaluate the expression above, we first focus on the integration over $x_{2k}$. Introducing $w_{2k} \equiv x_{2k} w_{2k+1}$, it is easy to factor out the $x_{2k}$ integral, that we call $K_{2k}$, as follows:
\begin{align}
C_{k,s} \equiv \delta_{|\varphi_1| > |\varphi_2|} &\oint \prod_{i=1}^{2k-1} \frac{\dd x_i}{2 \pi i x_i} \left[ \prod_{1 \le i<j \le 2k-1} \frac{1-x_i^{-1} x_j}{1-\varphi_1 x_i^{-1} x_j^{-1}}\right] \left[\prod_{i=1}^{2k-1} \frac{1}{1-\varphi_2 x_i^{-1}}\right] \notag \\[5pt]
&\times \frac{1}{1-\varphi_1^{s+2} w_{2k}^{-1}} K_{2k}\, ,
\end{align}
where
\begin{align}
K_{2k} \equiv &\oint \frac{\dd x_{2k}}{2 \pi i x_{2k}} \frac{1}{1-\varphi_2 x_{2k}^{-1}} \left[\prod_{i=1}^{2k-1} \frac{1-x_i^{-1} x_{2k}}{1-\varphi_1 x_i^{-1} x_{2k}^{-1}}\right] \notag \\[5pt]
&\times  \frac{1-\varphi_1^{s+1}x_{2k}^2 w_{2k}^{-1}}{1-\varphi_1^{s+1}\varphi_2 x_{2k} w_{2k}^{-1}} \left[\prod_{i=1}^{2k-1} \frac{1-\varphi_1^{s+1}x_i x_{2k} w_{2k}^{-1}}{1-\varphi_1^{s+2} x_i^{-1} x_{2k} w_{2k}^{-1}} \right] \notag \\[5pt]
&\times \left[1-\varphi_1^{-1}\varphi_2 x_{2k} \sum_{n=1}^{2k}\prod_{i=n}^{2k-1} \left(-\varphi_1^{-1}\varphi_2 x_i \right) \right] \,.
\end{align}
The contributing poles in $K_{2k}$ are placed at $x_{2k} = \varphi_1 x_m^{-1}$, with $m=1,\cdots,2k-1$, and $x_{2k} = \varphi_2$. For the sake of clarity, we deal separately with these two contributions, giving rise to
\begin{equation}
C_{k,s} \equiv C_{k,s}^{(1)} + C_{k,s}^{(2)}\, ,
\end{equation}
being
\begin{align}\label{eqn:Cks1}
C_{k,s}^{(1)} \equiv \delta_{|\varphi_1| > |\varphi_2|} &\oint \prod_{i=1}^{2k-1} \frac{\dd x_i}{2 \pi i x_i} \left[ \prod_{1 \le i<j \le 2k-1} \frac{1-x_i^{-1} x_j}{1-\varphi_1 x_i^{-1} x_j^{-1}}\right] \left[\prod_{i=1}^{2k-1} \frac{1}{1-\varphi_2 x_i^{-1}}\right] \notag \\[5pt]
&\times \sum_{m=1}^{2k-1} \frac{1-\varphi_1 x_m^{-2}}{1-\varphi_1^{-1}\varphi_2 x_m} \frac{1}{1-\varphi_1^{s+2}\varphi_2 x_m^{-1} w_{2k}^{-1}} \notag \\[5pt]
&\hspace{30pt} \times \left[\prod_{1\le i\le 2k-1}^{i\ne m} \frac{1-\varphi_1 x_i^{-1} x_m^{-1}}{1-x_i^{-1} x_m} \frac{1-\varphi_1^{s+2} x_i x_m^{-1} w_{2k}^{-1}}{1-\varphi_1^{s+3} x_i^{-1} x_m^{-1} w_{2k}^{-1}}\right] \notag \\[5pt] 
&\hspace{30pt} \times \left[1-\varphi_2 x_m^{-1} \sum_{n=1}^{2k} \prod_{i=n}^{2k-1} \left(-\varphi_1^{-1}\varphi_2 x_i \right)\right] \,,
\end{align}
and
\begin{align}\label{eqn:Cks2}
C_{k,s}^{(2)} \equiv \delta_{|\varphi_1| > |\varphi_2|} &\oint \prod_{i=1}^{2k-1} \frac{\dd x_i}{2 \pi i x_i} \left[ \prod_{1 \le i<j \le 2k-1} \frac{1-x_i^{-1} x_j}{1-\varphi_1 x_i^{-1} x_j^{-1}} \right] \frac{1}{1-\varphi_1^{s+2} w_{2k}^{-1}} \notag \\[5pt]
&\times \left[\prod_{i=1}^{2k-1} \frac{1}{1-\varphi_1\varphi_2^{-1}x_i^{-1}} \frac{1-\varphi_1^{s+1}\varphi_2 x_i w_{2k}^{-1}}{1-\varphi_1^{s+2}\varphi_2 x_i^{-1} w_{2k}^{-1}} \right] \notag \\[5pt]
&\times \left[ 1-\varphi_1^{-1}\varphi_2^2 \sum_{n=1}^{2k} \prod_{i=n}^{2k-1} \left(-\varphi_1^{-1}\varphi_2 x_i\right) \right]\, .
\end{align}
Our next step consists of integrating $x_{2k-1}$ in both $C_{k,s}^{(1)}$ and $C_{k,s}^{(2)}$. Regarding $C_{k,s}^{(1)}$ in \cref{eqn:Cks1}, it is convenient to apply the variable change proposed in \cref{eqn:xmtox2N-3VariableChange}. This induces exactly the same change quoted in \cref{eqn:inducedchange2}, as well as
\begin{equation}
x_m^{-1} \sum_{n=1}^{2k} \prod_{i=n}^{2k-1} \left(-\varphi_1^{-1}\varphi_2 x_i \right) \rightarrow -\varphi_1^{-1}\varphi_2 \sum_{n=1}^m \prod_{i=n}^{2k-2} \left(-\varphi_1^{-1}\varphi_2 x_i\right) + x_{2k}^{-1} \sum_{n=m}^{2k-1} \prod_{i=n}^{2k-2} \left(-\varphi_1^{-1}\varphi_2 x_i\right)\, .
\end{equation}
Using that $w_{2k-1} = x_{2k-1} w_{2k}$, and the factor decomposition in \cref{eqn:splitfactors2}, we can write down $C_{k,s}^{(1)}$ as
\begin{align}\label{eqn:Cks1interstep1}
C_{k,s}^{(1)} \equiv \delta_{|\varphi_1| > |\varphi_2|} &\oint \prod_{i=1}^{2k-2} \frac{\dd x_i}{2 \pi i x_i} \left[\prod_{1 \le i<j \le 2k-2} \frac{1-x_i^{-1} x_j}{1-\varphi_1 x_i^{-1} x_j^{-1}}\right] \notag \\[5pt]
&\times \left[\prod_{i=1}^{2k-2} \frac{1}{1-\varphi_2 x_i^{-1}} \frac{1-\varphi_1^{s+2}x_i w_{2k-1}^{-1}}{1-\varphi_1^{s+3} x_i^{-1} w_{2k-1}^{-1}}\right] \frac{1}{1-\varphi_1^{s+2}\varphi_2 w_{2k-1}^{-1}} K_{2k-1}^{(1)} \, ,
\end{align}
where 
\begin{align}\label{eqn:K2k-1(1)}
K_{2k-1}^{(1)} \equiv &\oint \frac{\dd x_{2k-1}}{2 \pi i x_{2k-1}} \frac{1}{1-\varphi_1^{-1}\varphi_2 x_{2k-1}} \frac{1-\varphi_1 x_{2k-1}^{-2}}{1-\varphi_2 x_{2k-1}^{-1}} \sum_{m=1}^{2k-1} \left[\prod_{i=m}^{2k-2}\left(-x_{2k-1}^{-1}x_i\right)\right] \notag \\[5pt]
&\times \left[1 + \varphi_1^{-1}\varphi_2^2 \sum_{n=1}^{m}\prod_{i=n}^{2k-2} \left(-\varphi_1^{-1}\varphi_2 x_i\right) - \varphi_2 x_{2k-1}^{-1} \sum_{n=m}^{2k-1}\prod_{i=n}^{2k-2} \left(-\varphi_1^{-1}\varphi_2 x_i\right)\right]\, ,
\end{align}
is the integral over $x_{2k-1}$. Then, changing variables as $x_{2k-1} \rightarrow x_{2k-1}^{-1}$, the pole structure of the integrand in \cref{eqn:K2k-1(1)} is reduced to a single contributing pole, namely at $x_{2k-1} = \varphi_1^{-1}\varphi_2$ (since we have assumed that $|\varphi_1| > |\varphi_2|$). After integration, we obtain
\begin{align}\label{eqn:K2k-1(1)Result}
K_{2k-1}^{(1)} \equiv \sum_{m=1}^{2k-1} &\left[\prod_{i=m}^{2k-2}\left(-\varphi_1^{-1}\varphi_2 x_i\right)\right] \notag \\[5pt]
\times &\left[1 + \varphi_1^{-1}\varphi_2^2 \sum_{n=1}^{m}\prod_{i=n}^{2k-2} \left(-\varphi_1^{-1}\varphi_2 x_i\right) - \varphi_1^{-1}\varphi_2^2 \sum_{n=m}^{2k-1}\prod_{i=n}^{2k-2} \left(-\varphi_1^{-1}\varphi_2 x_i\right)\right]\, .
\end{align}
The previous result can be further simplified as follows. Defining, in a similar way to \cref{eqn:andefinition}, 
\begin{equation}
b_n \equiv \prod_{i=n}^{2k-2} \left(-\varphi_1^{-1} \varphi_2 x_i \right)\, ,
\label{eqn:bndefinition}
\end{equation}
then the last two terms in the second line of \cref{eqn:K2k-1(1)Result} cancel out because
\begin{equation}
\sum_{m=1}^{2k-1} b_m \left( \sum_{n=1}^{m} b_n - \sum_{n=m}^{2k-1} b_n \right) = \sum_{1\le n\le m\le 2k-1} b_m b_n - \sum_{1\le m\le n\le 2k-1} b_m b_n = 0\, .
\label{eqn:bnsimplify}
\end{equation}
Therefore, plugging the result of the $K_{2k-1}^{(1)}$ integral into $C_{k,s}^{(1)}$ in \cref{eqn:Cks1interstep1}, we obtain
\begin{align}
C_{k,s}^{(1)} \equiv \delta_{|\varphi_1| > |\varphi_2|} &\oint \prod_{i=1}^{2k-2} \frac{\dd x_i}{2 \pi i x_i} \left[\prod_{1 \le i<j \le 2k-2} \frac{1-x_i^{-1} x_j}{1-\varphi_1 x_i^{-1} x_j^{-1}}\right] \notag \\[5pt]
&\times \left[\prod_{i=1}^{2k-2} \frac{1}{1-\varphi_2 x_i^{-1}} \frac{1-\varphi_1^{s+2}x_i w_{2k-1}^{-1}}{1-\varphi_1^{s+3} x_i^{-1} w_{2k-1}^{-1}}\right] \frac{1}{1-\varphi_1^{s+2}\varphi_2 w_{2k-1}^{-1}} \notag \\[5pt]
&\times \left[\sum_{m=1}^{2k-1} \prod_{i=m}^{2k-2}\left(-\varphi_1^{-1}\varphi_2 x_i\right)\right] \, .
\end{align}
Importantly, one can realize that this corresponds to
\begin{equation}
C_{k,s}^{(1)} \equiv C_{k-1,s+1} \,,
\end{equation}
when comparing with \cref{eqn:Cks}.

We now turn into the computation of $C_{k,s}^{(2)}$ in \cref{eqn:Cks2}. Following our previous strategy, we focus on the integration over $x_{2k-1}$. Introducing again $w_{2k-1} \equiv x_{2k-1} w_{2k}$, we can factor out the $x_{2k-1}$ integral, denoted $K_{2k-1}^{(2)}$, as
\begin{align}
C_{k,s}^{(2)} \equiv \delta_{|\varphi_1| > |\varphi_2|} &\oint \prod_{i=1}^{2k-2} \frac{\dd x_i}{2 \pi i x_i} \left[ \prod_{1 \le i<j \le 2k-2} \frac{1-x_i^{-1} x_j}{1-\varphi_1 x_i^{-1} x_j^{-1}} \right] \left[\prod_{i=1}^{2k-2} \frac{1}{1-\varphi_1\varphi_2^{-1} x_i^{-1}}\right] \notag \\[5pt]
&\times \frac{1}{1-\varphi_1^{s+2}\varphi_2 w_{2k-1}^{-1}} K_{2k-1}^{(2)} \,,
\end{align}
where
\begin{align}
K_{2k-1}^{(2)} \equiv &\oint \frac{\dd x_{2k-1}}{2 \pi i x_{2k-1}} \frac{1}{1-\varphi_1\varphi_2^{-1}x_{2k-1}^{-1}} \left[\prod_{i=1}^{2k-2} \frac{1-x_i^{-1}x_{2k-1}}{1-\varphi_1 x_i^{-1} x_{2k-1}^{-1}} \right] \notag \\[5pt]
&\times \frac{1-\varphi_1^{s+1}\varphi_2 x_{2k-1}^2 w_{2k-1}^{-1}}{1-\varphi_1^{s+2}x_{2k-1}w_{2k-1}^{-1}} \left[\prod_{i=1}^{2k-2} \frac{1-\varphi_1^{s+1}\varphi_2 x_i x_{2k-1} w_{2k-1}^{-1}}{1-\varphi_1^{s+2}\varphi_2 x_i^{-1} x_{2k-1} w_{2k-1}^{-1}}\right] \notag \\[5pt]
&\times \left[1 - \varphi_1^{-1}\varphi_2^2 + \varphi_1^{-2}\varphi_2^3 x_{2k-1} \sum_{n=1}^{2k-1}\prod_{i=n}^{2k-2}\left(-\varphi_1^{-1}\varphi_2 x_i\right)\right]\, .
\end{align}
The contributing poles to the integral $K_{2k-1}^{(2)}$ are placed at $x_{2k-1} = \varphi_1 x_m^{-1}$, with $m=1,\cdots,2k-2$. After integration, we obtain
\begin{align}
K_{2k-1}^{(2)} &\equiv \sum_{m=1}^{2k-2} \frac{1-\varphi_1 x_m^{-2}}{1-\varphi_2^{-1} x_m} \frac{1-\varphi_1^{s+2}\varphi_2 w_{2k-1}^{-1}}{1-\varphi_1^{s+3} x_m^{-1} w_{2k-1}^{-1}} \notag \\[5pt]
&\hspace{30pt} \times \left[\prod_{1\le i\le 2k-2}^{i\ne m} \frac{1-\varphi_1 x_i^{-1}x_m^{-1}}{1-x_i^{-1}x_m} \frac{1-\varphi_1^{s+2}\varphi_2 x_i x_m^{-1} w_{2k-1}^{-1}}{1-\varphi_1^{s+3}\varphi_2 x_i^{-1} x_m^{-1} w_{2k-1}^{-1}}\right] \notag \\[5pt]
&\hspace{30pt} \times \left[1 - \varphi_1^{-1}\varphi_2^2 + \varphi_1^{-1}\varphi_2^3\ x_m^{-1} \sum_{n=1}^{2k-1}\prod_{i=n}^{2k-2}\left(-\varphi_1^{-1}\varphi_2 x_i\right)\right] \,,
\end{align}
and thus $C_{k,s}^{(2)}$ reads
\begin{align}
C_{k,s}^{(2)} \equiv \delta_{|\varphi_1| > |\varphi_2|} &\oint \prod_{i=1}^{2k-2} \frac{\dd x_i}{2 \pi i x_i} \left[ \prod_{1 \le i<j \le 2k-2} \frac{1-x_i^{-1} x_j}{1-\varphi_1 x_i^{-1} x_j^{-1}} \right] \left[\prod_{i=1}^{2k-2} \frac{1}{1-\varphi_1\varphi_2^{-1} x_i^{-1}}\right] \notag \\[5pt]
&\times \sum_{m=1}^{2k-2} \frac{1-\varphi_1 x_m^{-2}}{1-\varphi_2^{-1} x_m} \frac{1}{1-\varphi_1^{s+3} x_m^{-1} w_{2k-1}^{-1}} \notag \\[5pt]
&\hspace{30pt} \times \left[\prod_{1\le i\le 2k-2}^{i\ne m} \frac{1-\varphi_1 x_i^{-1}x_m^{-1}}{1-x_i^{-1}x_m} \frac{1-\varphi_1^{s+2}\varphi_2 x_i x_m^{-1} w_{2k-1}^{-1}}{1-\varphi_1^{s+3}\varphi_2 x_i^{-1} x_m^{-1} w_{2k-1}^{-1}}\right] \notag \\[5pt]
&\hspace{30pt} \times \left[1 - \varphi_1^{-1}\varphi_2^2 + \varphi_1^{-1}\varphi_2^3\ x_m^{-1} \sum_{n=1}^{2k-1}\prod_{i=n}^{2k-2}\left(-\varphi_1^{-1}\varphi_2 x_i\right)\right] \,.
\end{align}
At this point, we apply once again the variable change presented in \cref{eqn:xmtox2N-3VariableChange}, yielding in this case
\begin{equation}
\prod_{1\le i < j \le 2k-2} \left(1-x_i^{-1}x_j\right) \longrightarrow \left[\prod_{i=m}^{2k-3} \left(-x_{2k-2}^{-1} x_i\right)\right] \left[\prod_{1\le i < j \le 2k-2} \left(1-x_i^{-1}x_j\right)\right] \,,
\label{eqn:inducedchange2}
\end{equation}
and
\begin{equation}
x_m^{-1} \sum_{n=1}^{2k-1} \prod_{i=n}^{2k-2} \left(-\varphi_1^{-1}\varphi_2 x_i \right) \rightarrow -\varphi_1^{-1}\varphi_2 \sum_{n=1}^m \prod_{i=n}^{2k-3} \left(-\varphi_1^{-1}\varphi_2 x_i\right) + x_{2k-2}^{-1} \sum_{n=m}^{2k-2} \prod_{i=n}^{2k-3} \left(-\varphi_1^{-1}\varphi_2 x_i\right) \,.
\end{equation}
As in previous cases, it is then convenient to decomposed
\begin{equation}
\prod_{1\le i < j \le 2k-2} \frac{1-x_i^{-1}x_j}{1-\varphi_1 x_i^{-1}x_j^{-1}} = \left[\prod_{1\le i < j \le 2k-3} \frac{1-x_i^{-1}x_j}{1-\varphi_1 x_i^{-1}x_j^{-1}}\right] \left[\prod_{i=1}^{2k-3} \frac{1-x_i^{-1} x_{2k-2}}{1-\varphi_1 x_i^{-1} x_{2k-2}^{-1}}\right] \,,
\end{equation}
in order to cancel out repeated factors in the integrand. With all these ingredients, we obtain
\begin{align}
C_{k,s}^{(2)} \equiv \delta_{|\varphi_1| > |\varphi_2|} &\oint \prod_{i=1}^{2k-3} \frac{\dd x_i}{2 \pi i x_i} \left[\prod_{1 \le i<j \le 2k-3} \frac{1-x_i^{-1} x_j}{1-\varphi_1 x_i^{-1} x_j^{-1}}\right] \notag \\[5pt]
&\times \left[\prod_{i=1}^{2k-3} \frac{1}{1-\varphi_1\varphi_2^{-1}x_i^{-1}}\frac{1-\varphi_1^{s+2}\varphi_2 x_i w_{2k-2}^{-1}}{1-\varphi_1^{s+3}\varphi_2 x_i^{-1} w_{2k-2}^{-1}} \right] \frac{1}{1-\varphi_1^{s+3}w_{2k-2}^{-1}} K_{2k-2}^{(2)}\, ,
\end{align}
where we have introduced $w_{2k-2} \equiv x_{2k-2} w_{2k-1}$, and defined 
\begin{align}\label{eqn:K2k-2(2)}
&K_{2k-2}^{(2)} \equiv \oint \frac{\dd x_{2k-2}}{2 \pi i x_{2k-2}}\frac{1}{1-\varphi_1\varphi_2^{-1}x_{2k-2}^{-1}} \frac{1-\varphi_1 x_{2k-2}^{-2}}{1-\varphi_2^{-1}x_{2k-2}} \sum_{m=1}^{2k-2} \left[\prod_{i=m}^{2k-3} \left(-x_{2k-2}^{-1} x_i\right) \right] \notag \\[5pt]
&\times \left[1 - \varphi_1^{-1}\varphi_2^2 - \varphi_1^{-2}\varphi_2^4 \sum_{n=1}^{m}\prod_{i=n}^{2k-3}\left(-\varphi_1^{-1}\varphi_2 x_i\right) + \varphi_1^{-1}\varphi_2^3\, x_{2k-2}^{-1} \sum_{n=m}^{2k-2}\prod_{i=n}^{2k-3} \left(-\varphi_1^{-1}\varphi_2 x_i\right) \right] \,,
\end{align}
that is, the $x_{2k-2}$ integral. Our next step is to perform the integration over $x_{2k-2}$. In order to simplify the pole structure of the integrand in \cref{eqn:K2k-2(2)}, we change variables as $x_{2k-2} \rightarrow x_{2k-2}^{-1}$, leaving us with a single contributing pole at $x_{2k-2} = \varphi_1^{-1}\varphi_2$. This leads to
\begin{align}\label{eqn:K2k-2-2Result}
K_{2k-2}^{(2)} &\equiv \varphi_1^{-1} \varphi_2^2 \sum_{m=1}^{2k-2} \left[\prod_{i=m}^{2k-3} \left(-\varphi_1^{-1}\varphi_2 x_i\right)\right] \notag \\[5pt]
&\times \left[1 - \varphi_1^{-1}\varphi_2^2 - \varphi_1^{-2}\varphi_2^4 \sum_{n=1}^{m}\prod_{i=n}^{2k-3}\left(-\varphi_1^{-1}\varphi_2 x_i\right) + \varphi_1^{-2}\varphi_2^4 \sum_{n=m}^{2k-2}\prod_{i=n}^{2k-3} \left(-\varphi_1^{-1}\varphi_2 x_i\right) \right]\, .
\end{align}
The previous result can be simplified by defining
\begin{equation}
c_n \equiv \prod_{i=n}^{2k-3} \left(-\varphi_1^{-1} \varphi_2 x_i \right)\, ,
\label{eqn:cndefinition}
\end{equation}
and noticing that the last two terms in the second line of \cref{eqn:K2k-2-2Result} cancel out as
\begin{equation}
\sum_{m=1}^{2k-2} c_m \left( \sum_{n=1}^{m} c_n - \sum_{n=m}^{2k-2} c_n \right) = \sum_{1\le n\le m\le 2k-2} c_m c_n - \sum_{1\le m\le n\le 2k-2} c_m c_n = 0\, .
\label{eqn:cnsimplify}
\end{equation}
Hence, the result for $C_{k,s}^{(2)}$ reads
\begin{align}
C_{k,s}^{(2)} &\equiv \varphi_1^{-1}\varphi_2^2 \left(1-\varphi_1^{-1}\varphi_2^2\right) \notag \\[5pt] 
&\times \delta_{|\varphi_1| > |\varphi_2|} \oint \prod_{i=1}^{2k-3} \frac{\dd x_i}{2 \pi i x_i} \left[\prod_{1 \le i<j \le 2k-3} \frac{1-x_i^{-1} x_j}{1-\varphi_1 x_i^{-1} x_j^{-1}}\right] \notag \\[5pt]
&\hspace{60pt} \times \left[\prod_{i=1}^{2k-3} \frac{1}{1-\varphi_1\varphi_2^{-1}x_i^{-1}} \frac{1-\varphi_1^{s+2}\varphi_2 x_i w_{2k-2}^{-1}}{1-\varphi_1^{s+3}\varphi_2 x_i^{-1} w_{2k-2}^{-1}}\right] \frac{1}{1-\varphi_1^{s+3}w_{2k-2}^{-1}} \notag \\[5pt]
&\hspace{60pt} \times \left[\sum_{m=1}^{2k-2} \prod_{i=m}^{2k-3} \left(-\varphi_1^{-1}\varphi_2 x_i\right)\right] \, .
\end{align}
In this form, comparing with \cref{eqn:Bks} and taking into account the result obtained in \cref{eqn:BksResult}, it is easy to check that 
\begin{equation}
C_{k,s}^{(2)} \equiv \varphi_1^{-1}\varphi_2^2 \left(1-\varphi_1^{-1}\varphi_2^2\right) B_{k-2,s+2} = 0\, .
\end{equation}
So, in summary, we end up with
\begin{equation}
C_{k,s} \equiv C_{k,s}^{(1)} + \underbrace{C_{k,s}^{(2)}}_0 = C_{k,s}^{(1)} = C_{k-1,s+1}\, .
\end{equation}
The expression above allows us to establish the relation $C_{k,s} = C_{1,s+k-1}$. The term on the right-hand side can be computed from $C_{1,s}$ by substituting $s \rightarrow s+k-1$. Hence, we just need to calculate the $C_{1,s}$ integral as follows:
\begin{align}
C_{1,s} \equiv \delta_{|\varphi_1| > |\varphi_2|} &\oint \prod_{i=1}^{2} \frac{\dd x_i}{2 \pi i x_i} \left[ \prod_{1 \le i<j \le 2} \frac{1-x_i^{-1} x_j}{1-\varphi_1 x_i^{-1} x_j^{-1}} \right] \left[\prod_{i=1}^{2} \frac{1}{1-\varphi_2 x_i^{-1}} \frac{1-\varphi_1^{s+1} x_i w_3^{-1}}{1-\varphi_1^{s+2}x_i^{-1} w_3^{-1}}\right] \notag \\[5pt]
&\times \frac{1}{1-\varphi_1^{s+1}\varphi_2 w_3^{-1}} \left[\sum_{n=1}^{3}\prod_{i=n}^{2}\left(-\varphi_1^{-1}\varphi_2 x_i\right)\right] \,,
\end{align}
with $w_3 \equiv \left(x_1 x_2\right)^{-1}$. If we factor out the $x_2$ integral
\begin{align}
C_{1,s} \equiv \delta_{|\varphi_1| > |\varphi_2|} &\oint \frac{\dd x_1}{2 \pi i x_1} \frac{1}{1-\varphi_2 x_1^{-1}} \frac{1}{1-\varphi_1^{s+2} x_1} \notag \\[5pt]
\times &\oint \frac{\dd x_2}{2 \pi i x_2} \frac{1-x_1^{-1}x_2}{1-\varphi_1 x_1^{-1} x_2^{-1}} \frac{1}{1-\varphi_2 x_2^{-1}} \notag \\[5pt]
&\hspace{30pt} \times \frac{\left(1-\varphi_1^{s+1}x_1^2 x_2\right)\left(1-\varphi_1^{s+1}x_1 x_2^2\right)\left(1-\varphi_1^{-1}\varphi_2 x_2 + \varphi_1^{-2}\varphi_2^2 x_1 x_2\right)}{\left(1-\varphi_1^{s+2}x_2\right)\left(1-\varphi_1^{s+1}\varphi_2 x_1 x_2\right)}\, ,
\end{align}
it is easy to check that the contributing poles are at $x_2 = \varphi_1 x_1^{-1}$ and $x_2 = \varphi_2$. After integration, we obtain
\begin{align}
&C_{1,s} \equiv \delta_{|\varphi_1| > |\varphi_2|} \frac{1}{1-\varphi_1^{s+2}\varphi_2} \oint \frac{\dd x_1}{2 \pi i x_1} \frac{1}{\varphi_1-\varphi_2 x_1} \notag \\[5pt]
&\times \left[ \varphi_1 \left(1-\varphi_1 x_1^{-2}\right) + \frac{\varphi_2^2\left(1-\varphi_1 x_1^{-2}\right)}{1-\varphi_2 x_1^{-1}} - \frac{\varphi_2 x_1\left(1-\varphi_1^{s+1}\varphi_2 x_1^2\right)\left(1-\varphi_1^{-1}\varphi_2^2 + \varphi_1^{-2}\varphi_2^3 x_1\right)}{1-\varphi_1^{s+2} x_1} \right] \,.
\end{align}
Finally, we address the integration over $x_1$. In this case, there is a contributing pole at $x_1 = 0$ for the first term in the squared brackets; and at $x_1 = 0$ and $x_1 = \varphi_2$, for the second term. Conversely, there are no contributing poles in the third term. The evaluation of the corresponding integrals yields
\begin{equation}
C_{1,s} \equiv \delta_{|\varphi_1| > |\varphi_2|} \frac{1}{1-\varphi_1^{s+2}\varphi_2} \,.
\label{eqn:C1sResult}
\end{equation}
Therefore, replacing $s \rightarrow s+k-1$ in the previous line, we finally obtain
\begin{equation}
C_{k,s} \equiv \delta_{|\varphi_1| > |\varphi_2|} \frac{1}{1-\varphi_1^{s+k+1}\varphi_2} \,.
\label{eqn:CksResult}
\end{equation}

\subsubsection{Fermionic case}
\label{appsubsubsec:ClassIIIFermion}

In this subsection, we compute the Hilbert series in \cref{eqn:HSIIIFermionShort} to derive the results in \cref{eqn:HSIIIFermionDistinct}.

We begin with the $G=SU(2N)$ case in \cref{eqn:HSIIIFermionGShort}. Using the Molien-Weyl formula for Grassmann odd building block fields, we have
\begin{equation}
\HS_\I^{G,\Psi} = \oint \left( \prod_{i=1}^{2N-1} \frac{\dd x_i}{2\pi i x_i} \right)
\left[ \prod_{1 \le i < j \le 2N} \left( 1 - \frac{x_i}{x_j} \right) \right]
\left[ \prod_{i=1}^{2N} \left( 1 + \psi x_i \right) \right] \,.
\end{equation}
We make the following integration variable change (where $r=2N-1$)
\begin{equation}
\mqty(x_1 \\ x_2 \\ \vdots \\ x_{r-1} \\ x_r)
= \mqty(y_1 \\ y_2 \\ \vdots \\ y_{r-1} \\ y_r \equiv \frac{1}{y_1 y_2 \cdots y_{r-1}}) y \,,
\end{equation}
which also leads to
\begin{equation}
x_{2N} = \frac{1}{x_1 \cdots x_r} = y^{-r} \,.
\end{equation}
Consequently, we obtain
\begin{align}
\HS_\I^{G,\Psi} &= \oint \left( \prod_{i=1}^{r-1} \frac{\dd y_i}{2\pi i y_i} \right)
\left[ \prod_{1 \le i < j \le r} \left( 1 - \frac{y_i}{y_j} \right) \right]
\notag\\[5pt]
&\qquad
\times \oint \frac{\dd y}{2\pi i y}
\left[ \prod_{i=1}^r \left( 1 - y_i y^{r+1} \right)  \right]
\left( 1 + \psi y^{-r} \right)
\left[ \prod_{i=1}^r \left( 1 + \psi y_i y \right)  \right] \,.
\label{eqn:HSIIIFermionG1}
\end{align}
The $y$ integral can be computed by evaluating the residue at $y=0$, which is given by the coefficient of the $y^r$ term in the following polynomial:
\begin{align}
&\left[ \prod_{i=1}^r \left( 1 - y_i y^{r+1} \right) \right]
\left( y^r + \psi \right)
\left[ \prod_{i=1}^r \left( 1 + \psi y_i y \right) \right] \Bigg|_{y^r}
\notag\\[5pt]
&\hspace{40pt}
= \left( y^r + \psi \right)
\left[ \prod_{i=1}^r \left( 1 + \psi y_i y \right) \right] \Bigg|_{y^r}
= 1 + \psi \left( \prod_{i=1}^r \psi y_i \right) = 1 + \psi^{2N} \,.
\label{eqn:yrCoeff}
\end{align}
It does not depend on the remaining integration variables $y_i$ in \cref{eqn:HSIIIFermionG1}. Therefore, the remaining integrals can be done trivially; they evaluate to 1, as they are the Haar measure integral of $SU(r)$. We finally obtain \cref{eqn:HSIIIFermionG}:
\begin{equation}
\HS_\I^{G,\Psi} = 1 + \psi^{2N} \,.
\end{equation}

Now let us move on to the $H=SU(2N-1)\times U(1)$ case in \cref{eqn:HSIIIFermionHShort}. Again using the Molien-Weyl formula for Grassmann odd building block fields (and sticking to $r=2N-1$), we have
\begin{align}
\HS_\I^{H,\Psi} &= \oint \left( \prod_{i=1}^{r-1} \frac{\dd x_i}{2\pi i x_i} \right)
\left[ \prod_{1 \le i < j \le r} \left( 1 - \frac{x_i}{x_j} \right) \right]
\oint \frac{\dd x}{2\pi i x}
\left[ \prod_{i=1}^r \left( 1 + \chi_1 x_i x \right) \right]
\left( 1 + \chi_2 x^{-r} \right)
\notag\\[5pt]
&= \oint \left( \prod_{i=1}^{r-1} \frac{\dd x_i}{2\pi i x_i} \right)
\left[ \prod_{1 \le i < j \le r} \left( 1 - \frac{x_i}{x_j} \right) \right]
\oint \frac{\dd x}{2\pi i} \frac{1}{x^{r+1}}
\left[ \prod_{i=1}^r \left( 1 + \chi_1 x_i x \right) \right]
\left( x^r + \chi_2 \right)
\notag\\[5pt]
&= \oint \left( \prod_{i=1}^{r-1} \frac{\dd x_i}{2\pi i x_i} \right)
\left[ \prod_{1 \le i < j \le r} \left( 1 - \frac{x_i}{x_j} \right) \right]
\left( 1 + \chi_1^r \chi_2 \right)
= 1 + \chi_1^{2N-1} \chi_2 \,,
\end{align}
where the $x$ integral has been evaluated much as in \cref{eqn:yrCoeff}, and the remaining integrals in $x_i$ correspond to the Haar measure integral of $SU(r)$, which evaluates to 1. This derives the result in \cref{eqn:HSIIIFermionH}.

\end{allowdisplaybreaks}


\begin{table}[t]
\renewcommand{\arraystretch}{1.3}
\setlength{\arrayrulewidth}{.2mm}
\setlength{\tabcolsep}{1.25em}
\centering
\begin{tabular}{cccc}
\toprule
$G=SO(10)$ Irrep  & Dimension  & $H=SU(5)$  & $H_\Phi=SO(9)$\\
Dynkin label & (name)  & Invariants & Invariants\\
\midrule
\dynkin{1, 0, 0, 0, 0} & \irrep{10} & 0 & 1\\
\dynkin{0, 0, 0, 0, 1} & \irrep{16} & 1 & 0\\
\dynkin{0, 1, 0, 0, 0} & \irrep{45} & 1 & 0\\
\dynkin{2, 0, 0, 0, 0} & \irrep{54} & 0 & 1\\
\dynkin{0, 0, 1, 0, 0} & \irrep{120} & 0 & 0\\
\dynkin{0, 0, 0, 2, 0} & \irrep{126} & 1 & 0\\
\dynkin{1, 0, 0, 1, 0} & \irrep{144} & 0 & 0\\
\dynkin{0, 0, 0, 1, 1} & \irrep{210} & 1 & 0\\
\dynkin{3, 0, 0, 0, 0} & \irrep[1]{210} & 0 & 1\\
\dynkin{1, 1, 0, 0, 0} & \irrep{320} & 0 & 0\\
\dynkin{0, 1, 0, 0, 1} & \irrep{560} & 1 & 0\\
\dynkin{4, 0, 0, 0, 0} & \irrep{660} & 0 & 1\\
\dynkin{0, 0, 0, 3, 0} & \irrep{672} & 1 & 0\\
\dynkin{2, 0, 0, 0, 1} & \irrep{720} & 0 & 0\\
\dynkin{0, 2, 0, 0, 0} & \irrep{770} & 1 & 0\\
\dynkin{1, 0, 1, 0, 0} & \irrep{945} & 0 & 0\\
\dynkin{1, 0, 0, 2, 0} & \irrep{1050} & 0 & 0\\
\dynkin{0, 0, 1, 1, 0} & \irrep{1200} & 0 & 0\\
\dynkin{2, 1, 0, 0, 0} & \irrep{1386} & 0 & 0\\
\dynkin{0, 0, 0, 1, 2} & \irrep{1440} & 1 & 0\\
\dynkin{1, 0, 0, 1, 1} & \irrep{1728} & 0 & 0\\
\dynkin{5, 0, 0, 0, 0} & \irrep{1782} & 0 & 1\\
\dynkin{3, 0, 0, 1, 0} & \irrep{2640} & 0 & 0\\
\dynkin{0, 0, 0, 4, 0} & \irrep{2772} & 1 & 0\\
\dynkin{0, 1, 1, 0, 0} & \irrep{2970} & 0 & 0\\
\dynkin{1, 1, 0, 1, 0} & \irrep{3696} & 0 & 0\\
\dynkin{0, 1, 0, 2, 0} & \irrep[1]{3696} & 1 & 0\\
\dynkin{0, 0, 2, 0, 0} & \irrep{4125} & 0 & 0\\
\dynkin{6, 0, 0, 0, 0} & \irrep{4290} & 0 & 1\\
\bottomrule
\end{tabular}
\caption{\emph{Experimental} check of the friendship relation between $H$ and $H_\Phi$ for the $N=5$ case of Class II all-order accidental symmetries. $G=SO(10)$, $H=SU(5)$, and $H_\Phi=SO(9)$. Non-trivial $G$-irreps are listed in the order of increasing irrep dimension, up to $4290$. None of them contains an invariant under both $H$ and $H_\Phi$, demonstrating their friendship relation.}
\label{tab:SO10Irreps}
\end{table}

\begin{table}[t]
\renewcommand{\arraystretch}{1.3}
\setlength{\arrayrulewidth}{.2mm}
\setlength{\tabcolsep}{1.25em}
\centering
\begin{tabular}{cccc}
\toprule
$G=SO(12)$ Irrep  & Dimension  & $H=SU(6)$  & $H_\Phi=SO(11)$\\
Dynkin label & (name)  & Invariants & Invariants\\
\midrule
\dynkin{1, 0, 0, 0, 0, 0} & \irrep{12} & 0 & 1\\
\dynkin{0, 0, 0, 0, 1, 0} & \irrep{32} & 0 & 0\\
\dynkin{0, 1, 0, 0, 0, 0} & \irrep{66} & 1 & 0\\
\dynkin{2, 0, 0, 0, 0, 0} & \irrep{77} & 0 & 1\\
\dynkin{0, 0, 1, 0, 0, 0} & \irrep{220} & 0 & 0\\
\dynkin{1, 0, 0, 0, 0, 1} & \irrep{352} & 0 & 0\\
\dynkin{3, 0, 0, 0, 0, 0} & \irrep[1]{352} & 0 & 1\\
\dynkin{0, 0, 0, 0, 2, 0} & \irrep{462} & 0 & 0\\
\dynkin{0, 0, 0, 1, 0, 0} & \irrep{495} & 1 & 0\\
\dynkin{1, 1, 0, 0, 0, 0} & \irrep{560} & 0 & 0\\
\dynkin{0, 0, 0, 0, 1, 1} & \irrep{792} & 0 & 0\\
\dynkin{4, 0, 0, 0, 0, 0} & \irrep{1287} & 0 & 1\\
\dynkin{0, 2, 0, 0, 0, 0} & \irrep{1638} & 1 & 0\\
\dynkin{0, 1, 0, 0, 1, 0} & \irrep{1728} & 0 & 0\\
\dynkin{1, 0, 1, 0, 0, 0} & \irrep{2079} & 0 & 0\\
\dynkin{2, 0, 0, 0, 1, 0} & \irrep{2112} & 0 & 0\\
\dynkin{2, 1, 0, 0, 0, 0} & \irrep{2860} & 0 & 0\\
\dynkin{5, 0, 0, 0, 0, 0} & \irrep{4004} & 0 & 1\\
\dynkin{0, 0, 0, 0, 3, 0} & \irrep{4224} & 0 & 0\\
\dynkin{1, 0, 0, 0, 2, 0} & \irrep{4752} & 0 & 0\\
\dynkin{1, 0, 0, 1, 0, 0} & \irrep{4928} & 0 & 0\\
\dynkin{0, 0, 1, 0, 0, 1} & \irrep[1]{4928} & 0 & 0\\
\dynkin{0, 1, 1, 0, 0, 0} & \irrep{8008} & 0 & 0\\
\dynkin{1, 0, 0, 0, 1, 1} & \irrep{8085} & 0 & 0\\
\dynkin{0, 0, 0, 1, 1, 0} & \irrep{8800} & 0 & 0\\
\dynkin{3, 0, 0, 0, 0, 1} & \irrep{9152} & 0 & 0\\
\dynkin{0, 0, 0, 0, 1, 2} & \irrep{9504} & 0 & 0\\
\dynkin{6, 0, 0, 0, 0, 0} & \irrep{11011} & 0 & 1\\
\dynkin{1, 2, 0, 0, 0, 0} & \irrep{11088} & 0 & 0\\
\dynkin{3, 1, 0, 0, 0, 0} & \irrep[1]{11088} & 0 & 0\\
\bottomrule
\end{tabular}
\caption{\emph{Experimental} check of the friendship relation between $H$ and $H_\Phi$ for the $N=6$ case of Class II all-order accidental symmetries. $G=SO(12)$, $H=SU(6)$, and $H_\Phi=SO(11)$. Non-trivial $G$-irreps are listed in the order of increasing irrep dimension, up to $11088$. None of them contains an invariant under both $H$ and $H_\Phi$, demonstrating their friendship relation.}
\label{tab:SO12Irreps}
\end{table}

\begin{table}[t]
\renewcommand{\arraystretch}{1.3}
\setlength{\arrayrulewidth}{.2mm}
\setlength{\tabcolsep}{1.25em}
\centering
\begin{tabular}{cccc}
\toprule
$G=SU(4)$ Irrep  & Dimension  & $H=SU(3)$  & $H_\Phi=Sp(4)$\\
Dynkin label & (name)  & Invariants & Invariants\\
\midrule
\dynkin{1, 0, 0} & \irrep{4} & 1 & 0\\
\dynkin{0, 1, 0} & \irrep{6} & 0 & 1\\
\dynkin{2, 0, 0} & \irrep{10} & 1 & 0\\
\dynkin{1, 0, 1} & \irrep{15} & 1 & 0\\
\dynkin{0, 1, 1} & \irrep{20} & 0 & 0\\
\dynkin{0, 2, 0} & \irrep[1]{20} & 0 & 1\\
\dynkin{0, 0, 3} & \irrep[2]{20} & 1 & 0\\
\dynkin{4, 0, 0} & \irrep{35} & 1 & 0\\
\dynkin{2, 0, 1} & \irrep{36} & 1 & 0\\
\dynkin{2, 1, 0} & \irrep{45} & 0 & 0\\
\dynkin{0, 3, 0} & \irrep{50} & 0 & 1\\
\dynkin{5, 0, 0} & \irrep{56} & 1 & 0\\
\dynkin{1, 2, 0} & \irrep{60} & 0 & 0\\
\dynkin{1, 1, 1} & \irrep{64} & 0 & 0\\
\dynkin{3, 0, 1} & \irrep{70} & 1 & 0\\
\dynkin{2, 0, 2} & \irrep{84} & 1 & 0\\
\dynkin{3, 1, 0} & \irrep[1]{84} & 0 & 0\\
\dynkin{6, 0, 0} & \irrep[2]{84} & 1 & 0\\
\dynkin{0, 4, 0} & \irrep{105} & 0 & 1\\
\dynkin{1, 0, 4} & \irrep{120} & 1 & 0\\
\dynkin{0, 0, 7} & \irrep[1]{120} & 1 & 0\\
\dynkin{2, 2, 0} & \irrep{126} & 0 & 0\\
\dynkin{1, 1, 2} & \irrep{140} & 0 & 0\\
\dynkin{0, 3, 1} & \irrep[1]{140} & 0 & 0\\
\dynkin{4, 1, 0} & \irrep[2]{140} & 0 & 0\\
\dynkin{3, 0, 2} & \irrep{160} & 1 & 0\\
\dynkin{8, 0, 0} & \irrep{165} & 1 & 0\\
\dynkin{1, 2, 1} & \irrep{175} & 0 & 0\\
\dynkin{5, 0, 1} & \irrep{189} & 1 & 0\\
\dynkin{0, 5, 0} & \irrep{196} & 0 & 1\\
\bottomrule
\end{tabular}
\caption{\emph{Experimental} check of the friendship relation between $H$ and $H_\Phi$ for the $N=2$ case of Class III all-order accidental symmetries. $G=SU(4)$, $H=SU(3)$, and $H_\Phi=Sp(4)$. Non-trivial $G$-irreps are listed in the order of increasing irrep dimension, up to $196$. None of them contains an invariant under both $H$ and $H_\Phi$, demonstrating their friendship relation.}
\label{tab:SU4Irreps}
\end{table}

\begin{table}[t]
\renewcommand{\arraystretch}{1.3}
\setlength{\arrayrulewidth}{.2mm}
\setlength{\tabcolsep}{1.25em}
\centering
\begin{tabular}{cccc}
\toprule
$G=SU(6)$ Irrep  & Dimension  & $H=SU(5)$  & $H_\Phi=Sp(6)$\\
Dynkin label & (name)  & Invariants & Invariants\\
\midrule
\dynkin{1, 0, 0, 0, 0} & \irrep{6} & 1 & 0\\
\dynkin{0, 1, 0, 0, 0} & \irrep{15} & 0 & 1\\
\dynkin{0, 0, 1, 0, 0} & \irrep{20} & 0 & 0\\
\dynkin{2, 0, 0, 0, 0} & \irrep{21} & 1 & 0\\
\dynkin{1, 0, 0, 0, 1} & \irrep{35} & 1 & 0\\
\dynkin{3, 0, 0, 0, 0} & \irrep{56} & 1 & 0\\
\dynkin{1, 1, 0, 0, 0} & \irrep{70} & 0 & 0\\
\dynkin{0, 1, 0, 0, 1} & \irrep{84} & 0 & 0\\
\dynkin{0, 0, 1, 0, 1} & \irrep{105} & 0 & 0\\
\dynkin{0, 0, 0, 2, 0} & \irrep[1]{105} & 0 & 1\\
\dynkin{2, 0, 0, 0, 1} & \irrep{120} & 1 & 0\\
\dynkin{0, 0, 0, 0, 4} & \irrep{126} & 1 & 0\\
\dynkin{0, 0, 2, 0, 0} & \irrep{175} & 0 & 0\\
\dynkin{0, 1, 0, 1, 0} & \irrep{189} & 0 & 1\\
\dynkin{0, 0, 1, 1, 0} & \irrep{210} & 0 & 0\\
\dynkin{0, 0, 0, 1, 2} & \irrep[1]{210} & 0 & 0\\
\dynkin{0, 0, 0, 0, 5} & \irrep{252} & 1 & 0\\
\dynkin{2, 0, 0, 1, 0} & \irrep{280} & 0 & 0\\
\dynkin{3, 0, 0, 0, 1} & \irrep{315} & 1 & 0\\
\dynkin{0, 0, 1, 0, 2} & \irrep{336} & 0 & 0\\
\dynkin{1, 1, 0, 0, 1} & \irrep{384} & 0 & 0\\
\dynkin{2, 0, 0, 0, 2} & \irrep{405} & 1 & 0\\
\dynkin{0, 0, 0, 2, 1} & \irrep{420} & 0 & 0\\
\dynkin{6, 0, 0, 0, 0} & \irrep{462} & 1 & 0\\
\dynkin{0, 3, 0, 0, 0} & \irrep{490} & 0 & 1\\
\dynkin{0, 0, 0, 1, 3} & \irrep{504} & 0 & 0\\
\dynkin{1, 0, 1, 0, 1} & \irrep{540} & 0 & 0\\
\dynkin{1, 0, 0, 2, 0} & \irrep{560} & 0 & 0\\
\dynkin{4, 0, 0, 0, 1} & \irrep{700} & 1 & 0\\
\dynkin{3, 0, 0, 1, 0} & \irrep{720} & 0 & 0\\
\bottomrule
\end{tabular}
\caption{\emph{Experimental} check of the friendship relation between $H$ and $H_\Phi$ for the $N=3$ case of Class III all-order accidental symmetries. $G=SU(6)$, $H=SU(5)$, and $H_\Phi=Sp(6)$. Non-trivial $G$-irreps are listed in the order of increasing irrep dimension, up to $720$. None of them contains an invariant under both $H$ and $H_\Phi$, demonstrating their friendship relation.}
\label{tab:SU6Irreps}
\end{table}

\begin{table}[t]
\renewcommand{\arraystretch}{1.3}
\setlength{\arrayrulewidth}{.2mm}
\setlength{\tabcolsep}{1.25em}
\centering
\begin{tabular}{cccc}
\toprule
$G=SU(8)$ Irrep  & Dimension  & $H=SU(7)$  & $H_\Phi=Sp(8)$\\
Dynkin label & (name)  & Invariants & Invariants\\
\midrule
\dynkin{1, 0, 0, 0, 0, 0, 0} & \irrep{8} & 1 & 0\\
\dynkin{0, 1, 0, 0, 0, 0, 0} & \irrep{28} & 0 & 1\\
\dynkin{2, 0, 0, 0, 0, 0, 0} & \irrep{36} & 1 & 0\\
\dynkin{0, 0, 1, 0, 0, 0, 0} & \irrep{56} & 0 & 0\\
\dynkin{1, 0, 0, 0, 0, 0, 1} & \irrep{63} & 1 & 0\\
\dynkin{0, 0, 0, 1, 0, 0, 0} & \irrep{70} & 0 & 1\\
\dynkin{3, 0, 0, 0, 0, 0, 0} & \irrep{120} & 1 & 0\\
\dynkin{1, 1, 0, 0, 0, 0, 0} & \irrep{168} & 0 & 0\\
\dynkin{0, 1, 0, 0, 0, 0, 1} & \irrep{216} & 0 & 0\\
\dynkin{2, 0, 0, 0, 0, 0, 1} & \irrep{280} & 1 & 0\\
\dynkin{4, 0, 0, 0, 0, 0, 0} & \irrep{330} & 1 & 0\\
\dynkin{0, 2, 0, 0, 0, 0, 0} & \irrep{336} & 0 & 1\\
\dynkin{1, 0, 1, 0, 0, 0, 0} & \irrep{378} & 0 & 0\\
\dynkin{0, 0, 1, 0, 0, 0, 1} & \irrep{420} & 0 & 0\\
\dynkin{0, 0, 0, 1, 0, 0, 1} & \irrep{504} & 0 & 0\\
\dynkin{2, 1, 0, 0, 0, 0, 0} & \irrep{630} & 0 & 0\\
\dynkin{0, 1, 0, 0, 0, 1, 0} & \irrep{720} & 0 & 1\\
\dynkin{0, 0, 0, 0, 0, 0, 5} & \irrep{792} & 1 & 0\\
\dynkin{3, 0, 0, 0, 0, 0, 1} & \irrep{924} & 1 & 0\\
\dynkin{2, 0, 0, 0, 0, 1, 0} & \irrep{945} & 0 & 0\\
\dynkin{0, 0, 0, 0, 1, 1, 0} & \irrep{1008} & 0 & 0\\
\dynkin{0, 0, 0, 0, 2, 0, 0} & \irrep{1176} & 0 & 0\\
\dynkin{2, 0, 0, 0, 0, 0, 2} & \irrep{1232} & 1 & 0\\
\dynkin{1, 1, 0, 0, 0, 0, 1} & \irrep{1280} & 0 & 0\\
\dynkin{0, 0, 1, 0, 0, 1, 0} & \irrep{1344} & 0 & 0\\
\dynkin{0, 0, 0, 1, 0, 1, 0} & \irrep{1512} & 0 & 1\\
\dynkin{0, 0, 0, 0, 1, 0, 2} & \irrep[1]{1512} & 0 & 0\\
\dynkin{0, 0, 0, 0, 0, 2, 1} & \irrep{1680} & 0 & 0\\
\dynkin{0, 0, 0, 0, 0, 0, 6} & \irrep{1716} & 1 & 0\\
\dynkin{0, 0, 0, 2, 0, 0, 0} & \irrep{1764} & 0 & 1\\
\bottomrule
\end{tabular}
\caption{\emph{Experimental} check of the friendship relation between $H$ and $H_\Phi$ for the $N=4$ case of Class III all-order accidental symmetries. $G=SU(8)$, $H=SU(7)$, and $H_\Phi=Sp(8)$. Non-trivial $G$-irreps are listed in the order of increasing irrep dimension, up to $1764$. None of them contains an invariant under both $H$ and $H_\Phi$, demonstrating their friendship relation.}
\label{tab:SU8Irreps}
\end{table}

\begin{table}[t]
\renewcommand{\arraystretch}{1.3}
\setlength{\arrayrulewidth}{.2mm}
\setlength{\tabcolsep}{1.25em}
\centering
\begin{tabular}{cccc}
\toprule
$G=SU(6)$ Irrep  & Dimension  & $H=SU(3)\times SU(2)$  & $H_\Phi=SU(5)$\\
Dynkin label & (name)  & Invariants & Invariants\\
\midrule
\dynkin{1, 0, 0, 0, 0} & \irrep{6}      & 0 & 1\\
\dynkin{0, 1, 0, 0, 0} & \irrep{15}     & 0 & 0\\
\dynkin{0, 0, 1, 0, 0} & \irrep{20}     & 0 & 0\\
\dynkin{2, 0, 0, 0, 0} & \irrep{21}     & 0 & 1\\
\dynkin{1, 0, 0, 0, 1} & \irrep{35}     & 0 & 1\\
\dynkin{3, 0, 0, 0, 0} & \irrep{56}     & 0 & 1\\
\dynkin{1, 1, 0, 0, 0} & \irrep{70}     & 0 & 0\\
\dynkin{0, 1, 0, 0, 1} & \irrep{84}     & 0 & 0\\
\dynkin{0, 0, 1, 0, 1} & \irrep{105}    & 0 & 0\\
\dynkin{0, 0, 0, 2, 0} & \irrep[1]{105} & 0 & 0\\
\dynkin{2, 0, 0, 0, 1} & \irrep{120}    & 0 & 1\\
\dynkin{0, 0, 0, 0, 4} & \irrep{126}    & 0 & 1\\
\dynkin{0, 0, 2, 0, 0} & \irrep{175}    & 0 & 0\\
\dynkin{0, 1, 0, 1, 0} & \irrep{189}    & 1 & 0\\
\dynkin{0, 0, 1, 1, 0} & \irrep{210}    & 0 & 0\\
\dynkin{0, 0, 0, 1, 2} & \irrep[1]{210} & 0 & 0\\
\dynkin{0, 0, 0, 0, 5} & \irrep{252}    & 0 & 1\\
\dynkin{2, 0, 0, 1, 0} & \irrep{280}    & 0 & 0\\
\dynkin{3, 0, 0, 0, 1} & \irrep{315}    & 0 & 1\\
\dynkin{0, 0, 1, 0, 2} & \irrep{336}    & 0 & 0\\
\dynkin{1, 1, 0, 0, 1} & \irrep{384}    & 0 & 0\\
\highlight{\dynkin{2, 0, 0, 0, 2}} & \highlight{\irrep{405}} & \highlight{1} & \highlight{1}\\
\dynkin{0, 0, 0, 2, 1} & \irrep{420}    & 0 & 0\\
\dynkin{6, 0, 0, 0, 0} & \irrep{462}    & 0 & 1\\
\dynkin{0, 3, 0, 0, 0} & \irrep{490}    & 1 & 0\\
\dynkin{0, 0, 0, 1, 3} & \irrep{504}    & 0 & 0\\
\dynkin{1, 0, 1, 0, 1} & \irrep{540}    & 0 & 0\\
\dynkin{1, 0, 0, 2, 0} & \irrep{560}    & 0 & 0\\
\dynkin{4, 0, 0, 0, 1} & \irrep{700}    & 0 & 1\\
\bottomrule
\end{tabular}
\caption{\emph{Experimental} check of the friendship relation between $H$ and $H_\Phi$ for the finite-order accidental symmetry example discussed in \cref{subsec:Nonderivative,subsubsec:VerifyRectangular}. $G=SU(6)$, $H=SU(3)\times SU(2)$, and $H_\Phi=SU(5)$. Non-trivial $G$-irreps are listed in the order of increasing irrep dimension. None of them contains an invariant under both $H$ and $H_\Phi$ until the \highlight{highlighted} irrep \highlight{(20002)} with irrep dimension $405$. This is the lowest dimension $G$-irrep that breaks the friendship between $H$ and $H_\Phi$.}
\label{tab:SU6SU3SU2Irreps}
\end{table}

\addcontentsline{toc}{section}{\protect\numberline{}References}%
\bibliographystyle{JHEP}
\bibliography{ref}
\end{spacing}

\end{document}